\documentclass[prl,aps,superscriptaddress,footinbib,twocolumn]{revtex4-2}
\usepackage[latin9]{inputenc}
\usepackage{color}
\usepackage{amsmath}
\usepackage{amssymb}
\usepackage{stmaryrd}
\usepackage{graphicx}
\usepackage[normalem]{ulem}
\usepackage{bm}
\usepackage{bbm}

\usepackage[unicode=true,bookmarks=true,bookmarksnumbered=false,bookmarksopen=false,breaklinks=false,pdfborder={0 0 1},backref=false,colorlinks=true]{hyperref}

\hypersetup{
 linkcolor=magenta, urlcolor=blue, citecolor=blue, pdfstartview={FitH}, unicode=true}

\makeatletter

\usepackage{amsfonts}\usepackage{tabularx}\usepackage{dcolumn}\usepackage{bm}\usepackage{graphicx}\usepackage{epstopdf}
\usepackage{times}
\def\ket#1{\left|#1\right\rangle}

\def\braket#1{\left\langle#1\right\rangle}

\newcommand{\beginExtendedData}{%
    \newcounter{ExFig}
    \makeatletter
    \@addtoreset{figure}{ExFig}
    \makeatother
    \stepcounter{ExFig}
    \renewcommand{\figurename}{Extended Data Fig.}
   }
\usepackage{siunitx}
\usepackage{braket}
\makeatother

\begin{document}
\title{Observation of topological prethermal strong zero modes}
\author{Feitong Jin}\thanks{These authors contributed equally}
\affiliation{School of Physics, ZJU-Hangzhou Global Scientific and Technological Innovation Center, and Zhejiang Key Laboratory of Micro-nano Quantum Chips and Quantum Control, Hangzhou, China}

\author{Si Jiang}\thanks{These authors contributed equally}
\affiliation{Center for Quantum Information, IIIS, Tsinghua University, Beijing 100084, China}
\affiliation{Shanghai Qi Zhi Institute, Shanghai 200232, China}

\author{Xuhao Zhu}\thanks{These authors contributed equally}
\author{Zehang Bao}
\author{Fanhao Shen}
\author{Ke Wang}
\author{Zitian Zhu}
\author{Shibo Xu}
\author{Zixuan Song}
\author{Jiachen Chen}
\author{Ziqi Tan}

\author{Yaozu Wu}

\author{Chuanyu Zhang}
\author{Yu Gao}

\author{Ning Wang}
\author{Yiren Zou}
\author{Aosai Zhang}
\author{Tingting Li}

\author{Jiarun Zhong}
\author{Zhengyi Cui}

\author{Yihang Han}
\author{Yiyang He}
\author{Han Wang}
\author{Jianan Yang}
\author{Yanzhe Wang}
\author{Jiayuan Shen}
\author{Gongyu Liu}

\author{Jinfeng Deng}
\author{Hang Dong}
\author{Pengfei Zhang}
\affiliation{School of Physics, ZJU-Hangzhou Global Scientific and Technological Innovation Center, and Zhejiang Key Laboratory of Micro-nano Quantum Chips and Quantum Control, Hangzhou, China}

\author{Weikang Li}
\affiliation{Center for Quantum Information, IIIS, Tsinghua University, Beijing 100084, China}
\affiliation{ Instituut-Lorentz, Universiteit Leiden, P.O. Box 9506, 2300 RA Leiden, The Netherlands}

\author{Dong Yuan}
\affiliation{Center for Quantum Information, IIIS, Tsinghua University, Beijing 100084, China}

\author{Zhide Lu}
\affiliation{Shanghai Qi Zhi Institute, Shanghai 200232, China}

\author{Zheng-Zhi Sun}
\affiliation{Center for Quantum Information, IIIS, Tsinghua University, Beijing 100084, China}
\affiliation{Hefei National Laboratory, Hefei 230088, China}

\author{Hekang Li}
\affiliation{School of Physics, ZJU-Hangzhou Global Scientific and Technological Innovation Center, and Zhejiang Key Laboratory of Micro-nano Quantum Chips and Quantum Control, Hangzhou, China}
\author{Junxiang Zhang}
\author{Chao Song}
\affiliation{School of Physics, ZJU-Hangzhou Global Scientific and Technological Innovation Center, and Zhejiang Key Laboratory of Micro-nano Quantum Chips and Quantum Control, Hangzhou, China}
\author{Zhen Wang}
\affiliation{School of Physics, ZJU-Hangzhou Global Scientific and Technological Innovation Center, and Zhejiang Key Laboratory of Micro-nano Quantum Chips and Quantum Control, Hangzhou, China}
\affiliation{Hefei National Laboratory, Hefei 230088, China}
\author{Qiujiang Guo}
\email{qguo@zju.edu.cn}
\affiliation{School of Physics, ZJU-Hangzhou Global Scientific and Technological Innovation Center, and Zhejiang Key Laboratory of Micro-nano Quantum Chips and Quantum Control, Hangzhou, China}
\affiliation{Hefei National Laboratory, Hefei 230088, China}

\author{Francisco Machado}
\affiliation{ITAMP, Harvard-Smithsonian Center for Astrophysics, Cambridge, Massachusetts, 02138, USA}
\affiliation{Department of Physics, Harvard University, Cambridge 02138 MA, USA}

\author{Jack Kemp}
\affiliation{Department of Physics, Harvard University, Cambridge 02138 MA, USA}

\author{Thomas Iadecola}
\affiliation{Department of Physics and Astronomy, Iowa State University, Ames, Iowa 50011, USA}
\affiliation{Ames National Laboratory, Ames, Iowa 50011, USA}

\author{Norman Y. Yao}
\affiliation{Department of Physics, Harvard University, Cambridge 02138 MA, USA}

\author{H. Wang}
\email{hhwang@zju.edu.cn}
\affiliation{School of Physics, ZJU-Hangzhou Global Scientific and Technological Innovation Center, and Zhejiang Key Laboratory of Micro-nano Quantum Chips and Quantum Control, Hangzhou, China}
\affiliation{Hefei National Laboratory, Hefei 230088, China}

\author{Dong-Ling Deng}
\email{dldeng@tsinghua.edu.cn}
\affiliation{Center for Quantum Information, IIIS, Tsinghua University, Beijing 100084, China}
\affiliation{Shanghai Qi Zhi Institute, Shanghai 200232, China}
\affiliation{Hefei National Laboratory, Hefei 230088, China}

\begin{abstract}
{
Symmetry-protected topological phases cannot be described by any local order parameter and are beyond the conventional symmetry-breaking paradigm for understanding quantum matter. They are characterized by topological boundary states robust against perturbations that respect the protecting symmetry. In a clean system without disorder, these edge
modes typically only occur for the ground states of systems
with a bulk energy gap and would not survive at finite temperatures due to mobile thermal excitations. Here, we report the observation of a distinct type of topological edge modes, which are protected by emergent symmetries and persist even up to infinite temperature, with an array of ${100}$ programmable superconducting qubits. In particular, through digital quantum simulation of the dynamics of a one-dimensional disorder-free ``cluster'' Hamiltonian, 
we observe robust long-lived topological edge modes over up to ${30}$ cycles at a wide range of temperatures.  By monitoring the propagation of thermal excitations, we show that despite the free mobility of these excitations, their interactions with the edge modes are substantially suppressed in the dimerized regime due to an emergent U(1)$\times$U(1) symmetry, resulting in an unusually prolonged lifetime of the topological edge modes even at infinite temperature. In addition, we exploit these topological edge modes as logical qubits and prepare a logical Bell state, which exhibits persistent coherence in the dimerized and off-resonant regime, despite the system being disorder-free and far from its ground state.
Our results establish a viable digital simulation approach to experimentally exploring a variety of finite-temperature topological phases and demonstrate a potential route to construct long-lived robust boundary qubits that survive to infinite temperature in disorder-free systems. 
}
\end{abstract}

\maketitle

\noindent
Symmetry and topology are fundamental to characterizing quantum phases of matter ~\cite{Chiu2016Classification,Wen2017RMP}. Their interplay gives rise to a rich variety of exotic phases~\cite{Chen2012Symmetry, Senthil2015Symmetry, Chiu2016Classification,Wen2017RMP} that cannot be described by the traditional Landau-Ginzburg symmetry-breaking paradigm \cite{Landau2013Statistical}. A prominent example is that of symmetry-protected topological (SPT) phases, which yield nonlocal order parameters but feature topological boundary states that are robust against local perturbations respecting the protecting symmetry \cite{Pollmann2012Symmetry,Fidkowski2011Topological,Chen2012Symmetry, Senthil2015Symmetry, Chiu2016Classification,Wen2017RMP,Ma2023Average}. Such robust boundary states provide an intriguing opportunity to store and process quantum information in a perturbation-resilient fashion \cite{Bravyi2010Topological}. In a clean system without disorder, these edge modes typically only occur at zero temperature, for the ground states of systems with a bulk energy gap. At finite temperature, they would interact strongly with thermal excitations in the bulk and decohere rapidly.  Realizing robust topological edge modes at finite temperatures is crucial in understanding ``hot'' SPT phases of matter and holds potential applications in building a noise-resilient quantum memory ~\cite{Brown2016Quantum}.

\begin{figure*}[ht]
    \centering\includegraphics[width=1\linewidth]{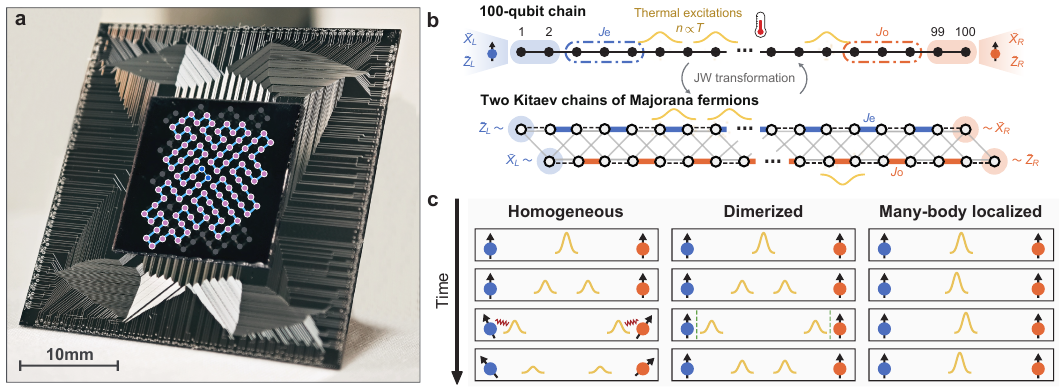}
    \caption{\textbf{125-qubit quantum processor and theoretical model.}
        {\bf a}, Photograph of the superconducting quantum processor. The $100$ qubits used to construct the one-dimensional ($1$D) chain are highlighted with circles, with two edge qubits marked in dark blue and the other qubits in pink. The couplers actively used to connect these qubits are highlighted with light blue lines.
        {\bf b}, Schematic illustration of the $1$D Hamiltonian in Eq.~(\ref{eq:1}) and its representation in the Majorana fermion picture. Three-body stabilizers $\{\sigma_{i-1}^z\sigma_i^x\sigma_{i+1}^z\}$ at even and odd sites, exemplified by blue and orange dashed frames, can have different strengths denoted by $J_{\rm e}$ and $J_{\rm o}$, receptively. Two spin-$1/2$ edge modes are situated at the two ends of the chain, labeled by $\Tilde{Z}_{\rm L}, \Tilde{X}_{\rm L}$ for the left edge and $\Tilde{Z}_{\rm R}, \Tilde{X}_{\rm R}$ for the right. 
        At finite temperatures, thermal excitations (yellow wave packets) emerge in the bulk of the chain, which flip the values of stabilizers. After the Jordan-Wigner (JW) transformation, the $1$D qubit chain is mapped into two Kitaev chains, where the upper (lower) chain inherits the even-site (odd-site) interaction strength $J_{\rm e}$ ($J_{\rm o}$), labeled by thick blue (orange) lines. Two edge modes are transformed into four Majorana fermions at the ends of two chains. Single-qubit $\sigma_i^x$ terms, represented by horizontal black dashed lines, become couplings on onsite Majorana pairs,
        and two-qubit $\sigma_i^x\sigma_{i+1}^x$ interactions, represented by gray lines, bridging the two chains.
        {\bf c}, Schematic of thermal excitation dynamics and their interactions with edges. Thermal excitations (yellow wave packets) can propagate through the chain under perturbations. In the homogeneous regime (left panel, $J_{\rm o}=J_{\rm e}$), edge-bulk interactions at the boundaries decohere and ruin the edge modes. Whereas, in the dimerized regime (middle panel, $J_{\rm o}\neq J_{\rm e}$), such interactions are markedly suppressed, resulting in long-lived robust edge modes at up to infinite temperature. In the many-body localized scenario (right panel), transport is forbidden and thermal excitations remain localized without influencing the boundaries.
}\label{fig:1}
\end{figure*}

A popular strategy to stabilize topological edge modes at finite temperature involves adding strong disorder so as to make the system many-body localized~\cite{Nandkishore2014Many-Body, Kjall2014Many-Body, Abanin2019Colloquium}. In such a scenario, the disorder localizes bulk thermal excitations, preventing them from scattering with and decohering the topological edge modes ~\cite{Huse2013Localization, Chandran2014Many-body, Bahri2015Localization}. Despite exciting progress along this direction \cite{Schreiber2015Observation, Choi2016Exploring, Smith2016Many}, the stability of many-body localization is still under active debate ~\cite{Morningstar2022PRB, Hyunsoo2023PRL, Long2023PRL, Leonard2023NP}, which overshadows the long-time behavior of localization-based SPT phases at finite temperatures. In addition, the presence of strong disorder slows down equilibration, making it difficult to unambiguously distinguish in experiment genuine late-time dynamics from early-time transient behavior \cite{Schreiber2015Observation, Choi2016Exploring, Smith2016Many}. 
An alternative strategy is to suppress the interactions between bulk excitations and edge modes by emergent symmetries, rather than localization \cite{Else2017Prethermal, Parker2019Topologically, Kemp2020Symmetry}. 
In this case, the system can be disorder-free and bulk excitations are mobile, but the emergent symmetries give rise to approximately conserved edge states that are effectively decoupled from the bulk. 
Such topological edge states form so-called prethermal strong zero modes featuring exponentially long coherence times even at infinite temperature \cite{Fendley2012Parafermionic, Fendley2016Strong, Kemp2017Long, Else2017Prethermal, Parker2019Topologically, Kemp2020Symmetry}. 
Pioneering experiments have observed signatures of topological edge modes at up to infinite temperature in periodically driven systems with strong disorder ~\cite{Zhang2022Digital, Dumitrescu2022Dynamical, Mi2022Noise}. Yet, observation of long-lived finite-temperature topological edge modes protected by emergent symmetries in disorder-free systems remains a notable challenge and has evaded experiments so far.

Here, we report such an observation with a newly developed high-performance $125$-qubit superconducting quantum processor (Fig.~\ref{fig:1}). We select $100$ neighboring qubits arranged in a one-dimensional (1D) chain (Fig.~\ref{fig:1}a), featuring median fidelities of simultaneous single- and two-qubit gates about 0.9995 and 0.995, respectively. This enables us to successfully implement the dynamics of a prototypical SPT Hamiltonian (Fig.~\ref{fig:1}b) in different regimes with quantum circuits of depth exceeding $270$ and gate counts above $17,000$. We prepare the system in different initial states with different energies, which correspond to different effective temperatures, and then evolve it under the SPT Hamiltonian with varying parameters. 
We observe that in the presence of thermal excitations, the lifetime of edge states is greatly enhanced in the dimerized regime with spatially periodic modulated couplings, in stark contrast to the fast decay in the homogeneous case. 
To reveal the underlying mechanism, we further measure the site-resolved dynamics of mobile excitations. 
Strikingly, despite thermal excitations moving back and forth, an approximate U(1)$\times$U(1) symmetry emerges in the dimerized case that suppresses the bulk-edge interactions, in sharp contrast to the many-body localized scenario where bulk excitations are localized by strong disorder (Fig.~\ref{fig:1}c). 
This prethermal mechanism is further confirmed by measuring the energy spectrum, where an extra gap gradually opens as the chain dimerizes, 
explaining the origin of the emergent symmetry. In addition, we prepare a logical Bell state, which is encoded by the topological edge modes, and demonstrate substantially prolonged coherence time at finite temperature in the dimerized and off-resonant regime. This shows that the edge modes open potential applications towards building a noise-resilient finite-temperature quantum memory.

\begin{figure*}[ht]
    \centering
    \includegraphics[width=1\linewidth]{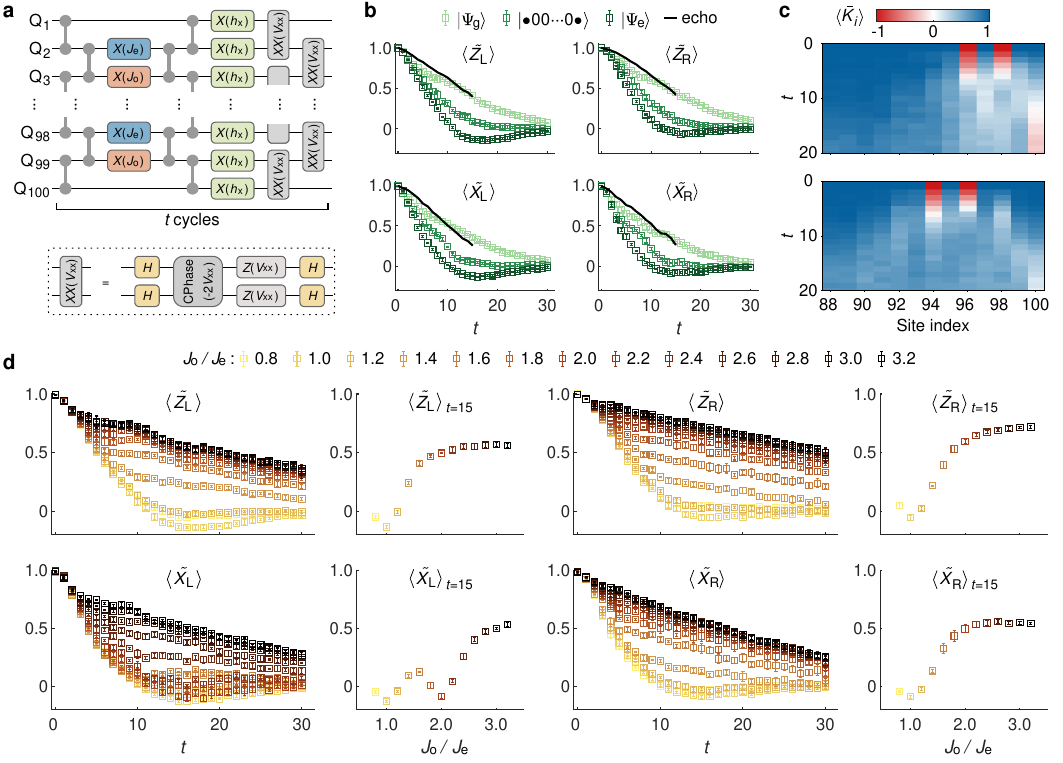}
    \caption{\textbf{Robust topological edge modes at up to infinite temperature.} 
        {\bf a}, Schematic of the experimental circuit for implementing $U(\delta t)$, which emulates a single-step evolution under the Hamiltonian in Eq.~(\ref{eq:1}). The system is initialized in either the cluster ground-state manifold $\{\ket{\Psi_{\rm g}}\}$ (thermal excitation number $n=0$, corresponding to zero temperature), the cluster excited-state manifold $\{\ket{\Psi_{\rm e}}\}$ ($n \neq 0$, finite temperature), or the product states $\ket{\bullet 00...0 \bullet}$ (infinite temperature), and then evolved with $U(\delta t)$ for $t$ cycles. Here, $J_{\rm o}$, $J_{\rm e}$, and $h_{x}$ are parameterized into the rotation angle $\theta$ around $x$-axis of the Bloch sphere [$X(\theta)$]. $V_{xx}$ is encoded in a combination of controlled-phase [CPhase($-2V_{xx}$)] and $Z$ phase gates [$Z(V_{xx}$)].
        {\bf b}, Measured time dynamics for edge operators $\Tilde{\braket{Z}}$ (upper panel) and $\Tilde{\braket{X}}$ (lower panel) in the homogeneous case $(J_{\rm o}=J_{\rm e}=\pi/5)$. Black lines show the results of echo circuits, which estimate the decay caused by circuit errors due to experimental imperfections.
        {\bf c}, Measured site-resolved dynamics of normalized expectation value $\bar{\braket{{K}_i}}$ for bulk stabilizers $\{\sigma_{i-1}^z\sigma_i^x\sigma_{i+1}^z\}$ and edge operator $\Tilde{X}_{\rm R}$ in the homogeneous case ($J_{\rm o}=J_{\rm e}=\pi/5$) near the right edge. The positions of nearest excitations to the right edge are $\{Q_{96}, Q_{98}\}$ (top panel) and $\{Q_{94}, Q_{96}\}$ (bottom panel).
        {\bf d}, Measured time dynamics of edge modes with fixed $J_{\rm e}=\pi/5$ and varying $J_{\rm o}$. 
        A resonant process involving $\tilde{X}_{\rm L}$ happens at $J_{\rm o}/J_{\rm e}=2$.
        Error bars in {\bf b} and {\bf d} stem from five repetitions of measurements.
        }
    \label{fig:2}
\end{figure*}

\begin{figure*}[ht]
    \centering
    \includegraphics[width=1\linewidth]{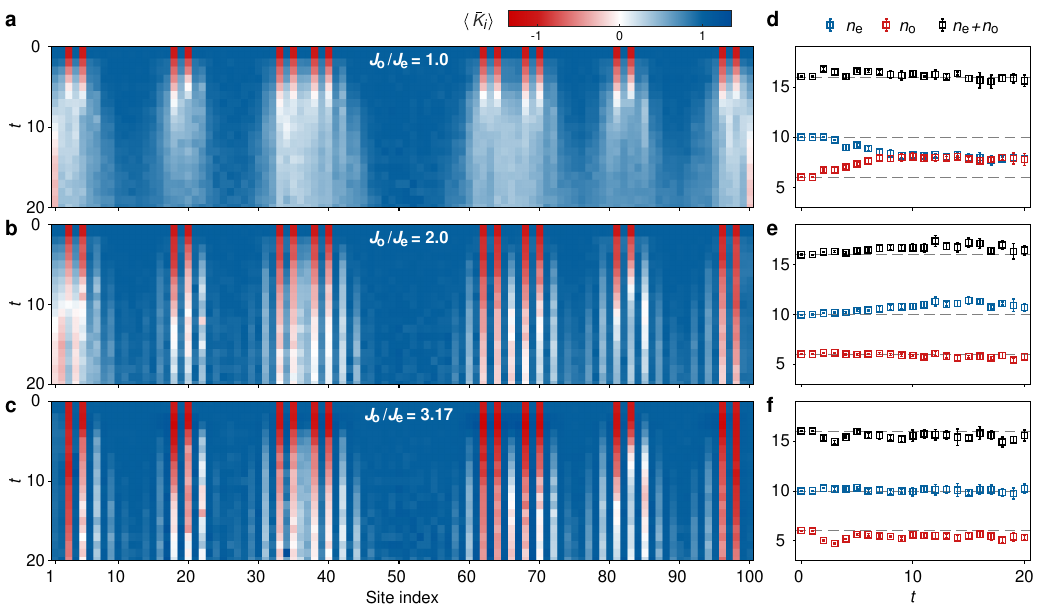}
    \caption{\textbf{Excitation dynamics and the emergent U(1)$\times$U(1) symmetry.} 
        {\bf a-c}, Measured site-resolved dynamics of normalized expectation value $\bar{\braket{{K}_i}}$ for the homogeneous (top panel, $J_{\rm o}=J_{\rm e}=\pi/5$), the dimerized but resonant (middle panel, $J_{\rm o}=2J_{\rm e}=2\pi/5$), and the dimerized and off-resonant (bottom panel, $J_{\rm o}=3.17J_{\rm e}=3.17\pi/5$) cases.
        {\bf d-f}, Measured time dynamics of the total excitation number $n$, and of the excitation number at even ($n_{\rm e}$) and odd ($n_{\rm o}$) sites, which are extracted from {\bf a-c}. In the homogeneous case (top), the values of $n_{\rm e}$ and $n_{\rm o}$ gradually converge, yet their sum remains approximately constant, reflecting the U(1) symmetry on the total excitation number $n$ in the bulk. In contrast, in the dimerized and off-resonant case (bottom), $n_{\rm e}$ and $n_{\rm o}$ are conserved independently, signifying an enlarged U(1)$\times$U(1) symmetry. In the dimerized but resonant case (middle), the exchange of excitations between two  Kitaev chains, which happens near the left edge, can be observed through the decrease of $n_{\rm o}$ and increase of $n_{\rm e}$. Gray dashed lines represent the initial values of $n_{\rm o}=6, n_{\rm e}=10$, and $n=16$. Error bars stem from five repetitions of measurements.
    }
    \label{fig:3}
\end{figure*}

\vspace{2mm}
\noindent\textbf{\large{}Hamiltonian and its implementation}\\
\noindent We consider a $1$D Hamiltonian with an even number of qubits denoted by $N$ (Fig.~\ref{fig:1}b),
\begin{eqnarray}
\label{eq:1}  
    && H = H_{0} + H_{1}, \nonumber  \\
    && H_{0} = - J_{\rm e} \sum_{i=1}^{\frac{N}{2}-1} \sigma_{2i-1}^z \sigma^x_{2i}\sigma^z_{2i+1} -J_{\rm o}\sum_{i=1}^{\frac{N}{2}-1} \sigma^z_{2i}\sigma^x_{2i+1}\sigma^z_{2i+2}, \nonumber \\
    && H_{1} =  h_x \sum_{i=1}^{N}\sigma^x_i+  V_{xx}\sum_{i=1}^{N-1} \sigma^x_i\sigma^x_{i+1},
\end{eqnarray}
where $\hbar$ is set to $1$, $\sigma^{x,z}_i$ are Pauli operators acting on the {\it i}-th qubit, $J_{\rm e}$ ($J_{\rm o}$) denotes the strength of three-body stabilizer terms centered around even (odd) sites, and $h_x$ and $V_{xx}$ are parameters characterizing the transverse field and interaction strength, respectively.
In the limit of $h_x, V_{xx}\to0$, $H=H_{0}$ and its eigenstates are the $1$D cluster stabilizer eigenstates~\cite{Briegel2001Persistent}. 
At zero temperature, the system will remain in the ground-state manifold $\{\ket{\Psi_{\rm g}}\}$ where all stabilizers $\{\sigma^z_{i-1} \sigma^x_{i}\sigma^z_{i+1} \}$ equal $+1$.
This ground-state degeneracy is four-fold, hosting two nontrivial spin-$1/2$ topological edge modes protected by a $\mathbb{Z}_2\times \mathbb{Z}_2$ symmetry, where each $\mathbb{Z}_2$ is generated by the products of $\sigma^x_i$ over even or odd sites (Supplementary Section 1A).
These SPT edge modes are characterized by logical operators $\Tilde{X}_{\rm L} = \sigma_1^x\sigma_2^z, \Tilde{Z}_{\rm L} = \sigma_1^z$ for the left edge and $\Tilde{X}_{\rm R}=\sigma_{N-1}^z\sigma_N^x$, $\Tilde{Z}_{\rm R}=\sigma_N^z$ for the right (Fig.~\ref{fig:1}b). 
As the temperature increases, the system occupies more excited states where some of the stabilizers are flipped to $-1$.
These local, thermal excitations remain stationary under $H_0$, however, when interactions $H_1$ are present, they can propagate along the system, reach the boundaries, and decohere the edge states.

\begin{figure*}[ht]
    \centering
    \includegraphics[width=1\linewidth]{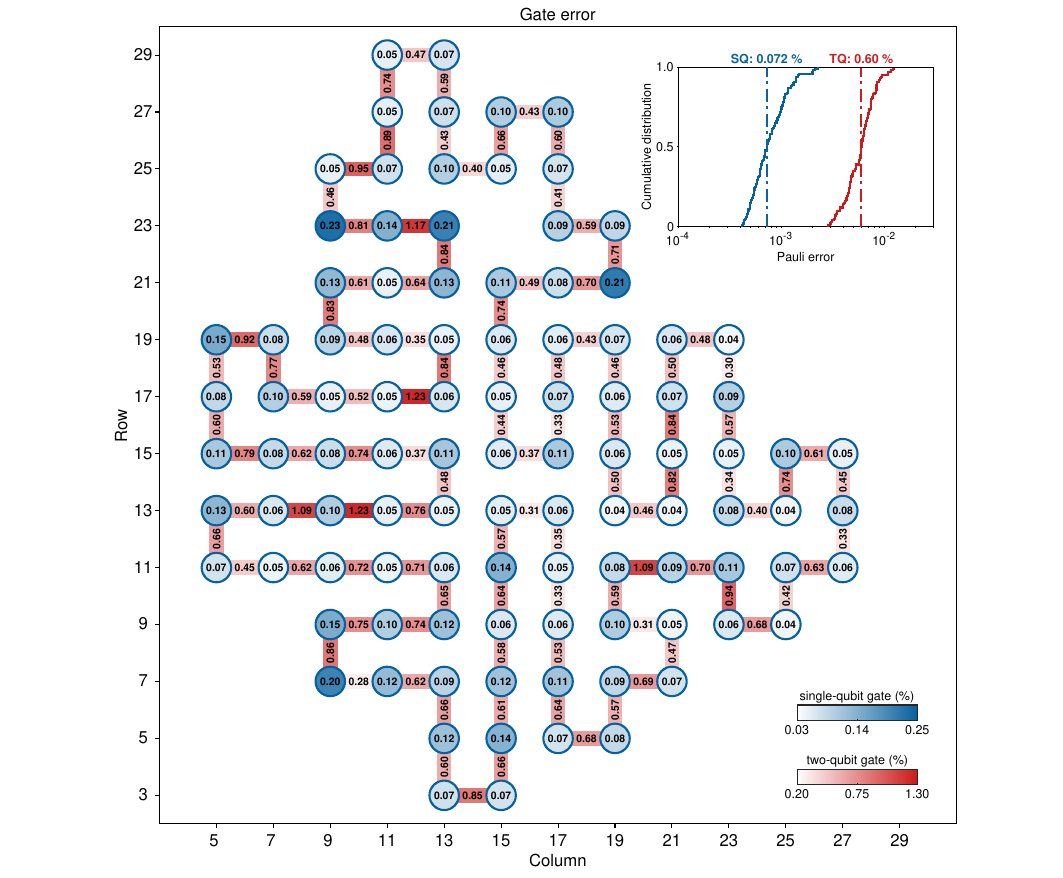}
    \caption{{\bf Pauli errors of single-qubit and two-qubit gates.} 
    Gate errors are benchmarked with simultaneous cross-entropy benchmarking (XEB). Errors of single-qubit gates (blue circles) are obtained by running single-qubit XEB sequences for all $100$ qubits simultaneously, while errors of two-qubit gates [red bars, including CZ and CPhase($-0.4$)] are averaged over all the two-qubit layers used in our experiments. For each two-qubit layer, we run two-qubit XEB sequences simultaneously for all the two-qubit gates in this layer. The maximum number of parallel two-qubit gates in our experiments is $50$. The inset shows the cumulative distribution of gate errors, with the dashed lines indicating the median values.}
    \label{fig:ex1}
\end{figure*}

We emulate many-body dynamics under the Hamiltonian (\ref{eq:1}) with $N=100$ superconducting qubits by means of first-order Trotter decomposition $U(\delta t) = U_1(\delta t)U_0(\delta t)$, where $U_1(\delta t) = {\rm e}^{-{\rm i} H_1 \delta t}$ and $ U_0(\delta t)={\rm e}^{-{\rm i} H_0 \delta t}$.
Implementing $U(\delta t)$ is challenging because three-body interactions do not arise naturally in superconducting platforms, leading to large circuit depths.
As shown in Fig.~\ref{fig:2}a, even a single time step $U(\delta t)$ demands a deep circuit with six layers of two-qubit gates and three layers of single-qubit gates, corresponding to a $288$-ns running time (Supplementary Sections 2B and 2C). Therefore, the high performance of the quantum processor (Supplementary Section 2A) is crucial to observing coherent dynamics under $U$ before the accumulated experimental errors dominate. In our experiments, we achieve low-error quantum gates at the $100$-qubit scale, with median simultaneous single- and two-qubit gate fidelities about  $0.9995$ and  $0.995$, respectively (Extended Data Fig.~\ref{fig:ex1}).
We set $\delta t=0.5$, $J_{\rm e} = \pi/5, h_x= 0.11$, and $V_{xx}=0.2$ and tune the odd-site stabilizer strength $J_{\rm o}$ to observe distinct behavior of the systems. Trotterization errors act as additional perturbations and make edge-bulk interactions more general (Methods and Supplementary Section 1E). 

\vspace{2mm}
\noindent\textbf{\large{}Robust edge modes at infinite temperature}

\begin{figure*}[ht]
    \centering
    \includegraphics[width=1\linewidth]{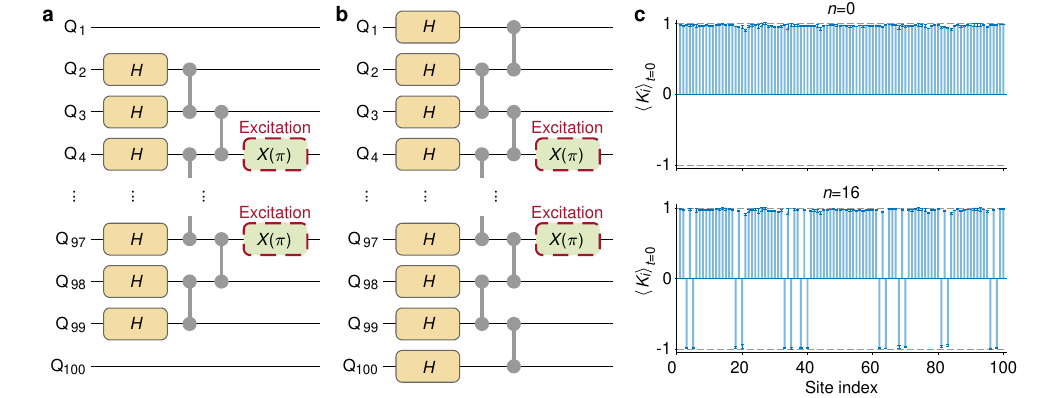}
    \caption{{\bf Initial state preparation.}
    {\bf a}, Quantum circuit for preparing the initial cluster state for measuring $\Tilde{Z}(t)$ in Fig.~\ref{fig:2}\textbf{b},\textbf{d}. 
    The first three layers, including one-layer Hadamard gates and two-layer CZ gates, prepare the ground state of $H_0$ with $\tilde{Z}_{\rm L}, \tilde{Z}_{\rm R}$ taking values $+1$. 
    The last layer applies single-qubit $\pi$ rotations around the $x$-axis of the Bloch sphere [$X(\pi)$ gates] on the bulk qubits, each flipping two stabilizers and hence inducing two bulk excitations.
    The initial state in the excited-state manifold $\{\ket{\Psi_{\rm e}}\}$ are obtained from applying the $X(\pi)$ gates on qubits $\{Q_4, Q_{19}, Q_{34}, Q_{39}, Q_{63}, Q_{69}, Q_{82}, Q_{97}\}$, which exhibits $16$ excitations in the bulk. 
    In the bottom panel of Fig.~\ref{fig:2}\textbf{c}, the $X(\pi)$ gates are applied to $\{Q_6, Q_{19}, Q_{34}, Q_{39}, Q_{63}, Q_{69}, Q_{82}, Q_{95}\}$
    to observe the effect of varying excitation positions on the edge modes.
    {\bf b}, Quantum circuit for preparing the initial state for measuring $\Tilde{X}(t)$ in Fig.~\ref{fig:2}\textbf{b},\textbf{d} and the excitation dynamics in Fig.~\ref{fig:3}, with $\tilde{X}_{\rm L}, \tilde{X}_{\rm R}$ taking values $+1$. 
    {\bf c}, Expectation values of the stabilizers $K_i$ ($\sigma_{i-1}^z\sigma_i^x\sigma_{i+1}^z$ in the bulk and $\Tilde{X}_{\rm L}, \Tilde{X}_{\rm R}$ at the edges) for initial states in $\{\Psi_{\rm g}\}$ (top) and in $\{\Psi_{\rm e}\}$ (bottom). The data are extracted from Extended Data Fig.~\ref{fig:ex3} at $t=0$. Each data point is averaged over fifteen repetitions of measurements (five for the homogeneous case, five for the dimerized but resonant case, and five for the dimerized and off-resonant case) from which the error bars stem. Gray dashed lines indicate the values of $\pm 1$.}
    \label{fig:ex2}
\end{figure*}

\noindent We first explore the influence of bulk excitations on edge modes in the homogeneous regime ($J_{\rm e}=J_{\rm o}$). 
We start by contrasting the experimentally measured time dependence of the edge modes when the system is initialized in cluster ground-state manifold $\{\ket{\Psi_{\rm g}}\}$ versus product states $|\bullet 00...0 \bullet\rangle$ in Fig.~\ref{fig:2}b 
(see Methods and Extended Fig.~\ref{fig:ex2} for initial state preparation).
The latter, manifesting as an effectively infinite-temperature state with poorly protected edge modes, decays much faster. 
Although $\ket{\Psi_{\rm g}}$ is not the exact ground state of the system in the presence of interaction term $H_1$, it resides in the low-energy-density regime, leading to limited effects of excitations on the edge modes.  
As such, the observed decay is attributed to external experimental imperfections, especially circuit errors.
This is verified by the agreement between the ground-state dynamics and echo circuit dynamics $U_{\text{echo}}(t)=(U^\dagger)^t U^t$~\cite{mi2022time}.

To expose the origin of faster decoherence for edge modes at finite temperatures, we further introduce excitations into the bulk in a controlled way by initializing the system in a cluster excited-state manifold $\{\ket{\Psi_{\rm e}}\}$ with $n=16$ excitations 
Notably, we observe that $\ket{\Psi_{\rm e}}$ with excitations near each end can show an even faster decay of the edge modes than the product state (Fig.~\ref{fig:2}b). To illustrate the effect of excitation positions, we further probe time-dependent expectation values of bulk stabilizers $\{K_i=\sigma_{i-1}^z\sigma_i^x\sigma_{i+1}^z\}^{N-1}_{i=2}$, and edge operators $\{K_1,K_N\}=\{\tilde{X}_{\rm L}$, $\tilde{X}_{\rm R}\}$. We define the normalized expectation value as $\bar{\braket{{{K}_i}}}={\Braket{\Psi_{\rm e}|{K}_i(t)|\Psi_{\rm e}}}/{\Braket{\Psi_{\rm g}|{K}_i(t)|\Psi_{\rm g}}}$ to underscore the decay caused by excitations.
In Fig.~\ref{fig:2}c, we show the measured $\bar{\braket{{{K}_i}}}$ dynamics near the right edge with two different initial excitation positions, and observe that the edge mode is maintained until excitations propagate to the edge, demonstrating that its rapid decay is due to the edge-bulk interactions.

Intriguingly, in the dimerized regime ($J_{\rm e}\neq J_{\rm o}$), the edge modes show distinct behaviors (Fig.~\ref{fig:2}d). 
Starting with $\ket{\Psi_{\rm e}}$, we measure the temporal dependence of edge operators for $J_{\rm o}/J_{\rm e}$ ranging from $0.8$ to $3.2$. It is evident from Fig.~\ref{fig:2}d that the lifetime of the edge modes is prolonged as $J_{\rm o}/J_{\rm e}$ deviates from $1$. 
Theoretically, the edge operators in the dimerized regime can be described as prethermal strong zero modes (Supplementary Section 1B), which induce almost exact four-fold degeneracy throughout the entire spectrum, leading to enhanced resilience against thermal excitations ~\cite{Kemp2017Long, Else2017Prethermal2, Kemp2020Symmetry}. 
It is noteworthy that such enhanced resilience breaks down for $\tilde{X}_{\rm L}$ at $J_{\rm o}/J_{\rm e}=2$, which originates from the first-order resonant process by the two-body interactions in $H_1$, and leads to a divergence for the prethermal strong zero modes.
This non-monotonicity observed in the lifetime of edge modes illustrates the intricacy of edge-bulk interactions, providing a clear distinction between the dimerization mechanism and the suppression of interaction strength.

\begin{figure}[ht]
    \centering
    \includegraphics[width=1\linewidth]{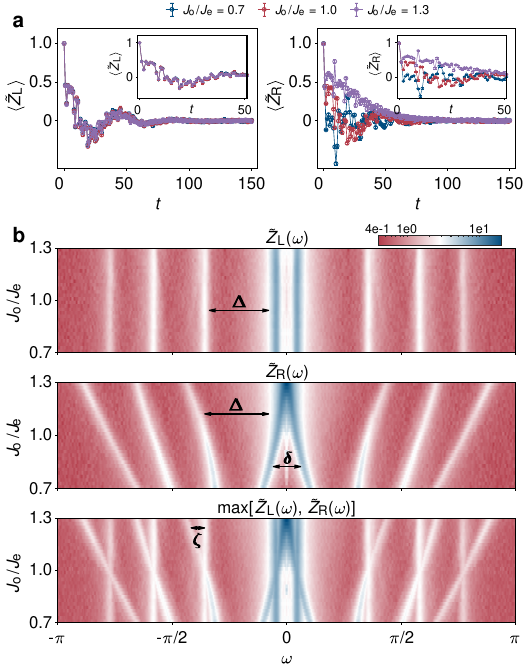}
    \caption{\textbf{Spectroscopy of energy spectrum.} 
        {\bf a}, Measured time-domain signals of $\Tilde{Z}_{\rm L}$ (left panel) and $\Tilde{Z}_{\rm R}$ (right panel) in the integrable limit ($V_{xx}=0$) on a chain of $N=8$. We fix $J_{\rm e}=\pi/2$, $h_x=7\pi/20$ and vary $J_{\rm o}/J_{\rm e}$ by tuning $J_{\rm o}$. The initial state is the product state $|00...0\rangle$. Whereas $\Tilde{Z}_{\rm R}$ oscillations are modulated by $J_{\rm o}$, the dynamics of $\Tilde{Z}_{\rm L}$ remain unaffected.
        {\bf b}, Fourier transforms of $\Tilde{Z}_{\rm L}$ (top panel) and $\Tilde{Z}_{\rm R}$ (middle panel) dynamics as functions of $\omega$ and $J_{\rm o}/J_{\rm e}$. The gap $\Delta\propto J_{\rm o}$ ($\propto J_{\rm e}$) separates the edge modes from the bulk excitation mode. $\delta$ ($\propto h_x$) indicates the hybridization between two edge modes. A complete spectrum (bottom panel) is obtained by combining $\Tilde{Z}_{\rm L}(\omega)$ and $\Tilde{Z}_{\rm R}(\omega)$, where $\zeta$ represents the gap between the bulk modes on different Kitaev chains.
    }
    \label{fig:4}
\end{figure}

\begin{figure*}[ht]
    \centering
    \includegraphics[width=1\linewidth]{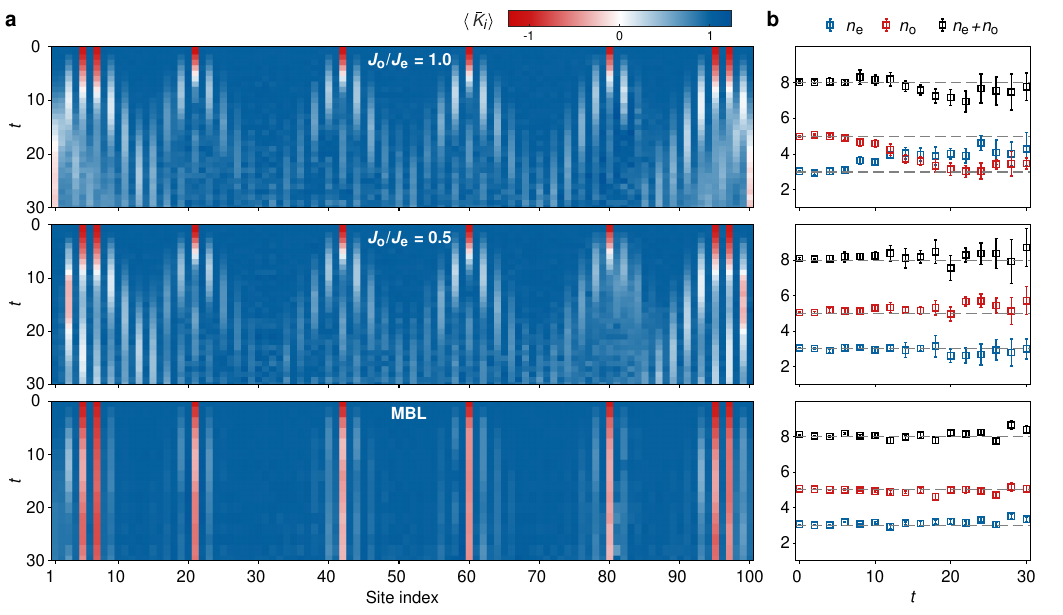}
    \caption{\textbf{Excitation dynamics with only two $V_{xx}$ interaction terms $\sigma^x_1\sigma^x_2$ and $\sigma^x_{99}\sigma^x_{100}$ at the two edges.}
    {\bf a}, Measured site-resolved dynamics of normalized expectation value $\bar{\braket{{K}_i}}$ for bulk stabilizers $\{\sigma_{i-1}^z\sigma_i^x\sigma_{i+1}^z\}$ and edge operators $\{\Tilde{Z}_{\rm L}$, $\Tilde{Z}_{\rm R}\}$ in the homogeneous (top panel, $J_{\rm o}=J_{\rm e}=\pi/2$), dimerized (middle panel, $J_{\rm o}=0.5J_{\rm e}=\pi/4$), and many-body localized regimes (bottom panel). 
    The bulk excitations are induced by applying $Z(\pi)$ gates on sites $\{Q_5, Q_7, Q_{21}, Q_{42}, Q_{60}, Q_{80}, Q_{95}, Q_{97}\}$.
    For the system in the homogeneous and dimerized regimes, we set $h_x=0.23$, $V_{xx}=0.2$, and error bars stem from five repetitions of measurements.
    For the system in the many-body localized regimes, we fix $V_{xx}=0.2$ and randomly choose $J_{\rm o}$, $J_{\rm e}, h_x$ from $[\pi/6,5\pi/6]$, $[\pi/6,5\pi/6]$, and $[0.18, 0.28]$, respectively.
    The data are averaged over $10$ random instances, and the error bars are the standard error of the statistical mean for these instances.
    {\bf b}, Measured time dynamics of the total excitation number $n$, excitation number at even ($n_{\rm e}$) and odd ($n_{\rm o}$) sites, which are calculated from {\bf a}. 
    }
    \label{fig:ex5}
\end{figure*}

\vspace{2mm}
\noindent\textbf{\large{}Excitation dynamics and emergent symmetry}\\
\noindent
To understand the dimerization mechanism for enhancing the lifetime of edge modes at finite temperatures, we examine site-resolved excitation dynamics and bulk-edge interactions for the whole chain.
We plot the measured dynamics of $\bar{\braket{{{K}_i}}}$ for $J_{\rm o}/J_{\rm e}=1.0$ (homogeneous), $2.0$ (dimerized but resonant), and $3.17$ (dimerized and off-resonant) in Fig.~\ref{fig:3}.
The excitation dynamics are clearly distinct in the three cases.
First, in the homogeneous case, excitations deep in the bulk propagate diffusely across even and odd sites (Fig.~\ref{fig:3}a). In contrast, in the two dimerized cases, excitations initially located at even (or odd) sites are constrained to move along sites of the same parity (Fig.~\ref{fig:3}b, c). 
Neighboring excitation pairs with different parities propagate freely without interacting with each other, while pairs with the same parity collide.
Second, excitations near the boundaries in the homogeneous case are absorbed by the edge states, while for the dimerized and off-resonant case (Fig.~\ref{fig:3}c), they are reflected at the boundaries without affecting the edge states (see also Extended Fig. \ref{fig:ex5}). 
Third, for the dimerized but resonant case (Fig.~\ref{fig:3}b), despite similar dynamics observed near the right boundary as in the off-resonant case, the excitations interact strongly with the left edge due to the resonance (Supplementary Section 1B).

The distinct behaviors of the three cases above can be better understood in the Majorana fermion picture~(Fig.~\ref{fig:1}b and Methods).  Through Jordan-Wigner transformation,
the cluster Hamiltonian $H_{0}$ is transformed into two Kitaev chains composed of Majorana fermions on even and odd sites, respectively. 
The stabilizers centered at even (odd) sites are mapped to inter-site coupling terms with strength $J_{\rm e}$ ($J_{\rm o}$) in the upper (lower) chain. 
The edge mode is mapped to two Majorana fermions at the end of each Kitaev chain, and the single- and two-body terms in $H_1$ are mapped to onsite and inter-chain coupling terms.
With $J_{\rm o}=J_{\rm e}$, the two Kitaev chains share the same coupling strength and can exchange excitations resonantly through $V_{xx}$ terms both in the bulk and at the boundaries, where the latter couple to the edge Majorana fermions and lead to the decay of the edge modes.
In the small-perturbation regime ($h_x, V_{xx}\ll J_{\rm o}, J_{\rm e}$), the system exhibits long-lived prethermal behavior with an approximate U(1) symmetry of the total excitation number $n$ in the bulk, which is observed for all three cases in our experiments~(Fig.~\ref{fig:3}).
However, despite the conserved $n$, the number of bulk excitations on even sites $n_{\rm e}$ and odd sites $n_{\rm o}$ rapidly equilibrate in the homogeneous case (Fig.~\ref{fig:3}d), manifesting the effect of the resonant inter-chain interactions. 
Dimerizing the coupling strengths makes the excitation exchange in the bulk off-resonant, but resonances can still arise at the boundaries for certain values of $J_{\rm o}/J_{\rm e}$.
For example, when $J_{\rm o}/J_{\rm e}=2.0$, the exchange of one excitation in the lower chain and two excitations in the upper chain through the $\sigma_2^x\sigma_3^x$ term in $H_1$ becomes resonant.
This results in the observed rapid decay of $\tilde{X}_{\rm L}$, and $n$ is no longer conserved~(Fig.~\ref{fig:3}e).
Such a resonance can be eliminated by choosing $J_{\rm o}/J_{\rm e}$ close to an incommensurable number.
Consequently, the excitation exchanges become off-resonant both in the bulk and at the boundaries, leading to two Kitaev chains effectively decoupled and exhibiting two separate approximate U(1) conservation laws for $n_{\rm e}$ and $n_{\rm o}$ in the system's prethermal regime~(Fig.~\ref{fig:3}f). 
This dimerization-induced U(1)$\times$U(1) symmetry together with the $\mathbb{Z}_2\times \mathbb{Z}_2$ symmetry inherent in the system gives rise to robust edge modes persisting up to infinite temperature (Supplementary Section 1C).

\vspace{2mm}
\noindent\textbf{\large{}Energy spectrum}

\noindent Recent theoretical progresses suggest that prethermalization is a generic phenomenon in gapped local many-body systems, where quantum dynamics is restricted to each symmetry sector protected by the energy gaps~\cite{Yin2023Prethermalization}. 
Such a prediction is also observed in our experiments, as the emergent U(1)$\times$U(1) symmetry and the robust edge modes are manifestations of energy gaps in the spectrum. 
Utilizing energy spectroscopy technique~\cite{Roushan2017Spectroscopic, Mi2022Noise}, we measure the spectrum of a smaller SPT chain with $N=8$ qubits on another processor~\cite{Xu2023CPL} in parallel, which has a similar design but better coherence performance.
The spectrum in the integrable limit ($V_{xx}=0$) is obtained through the Fourier transform of the dynamics of $\tilde{Z}_{\rm L}$ and $\tilde{Z}_{\rm R}$ measured in experiments (Supplementary Sections 1F and 2D).
In Fig.~\ref{fig:4}a, we display three representative time-domain signals of $\Tilde{Z}_{\rm L}$ and $\Tilde{Z}_{\rm R}$ for $J_{\rm o}/J_{\rm e}=0.7, 1.0$, and $1.3$. 
Notably, varying $J_{\rm o}$ only influences the dynamics of $\Tilde{Z}_{\rm R}$ while $\Tilde{Z}_{\rm L}$ is unaffected. 
This observation aligns with the theoretical prediction in the Majorana fermion picture, where the two Kitaev chains remain decoupled at $V_{xx}=0$, and $\Tilde{Z}_{\rm L}$ and $\Tilde{Z}_{\rm R}$ are mapped into Majorana edge modes in different chains.

In Fig.~\ref{fig:4}b, we show the frequency-domain signals of $\tilde{Z}_{\rm L}$, $\tilde{Z}_{\rm R}$, and their combination with varying $J_{\rm o}/J_{\rm e}$, which provides substantial information to understand the origin of emergent symmetries.
First, as two Kitaev chains are decoupled at $V_{xx}=0$, $\tilde{Z}_{\rm L}(\omega)$ [$\tilde{Z}_{\rm R}(\omega)$] gives rise to the spectrum for the upper [lower] chain, and their combination reveals the full spectrum of the entire system.
The peaks correspond to Bogoliubov fermionic modes in each chain, where peaks near $\omega = 0$ are attributed to the edge modes, and the remaining peaks characterize the bulk excitation modes.
In our finite-size system, the edge modes are hybridized by a gap $\delta$ caused by the $h_x\sigma_i^x$ terms in $H_1$.
As we increase $J_{\rm o}$, which consequently decreases the correlation length in the lower Kitaev chain (Fig. \ref{fig:1}b), such a gap in $\Tilde{Z}_{\rm R}(\omega)$ gradually closes.
Second, we observe a gap $\Delta\propto J_{\rm o}$ ($\propto J_{\rm e}$) separating the edge mode from the bulk excitation mode, impeding transitions between edges and bulk caused by onsite interactions ($h_x\sigma_i^x$).
When two chains are decoupled, $\Delta$ gives rise to the approximate U(1) symmetry in each chain.
However, these U(1) symmetries can be destroyed when inter-chain interactions are present. 
We observe that the energy for bulk modes in two Kitaev chains becomes exactly equal when $J_{\rm o}/J_{\rm e}=1$ (Fig. \ref{fig:4}b, bottom panel), explaining the strongly resonant excitation exchange process that happened in the homogeneous case. 
Strikingly, when the system is dimerized ($J_{\rm o}\neq J_{\rm e}$), an extra gap $\zeta\propto |J_{\rm e} - J_{\rm o}|$ appears, signifying the energy required to exchange one pair of excitation between the chains. 
Such a gap bolsters the emergent U(1)$\times$ U(1) symmetry and suppresses the excitation exchange process at boundaries, resulting in robust long-lived edge modes at up to infinite temperature.

\begin{figure}[t]
    \centering
    \includegraphics[width=1\linewidth]{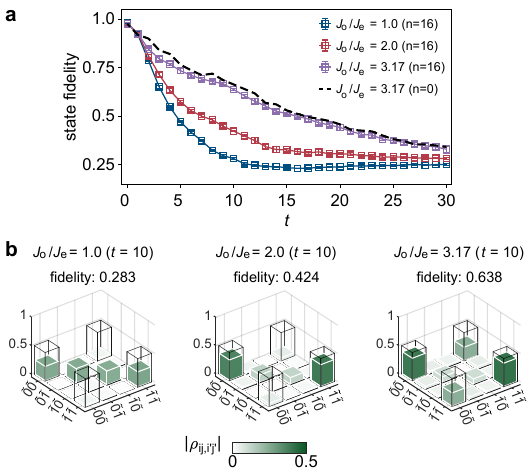}
    \caption{\textbf{Fidelity dynamics of the logical Bell state at finite temperature.} 
        {\bf a}, Measured fidelity dynamics of logical Bell state in the homogeneous ($J_{\rm o}=J_{\rm e}=\pi/5$), the dimerized but resonant ($J_{\rm o}=2J_{\rm e}=2\pi/5$) and the dimerized and off-resonant ($J_{\rm o}=3.17J_{\rm e}=3.17\pi/5$) cases. The data shown in solid (dashed) lines is obtained from initial states being within $\{\ket{\Psi_{\rm e}}\}$ ($\{\ket{\Psi_{\rm g}}\}$).
        The initial state preparation circuit is illustrated in Extended data Fig.\ref{fig:ex4}a. Error bars stem from five repetitions of measurements.
        {\bf b}, Measured density matrices (green bars) of logical Bell state after a time evolution of $t=10$ in the three different cases with initial states being within $\{\ket{\Psi_{\rm e}}\}$. The ideal Bell state density matrix is shown with the hollow frame.}
    \label{fig:5}
\end{figure}

\begin{figure*}[ht]
    \centering
    \includegraphics[width=1\linewidth]{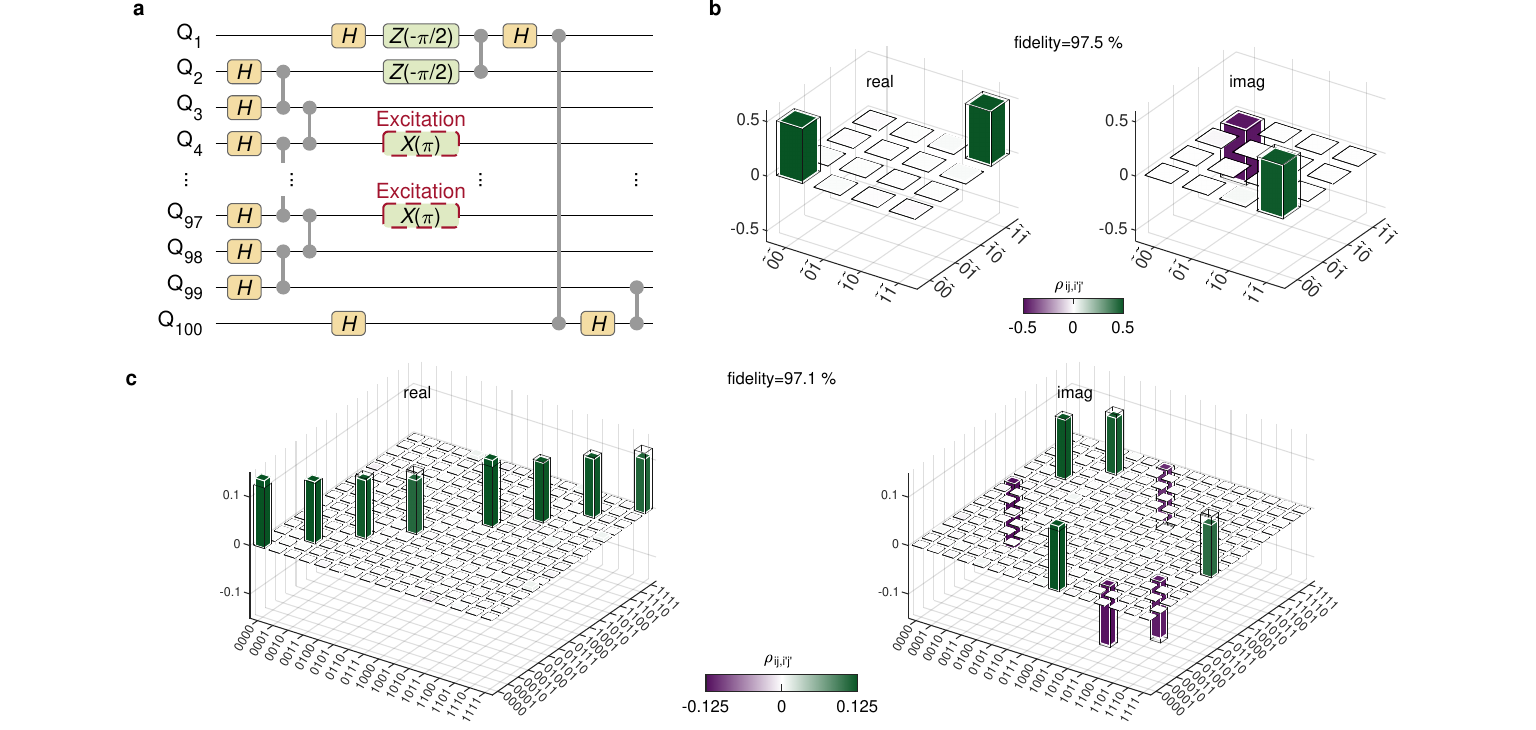}
    \caption{\textbf{Logical Bell state preparation.}
    {\bf a}, Quantum circuit for preparing the logical Bell state $\Tilde{\ket{0}}_{\rm L}\Tilde{\ket{0}}_{\rm R}+\rm{i}\Tilde{\ket{1}}_{\rm L}\Tilde{\ket{1}}_{\rm R}$. 
    The CZ gate applying on two edge qubits $Q_1$ and $Q_{100}$ is local as the two edge qubits are geometrically near to each other on our processor (see Fig.~\ref{fig:1} of the main text). 
    {\bf b}, Measured density matrix of the prepared logical Bell state, which is extracted from logical state tomography. Its fidelity is about $97.5\%$. 
    {\bf c}, Measured density matrix in full computational space of $Q_1$, $Q_2$, $Q_{99}$, $Q_{100}$,  with the fidelity of $97.1\%$ to the ideal density matrix. In {\bf b} and {\bf c}, solid bars are experimental data and hollow frames are the ideal density matrix.
    }
    \label{fig:ex4}
\end{figure*}

\vspace{2mm}
\noindent\textbf{\large{}Protection of logical Bell state}
\noindent
\\ The observed long-lived topological edge modes in experiments offer a potential application to store quantum information at finite temperatures, contrasting with the Ising chains where a classical bit might be preserved via edge spin polarization~\cite{Kemp2017Long, Mi2022Noise}. To this end, we prepare a logical Bell state encoded by these edge states and show its robustness to thermal excitations. 
Owing to the geometrically adjacent two edges on the processor (blue circles in Fig.~\ref{fig:1}a), we can initialize the system with edge modes being a logical Bell state $\Tilde{\ket{0}}_{\rm L}\Tilde{\ket{0}}_{\rm R}+\rm{i}\Tilde{\ket{1}}_{\rm L}\Tilde{\ket{1}}_{\rm R}$ by local two-qubit gates (see Extended Fig.~\ref{fig:ex4} for the details of preparation circuit and logical Bell state fidelity).

The solid lines in Fig.~\ref{fig:5}a illustrate the measured fidelity dynamics of the logical Bell state for initial states being within $\{\ket{\Psi_{\rm e}}\}$ with $J_{\rm o}/J_{\rm e}=1.0$, $2.0$, and $3.17$, respectively.  
 As expected, the fidelity in the homogeneous scenario decays the most rapidly to the lower bound of $0.25$, followed by the dimerized but resonant system. 
 The lifetime of the logical Bell state in the dimerized and off-resonant system is largely prolonged, almost reaching that of the ground-state case (dashed line).
 Furthermore, we carry out state tomography (Supplementary Section 2E) on the logical space of each system after a time evolution of $t=10$. 
 As shown in Fig.~\ref{fig:5}b, the logical Bell state in the homogeneous system is completely decohered, corresponding to an identity matrix of maximally mixed state. 
 The logical Bell state in the dimerized but resonant scenario also exhibits rapid decoherence with vanishing off-diagonal terms.
 In contrast, the density matrix is largely preserved for the dimerized and off-resonant case,  thus contributing significantly to the protection of entanglement at infinite temperature.

\vspace{2mm}
\noindent\textbf{\large{}Discussions}
\noindent
\\ The robust edge modes observed in our experiments are attributed to the emergent symmetries within the prethermal regime, thereby eliminating the necessity for strong disorder.
We established that these symmetries arise from distinct gaps in the energy spectrum, a common phenomenon in gapped quantum many-body systems. 
This dimerization-induced prethermalization mechanism is neither restricted to $1$D systems nor SPT phases. 
Recent works predict the robust storage of quantum information at finite temperatures using nonlocal operators in toric codes~\cite{Else2017Prethermal} and two-dimensional subsystem codes~\cite{Wildeboer2022Symmetry}, and local corner modes in higher-order SPT phases~\cite{Jiang2024Prethermal}
Our work opens new possibilities for quantum information storage resilient to thermal excitations on noisy intermediate-scale quantum devices. 
In addition, it has been shown that periodic and quasi-periodic driven systems possessed additional dynamical symmetries, which can substitute the intrinsic symmetries in the Hamiltonian~\cite{Else2020Long-Lived, Friedman2022Topological,  Dumitrescu2022Dynamical}.
In particular, it could be possible to extend our study to realize novel dynamical SPT phases that possess resilient edge modes against both perturbations and thermal excitations, without relying on any intrinsic symmetry or localization. 
\bibliography{mainRef.bib}

\begin{thebibliography}{47}%
\makeatletter
\providecommand \@ifxundefined [1]{%
 \@ifx{#1\undefined}
}%
\providecommand \@ifnum [1]{%
 \ifnum #1\expandafter \@firstoftwo
 \else \expandafter \@secondoftwo
 \fi
}%
\providecommand \@ifx [1]{%
 \ifx #1\expandafter \@firstoftwo
 \else \expandafter \@secondoftwo
 \fi
}%
\providecommand \natexlab [1]{#1}%
\providecommand \enquote  [1]{``#1''}%
\providecommand \bibnamefont  [1]{#1}%
\providecommand \bibfnamefont [1]{#1}%
\providecommand \citenamefont [1]{#1}%
\providecommand \href@noop [0]{\@secondoftwo}%
\providecommand \href [0]{\begingroup \@sanitize@url \@href}%
\providecommand \@href[1]{\@@startlink{#1}\@@href}%
\providecommand \@@href[1]{\endgroup#1\@@endlink}%
\providecommand \@sanitize@url [0]{\catcode `\\12\catcode `\$12\catcode
  `\&12\catcode `\#12\catcode `\^12\catcode `\_12\catcode `\%12\relax}%
\providecommand \@@startlink[1]{}%
\providecommand \@@endlink[0]{}%
\providecommand \url  [0]{\begingroup\@sanitize@url \@url }%
\providecommand \@url [1]{\endgroup\@href {#1}{\urlprefix }}%
\providecommand \urlprefix  [0]{URL }%
\providecommand \Eprint [0]{\href }%
\providecommand \doibase [0]{https://doi.org/}%
\providecommand \selectlanguage [0]{\@gobble}%
\providecommand \bibinfo  [0]{\@secondoftwo}%
\providecommand \bibfield  [0]{\@secondoftwo}%
\providecommand \translation [1]{[#1]}%
\providecommand \BibitemOpen [0]{}%
\providecommand \bibitemStop [0]{}%
\providecommand \bibitemNoStop [0]{.\EOS\space}%
\providecommand \EOS [0]{\spacefactor3000\relax}%
\providecommand \BibitemShut  [1]{\csname bibitem#1\endcsname}%
\let\auto@bib@innerbib\@empty
\bibitem [{\citenamefont {Chiu}\ \emph {et~al.}(2016)\citenamefont {Chiu},
  \citenamefont {Teo}, \citenamefont {Schnyder},\ and\ \citenamefont
  {Ryu}}]{Chiu2016Classification}%
  \BibitemOpen
  \bibfield  {author} {\bibinfo {author} {\bibfnamefont {C.-K.}\ \bibnamefont
  {Chiu}}, \bibinfo {author} {\bibfnamefont {J.~C.~Y.}\ \bibnamefont {Teo}},
  \bibinfo {author} {\bibfnamefont {A.~P.}\ \bibnamefont {Schnyder}},\ and\
  \bibinfo {author} {\bibfnamefont {S.}~\bibnamefont {Ryu}},\ }\bibfield
  {title} {\bibinfo {title} {Classification of topological quantum matter with
  symmetries},\ }\href {https://doi.org/10.1103/RevModPhys.88.035005}
  {\bibfield  {journal} {\bibinfo  {journal} {Rev. Mod. Phys.}\ }\textbf
  {\bibinfo {volume} {88}},\ \bibinfo {pages} {035005} (\bibinfo {year}
  {2016})}\BibitemShut {NoStop}%
\bibitem [{\citenamefont {Wen}(2017)}]{Wen2017RMP}%
  \BibitemOpen
  \bibfield  {author} {\bibinfo {author} {\bibfnamefont {X.-G.}\ \bibnamefont
  {Wen}},\ }\bibfield  {title} {\bibinfo {title} {Colloquium: Zoo of
  quantum-topological phases of matter},\ }\href
  {https://doi.org/10.1103/RevModPhys.89.041004} {\bibfield  {journal}
  {\bibinfo  {journal} {Rev. Mod. Phys.}\ }\textbf {\bibinfo {volume} {89}},\
  \bibinfo {pages} {041004} (\bibinfo {year} {2017})}\BibitemShut {NoStop}%
\bibitem [{\citenamefont {Chen}\ \emph {et~al.}(2012)\citenamefont {Chen},
  \citenamefont {Gu}, \citenamefont {Liu},\ and\ \citenamefont
  {Wen}}]{Chen2012Symmetry}%
  \BibitemOpen
  \bibfield  {author} {\bibinfo {author} {\bibfnamefont {X.}~\bibnamefont
  {Chen}}, \bibinfo {author} {\bibfnamefont {Z.-C.}\ \bibnamefont {Gu}},
  \bibinfo {author} {\bibfnamefont {Z.-X.}\ \bibnamefont {Liu}},\ and\ \bibinfo
  {author} {\bibfnamefont {X.-G.}\ \bibnamefont {Wen}},\ }\bibfield  {title}
  {\bibinfo {title} {Symmetry-protected topological orders in interacting
  bosonic systems},\ }\href {https://doi.org/10.1126/science.1227224}
  {\bibfield  {journal} {\bibinfo  {journal} {Science}\ }\textbf {\bibinfo
  {volume} {338}},\ \bibinfo {pages} {1604} (\bibinfo {year}
  {2012})}\BibitemShut {NoStop}%
\bibitem [{\citenamefont {Senthil}(2015)}]{Senthil2015Symmetry}%
  \BibitemOpen
  \bibfield  {author} {\bibinfo {author} {\bibfnamefont {T.}~\bibnamefont
  {Senthil}},\ }\bibfield  {title} {\bibinfo {title} {Symmetry-protected
  topological phases of quantum matter},\ }\href
  {https://doi.org/10.1146/annurev-conmatphys-031214-014740} {\bibfield
  {journal} {\bibinfo  {journal} {Annu. Rev. Condens. Matter Phys.}\ }\textbf
  {\bibinfo {volume} {6}},\ \bibinfo {pages} {299} (\bibinfo {year}
  {2015})}\BibitemShut {NoStop}%
\bibitem [{\citenamefont {Landau}\ and\ \citenamefont
  {Lifshitz}(2013)}]{Landau2013Statistical}%
  \BibitemOpen
  \bibfield  {author} {\bibinfo {author} {\bibfnamefont {L.~D.}\ \bibnamefont
  {Landau}}\ and\ \bibinfo {author} {\bibfnamefont {E.~M.}\ \bibnamefont
  {Lifshitz}},\ }\href@noop {} {\emph {\bibinfo {title} {Statistical Physics:
  Volume 5}}},\ Vol.~\bibinfo {volume} {5}\ (\bibinfo  {publisher} {Elsevier},\
  \bibinfo {year} {2013})\BibitemShut {NoStop}%
\bibitem [{\citenamefont {Pollmann}\ \emph {et~al.}(2012)\citenamefont
  {Pollmann}, \citenamefont {Berg}, \citenamefont {Turner},\ and\ \citenamefont
  {Oshikawa}}]{Pollmann2012Symmetry}%
  \BibitemOpen
  \bibfield  {author} {\bibinfo {author} {\bibfnamefont {F.}~\bibnamefont
  {Pollmann}}, \bibinfo {author} {\bibfnamefont {E.}~\bibnamefont {Berg}},
  \bibinfo {author} {\bibfnamefont {A.~M.}\ \bibnamefont {Turner}},\ and\
  \bibinfo {author} {\bibfnamefont {M.}~\bibnamefont {Oshikawa}},\ }\bibfield
  {title} {\bibinfo {title} {Symmetry protection of topological phases in
  one-dimensional quantum spin systems},\ }\href
  {https://doi.org/10.1103/PhysRevB.85.075125} {\bibfield  {journal} {\bibinfo
  {journal} {Phys. Rev. B}\ }\textbf {\bibinfo {volume} {85}},\ \bibinfo
  {pages} {075125} (\bibinfo {year} {2012})}\BibitemShut {NoStop}%
\bibitem [{\citenamefont {Fidkowski}\ and\ \citenamefont
  {Kitaev}(2011)}]{Fidkowski2011Topological}%
  \BibitemOpen
  \bibfield  {author} {\bibinfo {author} {\bibfnamefont {L.}~\bibnamefont
  {Fidkowski}}\ and\ \bibinfo {author} {\bibfnamefont {A.}~\bibnamefont
  {Kitaev}},\ }\bibfield  {title} {\bibinfo {title} {Topological phases of
  fermions in one dimension},\ }\href
  {https://doi.org/10.1103/PhysRevB.83.075103} {\bibfield  {journal} {\bibinfo
  {journal} {Phys. Rev. B}\ }\textbf {\bibinfo {volume} {83}},\ \bibinfo
  {pages} {075103} (\bibinfo {year} {2011})}\BibitemShut {NoStop}%
\bibitem [{\citenamefont {Ma}\ and\ \citenamefont
  {Wang}(2023)}]{Ma2023Average}%
  \BibitemOpen
  \bibfield  {author} {\bibinfo {author} {\bibfnamefont {R.}~\bibnamefont
  {Ma}}\ and\ \bibinfo {author} {\bibfnamefont {C.}~\bibnamefont {Wang}},\
  }\bibfield  {title} {\bibinfo {title} {Average symmetry-protected topological
  phases},\ }\href {https://doi.org/10.1103/PhysRevX.13.031016} {\bibfield
  {journal} {\bibinfo  {journal} {Phys. Rev. X}\ }\textbf {\bibinfo {volume}
  {13}},\ \bibinfo {pages} {031016} (\bibinfo {year} {2023})}\BibitemShut
  {NoStop}%
\bibitem [{\citenamefont {Bravyi}\ \emph {et~al.}(2010)\citenamefont {Bravyi},
  \citenamefont {Hastings},\ and\ \citenamefont
  {Michalakis}}]{Bravyi2010Topological}%
  \BibitemOpen
  \bibfield  {author} {\bibinfo {author} {\bibfnamefont {S.}~\bibnamefont
  {Bravyi}}, \bibinfo {author} {\bibfnamefont {M.~B.}\ \bibnamefont
  {Hastings}},\ and\ \bibinfo {author} {\bibfnamefont {S.}~\bibnamefont
  {Michalakis}},\ }\bibfield  {title} {\bibinfo {title} {Topological quantum
  order: Stability under local perturbations},\ }\bibfield  {journal} {\bibinfo
   {journal} {J. Math. Phys.}\ }\textbf {\bibinfo {volume} {51}},\ \href
  {https://doi.org/10.1063/1.3490195} {10.1063/1.3490195} (\bibinfo {year}
  {2010})\BibitemShut {NoStop}%
\bibitem [{\citenamefont {Brown}\ \emph {et~al.}(2016)\citenamefont {Brown},
  \citenamefont {Loss}, \citenamefont {Pachos}, \citenamefont {Self},\ and\
  \citenamefont {Wootton}}]{Brown2016Quantum}%
  \BibitemOpen
  \bibfield  {author} {\bibinfo {author} {\bibfnamefont {B.~J.}\ \bibnamefont
  {Brown}}, \bibinfo {author} {\bibfnamefont {D.}~\bibnamefont {Loss}},
  \bibinfo {author} {\bibfnamefont {J.~K.}\ \bibnamefont {Pachos}}, \bibinfo
  {author} {\bibfnamefont {C.~N.}\ \bibnamefont {Self}},\ and\ \bibinfo
  {author} {\bibfnamefont {J.~R.}\ \bibnamefont {Wootton}},\ }\bibfield
  {title} {\bibinfo {title} {Quantum memories at finite temperature},\ }\href
  {https://doi.org/10.1103/RevModPhys.88.045005} {\bibfield  {journal}
  {\bibinfo  {journal} {Rev. Mod. Phys.}\ }\textbf {\bibinfo {volume} {88}},\
  \bibinfo {pages} {045005} (\bibinfo {year} {2016})}\BibitemShut {NoStop}%
\bibitem [{\citenamefont {Nandkishore}\ and\ \citenamefont
  {Huse}(2014)}]{Nandkishore2014Many-Body}%
  \BibitemOpen
  \bibfield  {author} {\bibinfo {author} {\bibfnamefont {R.}~\bibnamefont
  {Nandkishore}}\ and\ \bibinfo {author} {\bibfnamefont {D.}~\bibnamefont
  {Huse}},\ }\bibfield  {title} {\bibinfo {title} {Many-body localization and
  thermalization in quantum statistical mechanics},\ }\href
  {https://doi.org/10.1146/annurev-conmatphys-031214-014726} {\bibfield
  {journal} {\bibinfo  {journal} {Annu. Rev. Condens. Matter Phys.}\ }\textbf
  {\bibinfo {volume} {6}},\ \bibinfo {pages} {15} (\bibinfo {year}
  {2014})}\BibitemShut {NoStop}%
\bibitem [{\citenamefont {Kj\"all}\ \emph {et~al.}(2014)\citenamefont
  {Kj\"all}, \citenamefont {Bardarson},\ and\ \citenamefont
  {Pollmann}}]{Kjall2014Many-Body}%
  \BibitemOpen
  \bibfield  {author} {\bibinfo {author} {\bibfnamefont {J.~A.}\ \bibnamefont
  {Kj\"all}}, \bibinfo {author} {\bibfnamefont {J.~H.}\ \bibnamefont
  {Bardarson}},\ and\ \bibinfo {author} {\bibfnamefont {F.}~\bibnamefont
  {Pollmann}},\ }\bibfield  {title} {\bibinfo {title} {Many-body localization
  in a disordered quantum ising chain},\ }\href
  {https://doi.org/10.1103/PhysRevLett.113.107204} {\bibfield  {journal}
  {\bibinfo  {journal} {Phys. Rev. Lett.}\ }\textbf {\bibinfo {volume} {113}},\
  \bibinfo {pages} {107204} (\bibinfo {year} {2014})}\BibitemShut {NoStop}%
\bibitem [{\citenamefont {Abanin}\ \emph {et~al.}(2019)\citenamefont {Abanin},
  \citenamefont {Altman}, \citenamefont {Bloch},\ and\ \citenamefont
  {Serbyn}}]{Abanin2019Colloquium}%
  \BibitemOpen
  \bibfield  {author} {\bibinfo {author} {\bibfnamefont {D.~A.}\ \bibnamefont
  {Abanin}}, \bibinfo {author} {\bibfnamefont {E.}~\bibnamefont {Altman}},
  \bibinfo {author} {\bibfnamefont {I.}~\bibnamefont {Bloch}},\ and\ \bibinfo
  {author} {\bibfnamefont {M.}~\bibnamefont {Serbyn}},\ }\bibfield  {title}
  {\bibinfo {title} {Colloquium: Many-body localization, thermalization, and
  entanglement},\ }\href {https://doi.org/10.1103/RevModPhys.91.021001}
  {\bibfield  {journal} {\bibinfo  {journal} {Rev. Mod. Phys.}\ }\textbf
  {\bibinfo {volume} {91}},\ \bibinfo {pages} {021001} (\bibinfo {year}
  {2019})}\BibitemShut {NoStop}%
\bibitem [{\citenamefont {Huse}\ \emph {et~al.}(2013)\citenamefont {Huse},
  \citenamefont {Nandkishore}, \citenamefont {Oganesyan}, \citenamefont {Pal},\
  and\ \citenamefont {Sondhi}}]{Huse2013Localization}%
  \BibitemOpen
  \bibfield  {author} {\bibinfo {author} {\bibfnamefont {D.~A.}\ \bibnamefont
  {Huse}}, \bibinfo {author} {\bibfnamefont {R.}~\bibnamefont {Nandkishore}},
  \bibinfo {author} {\bibfnamefont {V.}~\bibnamefont {Oganesyan}}, \bibinfo
  {author} {\bibfnamefont {A.}~\bibnamefont {Pal}},\ and\ \bibinfo {author}
  {\bibfnamefont {S.~L.}\ \bibnamefont {Sondhi}},\ }\bibfield  {title}
  {\bibinfo {title} {Localization-protected quantum order},\ }\href
  {https://doi.org/10.1103/PhysRevB.88.014206} {\bibfield  {journal} {\bibinfo
  {journal} {Phys. Rev. B}\ }\textbf {\bibinfo {volume} {88}},\ \bibinfo
  {pages} {014206} (\bibinfo {year} {2013})}\BibitemShut {NoStop}%
\bibitem [{\citenamefont {Chandran}\ \emph {et~al.}(2014)\citenamefont
  {Chandran}, \citenamefont {Khemani}, \citenamefont {Laumann},\ and\
  \citenamefont {Sondhi}}]{Chandran2014Many-body}%
  \BibitemOpen
  \bibfield  {author} {\bibinfo {author} {\bibfnamefont {A.}~\bibnamefont
  {Chandran}}, \bibinfo {author} {\bibfnamefont {V.}~\bibnamefont {Khemani}},
  \bibinfo {author} {\bibfnamefont {C.~R.}\ \bibnamefont {Laumann}},\ and\
  \bibinfo {author} {\bibfnamefont {S.~L.}\ \bibnamefont {Sondhi}},\ }\bibfield
   {title} {\bibinfo {title} {Many-body localization and symmetry-protected
  topological order},\ }\href {https://doi.org/10.1103/PhysRevB.89.144201}
  {\bibfield  {journal} {\bibinfo  {journal} {Phys. Rev. B}\ }\textbf {\bibinfo
  {volume} {89}},\ \bibinfo {pages} {144201} (\bibinfo {year}
  {2014})}\BibitemShut {NoStop}%
\bibitem [{\citenamefont {Bahri}\ \emph {et~al.}(2015)\citenamefont {Bahri},
  \citenamefont {Vosk}, \citenamefont {Altman},\ and\ \citenamefont
  {Vishwanath}}]{Bahri2015Localization}%
  \BibitemOpen
  \bibfield  {author} {\bibinfo {author} {\bibfnamefont {Y.}~\bibnamefont
  {Bahri}}, \bibinfo {author} {\bibfnamefont {R.}~\bibnamefont {Vosk}},
  \bibinfo {author} {\bibfnamefont {E.}~\bibnamefont {Altman}},\ and\ \bibinfo
  {author} {\bibfnamefont {A.}~\bibnamefont {Vishwanath}},\ }\bibfield  {title}
  {\bibinfo {title} {Localization and topology protected quantum coherence at
  the edge of hot matter},\ }\href
  {https://doi.org/https://doi.org/10.1038/ncomms8341} {\bibfield  {journal}
  {\bibinfo  {journal} {Nat. Commun.}\ }\textbf {\bibinfo {volume} {6}},\
  \bibinfo {pages} {7341} (\bibinfo {year} {2015})}\BibitemShut {NoStop}%
\bibitem [{\citenamefont {Schreiber}\ \emph {et~al.}(2015)\citenamefont
  {Schreiber}, \citenamefont {Hodgman}, \citenamefont {Bordia}, \citenamefont
  {L{\"u}schen}, \citenamefont {Fischer}, \citenamefont {Vosk}, \citenamefont
  {Altman}, \citenamefont {Schneider},\ and\ \citenamefont
  {Bloch}}]{Schreiber2015Observation}%
  \BibitemOpen
  \bibfield  {author} {\bibinfo {author} {\bibfnamefont {M.}~\bibnamefont
  {Schreiber}}, \bibinfo {author} {\bibfnamefont {S.~S.}\ \bibnamefont
  {Hodgman}}, \bibinfo {author} {\bibfnamefont {P.}~\bibnamefont {Bordia}},
  \bibinfo {author} {\bibfnamefont {H.~P.}\ \bibnamefont {L{\"u}schen}},
  \bibinfo {author} {\bibfnamefont {M.~H.}\ \bibnamefont {Fischer}}, \bibinfo
  {author} {\bibfnamefont {R.}~\bibnamefont {Vosk}}, \bibinfo {author}
  {\bibfnamefont {E.}~\bibnamefont {Altman}}, \bibinfo {author} {\bibfnamefont
  {U.}~\bibnamefont {Schneider}},\ and\ \bibinfo {author} {\bibfnamefont
  {I.}~\bibnamefont {Bloch}},\ }\bibfield  {title} {\bibinfo {title}
  {Observation of many-body localization of interacting fermions in a
  quasirandom optical lattice},\ }\href
  {https://doi.org/10.1126/science.aaa7432} {\bibfield  {journal} {\bibinfo
  {journal} {Science}\ }\textbf {\bibinfo {volume} {349}},\ \bibinfo {pages}
  {842} (\bibinfo {year} {2015})}\BibitemShut {NoStop}%
\bibitem [{\citenamefont {Choi}\ \emph {et~al.}(2016)\citenamefont {Choi},
  \citenamefont {Hild}, \citenamefont {Zeiher}, \citenamefont {Schau{\ss}},
  \citenamefont {Rubio-Abadal}, \citenamefont {Yefsah}, \citenamefont
  {Khemani}, \citenamefont {Huse}, \citenamefont {Bloch},\ and\ \citenamefont
  {Gross}}]{Choi2016Exploring}%
  \BibitemOpen
  \bibfield  {author} {\bibinfo {author} {\bibfnamefont {J.-y.}\ \bibnamefont
  {Choi}}, \bibinfo {author} {\bibfnamefont {S.}~\bibnamefont {Hild}}, \bibinfo
  {author} {\bibfnamefont {J.}~\bibnamefont {Zeiher}}, \bibinfo {author}
  {\bibfnamefont {P.}~\bibnamefont {Schau{\ss}}}, \bibinfo {author}
  {\bibfnamefont {A.}~\bibnamefont {Rubio-Abadal}}, \bibinfo {author}
  {\bibfnamefont {T.}~\bibnamefont {Yefsah}}, \bibinfo {author} {\bibfnamefont
  {V.}~\bibnamefont {Khemani}}, \bibinfo {author} {\bibfnamefont {D.~A.}\
  \bibnamefont {Huse}}, \bibinfo {author} {\bibfnamefont {I.}~\bibnamefont
  {Bloch}},\ and\ \bibinfo {author} {\bibfnamefont {C.}~\bibnamefont {Gross}},\
  }\bibfield  {title} {\bibinfo {title} {Exploring the many-body localization
  transition in two dimensions},\ }\href
  {https://doi.org/10.1126/science.aaf8834} {\bibfield  {journal} {\bibinfo
  {journal} {Science}\ }\textbf {\bibinfo {volume} {352}},\ \bibinfo {pages}
  {1547} (\bibinfo {year} {2016})}\BibitemShut {NoStop}%
\bibitem [{\citenamefont {Smith}\ \emph {et~al.}(2016)\citenamefont {Smith},
  \citenamefont {Lee}, \citenamefont {Richerme}, \citenamefont {Neyenhuis},
  \citenamefont {Hess}, \citenamefont {Hauke}, \citenamefont {Heyl},
  \citenamefont {Huse},\ and\ \citenamefont {Monroe}}]{Smith2016Many}%
  \BibitemOpen
  \bibfield  {author} {\bibinfo {author} {\bibfnamefont {J.}~\bibnamefont
  {Smith}}, \bibinfo {author} {\bibfnamefont {A.}~\bibnamefont {Lee}}, \bibinfo
  {author} {\bibfnamefont {P.}~\bibnamefont {Richerme}}, \bibinfo {author}
  {\bibfnamefont {B.}~\bibnamefont {Neyenhuis}}, \bibinfo {author}
  {\bibfnamefont {P.~W.}\ \bibnamefont {Hess}}, \bibinfo {author}
  {\bibfnamefont {P.}~\bibnamefont {Hauke}}, \bibinfo {author} {\bibfnamefont
  {M.}~\bibnamefont {Heyl}}, \bibinfo {author} {\bibfnamefont {D.~A.}\
  \bibnamefont {Huse}},\ and\ \bibinfo {author} {\bibfnamefont
  {C.}~\bibnamefont {Monroe}},\ }\bibfield  {title} {\bibinfo {title}
  {Many-body localization in a quantum simulator with programmable random
  disorder},\ }\href {https://doi.org/https://doi.org/10.1038/nphys3783}
  {\bibfield  {journal} {\bibinfo  {journal} {Nat. Phys.}\ }\textbf {\bibinfo
  {volume} {12}},\ \bibinfo {pages} {907} (\bibinfo {year} {2016})}\BibitemShut
  {NoStop}%
\bibitem [{\citenamefont {Morningstar}\ \emph {et~al.}(2022)\citenamefont
  {Morningstar}, \citenamefont {Colmenarez}, \citenamefont {Khemani},
  \citenamefont {Luitz},\ and\ \citenamefont {Huse}}]{Morningstar2022PRB}%
  \BibitemOpen
  \bibfield  {author} {\bibinfo {author} {\bibfnamefont {A.}~\bibnamefont
  {Morningstar}}, \bibinfo {author} {\bibfnamefont {L.}~\bibnamefont
  {Colmenarez}}, \bibinfo {author} {\bibfnamefont {V.}~\bibnamefont {Khemani}},
  \bibinfo {author} {\bibfnamefont {D.~J.}\ \bibnamefont {Luitz}},\ and\
  \bibinfo {author} {\bibfnamefont {D.~A.}\ \bibnamefont {Huse}},\ }\bibfield
  {title} {\bibinfo {title} {Avalanches and many-body resonances in many-body
  localized systems},\ }\href {https://doi.org/10.1103/PhysRevB.105.174205}
  {\bibfield  {journal} {\bibinfo  {journal} {Phys. Rev. B}\ }\textbf {\bibinfo
  {volume} {105}},\ \bibinfo {pages} {174205} (\bibinfo {year}
  {2022})}\BibitemShut {NoStop}%
\bibitem [{\citenamefont {Ha}\ \emph {et~al.}(2023)\citenamefont {Ha},
  \citenamefont {Morningstar},\ and\ \citenamefont {Huse}}]{Hyunsoo2023PRL}%
  \BibitemOpen
  \bibfield  {author} {\bibinfo {author} {\bibfnamefont {H.}~\bibnamefont
  {Ha}}, \bibinfo {author} {\bibfnamefont {A.}~\bibnamefont {Morningstar}},\
  and\ \bibinfo {author} {\bibfnamefont {D.~A.}\ \bibnamefont {Huse}},\
  }\bibfield  {title} {\bibinfo {title} {Many-body resonances in the avalanche
  instability of many-body localization},\ }\href
  {https://doi.org/10.1103/PhysRevLett.130.250405} {\bibfield  {journal}
  {\bibinfo  {journal} {Phys. Rev. Lett.}\ }\textbf {\bibinfo {volume} {130}},\
  \bibinfo {pages} {250405} (\bibinfo {year} {2023})}\BibitemShut {NoStop}%
\bibitem [{\citenamefont {Long}\ \emph {et~al.}(2023)\citenamefont {Long},
  \citenamefont {Crowley}, \citenamefont {Khemani},\ and\ \citenamefont
  {Chandran}}]{Long2023PRL}%
  \BibitemOpen
  \bibfield  {author} {\bibinfo {author} {\bibfnamefont {D.~M.}\ \bibnamefont
  {Long}}, \bibinfo {author} {\bibfnamefont {P.~J.~D.}\ \bibnamefont
  {Crowley}}, \bibinfo {author} {\bibfnamefont {V.}~\bibnamefont {Khemani}},\
  and\ \bibinfo {author} {\bibfnamefont {A.}~\bibnamefont {Chandran}},\
  }\bibfield  {title} {\bibinfo {title} {Phenomenology of the prethermal
  many-body localized regime},\ }\href
  {https://doi.org/10.1103/PhysRevLett.131.106301} {\bibfield  {journal}
  {\bibinfo  {journal} {Phys. Rev. Lett.}\ }\textbf {\bibinfo {volume} {131}},\
  \bibinfo {pages} {106301} (\bibinfo {year} {2023})}\BibitemShut {NoStop}%
\bibitem [{\citenamefont {L\'{e}onard}\ \emph {et~al.}(2023)\citenamefont
  {L\'{e}onard}, \citenamefont {Kim}, \citenamefont {Rispoli}, \citenamefont
  {Lukin}, \citenamefont {Schittko}, \citenamefont {Kwan}, \citenamefont
  {Demler}, \citenamefont {Sels},\ and\ \citenamefont
  {Greiner}}]{Leonard2023NP}%
  \BibitemOpen
  \bibfield  {author} {\bibinfo {author} {\bibfnamefont {J.}~\bibnamefont
  {L\'{e}onard}}, \bibinfo {author} {\bibfnamefont {S.}~\bibnamefont {Kim}},
  \bibinfo {author} {\bibfnamefont {M.}~\bibnamefont {Rispoli}}, \bibinfo
  {author} {\bibfnamefont {A.}~\bibnamefont {Lukin}}, \bibinfo {author}
  {\bibfnamefont {R.}~\bibnamefont {Schittko}}, \bibinfo {author}
  {\bibfnamefont {J.}~\bibnamefont {Kwan}}, \bibinfo {author} {\bibfnamefont
  {E.}~\bibnamefont {Demler}}, \bibinfo {author} {\bibfnamefont
  {D.}~\bibnamefont {Sels}},\ and\ \bibinfo {author} {\bibfnamefont
  {M.}~\bibnamefont {Greiner}},\ }\bibfield  {title} {\bibinfo {title} {Probing
  the onset of quantum avalanches in a many-body localized system},\ }\href
  {https://doi.org/10.1038/s41567-022-01887-3} {\bibfield  {journal} {\bibinfo
  {journal} {Nat. Phys.}\ }\textbf {\bibinfo {volume} {19}},\ \bibinfo {pages}
  {481} (\bibinfo {year} {2023})}\BibitemShut {NoStop}%
\bibitem [{\citenamefont {Else}\ \emph
  {et~al.}(2017{\natexlab{a}})\citenamefont {Else}, \citenamefont {Fendley},
  \citenamefont {Kemp},\ and\ \citenamefont {Nayak}}]{Else2017Prethermal}%
  \BibitemOpen
  \bibfield  {author} {\bibinfo {author} {\bibfnamefont {D.~V.}\ \bibnamefont
  {Else}}, \bibinfo {author} {\bibfnamefont {P.}~\bibnamefont {Fendley}},
  \bibinfo {author} {\bibfnamefont {J.}~\bibnamefont {Kemp}},\ and\ \bibinfo
  {author} {\bibfnamefont {C.}~\bibnamefont {Nayak}},\ }\bibfield  {title}
  {\bibinfo {title} {Prethermal strong zero modes and topological qubits},\
  }\href {https://doi.org/10.1103/PhysRevX.7.041062} {\bibfield  {journal}
  {\bibinfo  {journal} {Phys. Rev. X}\ }\textbf {\bibinfo {volume} {7}},\
  \bibinfo {pages} {041062} (\bibinfo {year} {2017}{\natexlab{a}})}\BibitemShut
  {NoStop}%
\bibitem [{\citenamefont {Parker}\ \emph {et~al.}(2019)\citenamefont {Parker},
  \citenamefont {Vasseur},\ and\ \citenamefont
  {Scaffidi}}]{Parker2019Topologically}%
  \BibitemOpen
  \bibfield  {author} {\bibinfo {author} {\bibfnamefont {D.~E.}\ \bibnamefont
  {Parker}}, \bibinfo {author} {\bibfnamefont {R.}~\bibnamefont {Vasseur}},\
  and\ \bibinfo {author} {\bibfnamefont {T.}~\bibnamefont {Scaffidi}},\
  }\bibfield  {title} {\bibinfo {title} {Topologically protected long edge
  coherence times in symmetry-broken phases},\ }\href
  {https://doi.org/10.1103/PhysRevLett.122.240605} {\bibfield  {journal}
  {\bibinfo  {journal} {Phys. Rev. Lett.}\ }\textbf {\bibinfo {volume} {122}},\
  \bibinfo {pages} {240605} (\bibinfo {year} {2019})}\BibitemShut {NoStop}%
\bibitem [{\citenamefont {Kemp}\ \emph {et~al.}(2020)\citenamefont {Kemp},
  \citenamefont {Yao},\ and\ \citenamefont {Laumann}}]{Kemp2020Symmetry}%
  \BibitemOpen
  \bibfield  {author} {\bibinfo {author} {\bibfnamefont {J.}~\bibnamefont
  {Kemp}}, \bibinfo {author} {\bibfnamefont {N.~Y.}\ \bibnamefont {Yao}},\ and\
  \bibinfo {author} {\bibfnamefont {C.~R.}\ \bibnamefont {Laumann}},\
  }\bibfield  {title} {\bibinfo {title} {Symmetry-enhanced boundary qubits at
  infinite temperature},\ }\href
  {https://doi.org/10.1103/physrevlett.125.200506} {\bibfield  {journal}
  {\bibinfo  {journal} {Phys. Rev. Lett.}\ }\textbf {\bibinfo {volume} {125}},\
  \bibinfo {pages} {200506} (\bibinfo {year} {2020})}\BibitemShut {NoStop}%
\bibitem [{\citenamefont {Fendley}(2012)}]{Fendley2012Parafermionic}%
  \BibitemOpen
  \bibfield  {author} {\bibinfo {author} {\bibfnamefont {P.}~\bibnamefont
  {Fendley}},\ }\bibfield  {title} {\bibinfo {title} {Parafermionic edge zero
  modes in zn-invariant spin chains},\ }\href
  {https://doi.org/10.1088/1742-5468/2012/11/P11020} {\bibfield  {journal}
  {\bibinfo  {journal} {J. Stat. Mech. Theor. Exp.}\ }\textbf {\bibinfo
  {volume} {2012}},\ \bibinfo {pages} {P11020} (\bibinfo {year}
  {2012})}\BibitemShut {NoStop}%
\bibitem [{\citenamefont {Fendley}(2016)}]{Fendley2016Strong}%
  \BibitemOpen
  \bibfield  {author} {\bibinfo {author} {\bibfnamefont {P.}~\bibnamefont
  {Fendley}},\ }\bibfield  {title} {\bibinfo {title} {Strong zero modes and
  eigenstate phase transitions in the xyz/interacting majorana chain},\ }\href
  {https://doi.org/10.1088/1751-8113/49/30/30LT01} {\bibfield  {journal}
  {\bibinfo  {journal} {J. Phys. A Math. Theor.}\ }\textbf {\bibinfo {volume}
  {49}},\ \bibinfo {pages} {30LT01} (\bibinfo {year} {2016})}\BibitemShut
  {NoStop}%
\bibitem [{\citenamefont {Kemp}\ \emph {et~al.}(2017)\citenamefont {Kemp},
  \citenamefont {Yao}, \citenamefont {Laumann},\ and\ \citenamefont
  {Fendley}}]{Kemp2017Long}%
  \BibitemOpen
  \bibfield  {author} {\bibinfo {author} {\bibfnamefont {J.}~\bibnamefont
  {Kemp}}, \bibinfo {author} {\bibfnamefont {N.~Y.}\ \bibnamefont {Yao}},
  \bibinfo {author} {\bibfnamefont {C.~R.}\ \bibnamefont {Laumann}},\ and\
  \bibinfo {author} {\bibfnamefont {P.}~\bibnamefont {Fendley}},\ }\bibfield
  {title} {\bibinfo {title} {Long coherence times for edge spins},\ }\href
  {https://doi.org/10.1088/1742-5468/aa73f0} {\bibfield  {journal} {\bibinfo
  {journal} {J. Stat. Mech.}\ }\textbf {\bibinfo {volume} {2017}},\ \bibinfo
  {pages} {063105} (\bibinfo {year} {2017})}\BibitemShut {NoStop}%
\bibitem [{\citenamefont {Zhang}\ \emph {et~al.}(2022)\citenamefont {Zhang},
  \citenamefont {Jiang}, \citenamefont {Deng}, \citenamefont {Wang},
  \citenamefont {Chen}, \citenamefont {Zhang}, \citenamefont {Ren},
  \citenamefont {Dong}, \citenamefont {Xu}, \citenamefont {Gao} \emph
  {et~al.}}]{Zhang2022Digital}%
  \BibitemOpen
  \bibfield  {author} {\bibinfo {author} {\bibfnamefont {X.}~\bibnamefont
  {Zhang}}, \bibinfo {author} {\bibfnamefont {W.}~\bibnamefont {Jiang}},
  \bibinfo {author} {\bibfnamefont {J.}~\bibnamefont {Deng}}, \bibinfo {author}
  {\bibfnamefont {K.}~\bibnamefont {Wang}}, \bibinfo {author} {\bibfnamefont
  {J.}~\bibnamefont {Chen}}, \bibinfo {author} {\bibfnamefont {P.}~\bibnamefont
  {Zhang}}, \bibinfo {author} {\bibfnamefont {W.}~\bibnamefont {Ren}}, \bibinfo
  {author} {\bibfnamefont {H.}~\bibnamefont {Dong}}, \bibinfo {author}
  {\bibfnamefont {S.}~\bibnamefont {Xu}}, \bibinfo {author} {\bibfnamefont
  {Y.}~\bibnamefont {Gao}}, \emph {et~al.},\ }\bibfield  {title} {\bibinfo
  {title} {Digital quantum simulation of floquet symmetry-protected topological
  phases},\ }\href {https://doi.org/https://doi.org/10.1038/s41586-022-04854-3}
  {\bibfield  {journal} {\bibinfo  {journal} {Nature}\ }\textbf {\bibinfo
  {volume} {607}},\ \bibinfo {pages} {468} (\bibinfo {year}
  {2022})}\BibitemShut {NoStop}%
\bibitem [{\citenamefont {Dumitrescu}\ \emph {et~al.}(2022)\citenamefont
  {Dumitrescu}, \citenamefont {Bohnet}, \citenamefont {Gaebler}, \citenamefont
  {Hankin}, \citenamefont {Hayes}, \citenamefont {Kumar}, \citenamefont
  {Neyenhuis}, \citenamefont {Vasseur},\ and\ \citenamefont
  {Potter}}]{Dumitrescu2022Dynamical}%
  \BibitemOpen
  \bibfield  {author} {\bibinfo {author} {\bibfnamefont {P.~T.}\ \bibnamefont
  {Dumitrescu}}, \bibinfo {author} {\bibfnamefont {J.~G.}\ \bibnamefont
  {Bohnet}}, \bibinfo {author} {\bibfnamefont {J.~P.}\ \bibnamefont {Gaebler}},
  \bibinfo {author} {\bibfnamefont {A.}~\bibnamefont {Hankin}}, \bibinfo
  {author} {\bibfnamefont {D.}~\bibnamefont {Hayes}}, \bibinfo {author}
  {\bibfnamefont {A.}~\bibnamefont {Kumar}}, \bibinfo {author} {\bibfnamefont
  {B.}~\bibnamefont {Neyenhuis}}, \bibinfo {author} {\bibfnamefont
  {R.}~\bibnamefont {Vasseur}},\ and\ \bibinfo {author} {\bibfnamefont {A.~C.}\
  \bibnamefont {Potter}},\ }\bibfield  {title} {\bibinfo {title} {Dynamical
  topological phase realized in a trapped-ion quantum simulator},\ }\href
  {https://doi.org/https://doi.org/10.1038/s41586-022-04853-4} {\bibfield
  {journal} {\bibinfo  {journal} {Nature}\ }\textbf {\bibinfo {volume} {607}},\
  \bibinfo {pages} {463} (\bibinfo {year} {2022})}\BibitemShut {NoStop}%
\bibitem [{\citenamefont {Mi}\ \emph {et~al.}(2022{\natexlab{a}})\citenamefont
  {Mi}, \citenamefont {Sonner}, \citenamefont {Niu} \emph
  {et~al.}}]{Mi2022Noise}%
  \BibitemOpen
  \bibfield  {author} {\bibinfo {author} {\bibfnamefont {X.}~\bibnamefont
  {Mi}}, \bibinfo {author} {\bibfnamefont {M.}~\bibnamefont {Sonner}}, \bibinfo
  {author} {\bibfnamefont {M.~Y.}\ \bibnamefont {Niu}}, \emph {et~al.},\
  }\bibfield  {title} {\bibinfo {title} {Noise-resilient edge modes on a chain
  of superconducting qubits},\ }\href {https://doi.org/10.1126/science.abq5769}
  {\bibfield  {journal} {\bibinfo  {journal} {Science}\ }\textbf {\bibinfo
  {volume} {378}},\ \bibinfo {pages} {785} (\bibinfo {year}
  {2022}{\natexlab{a}})}\BibitemShut {NoStop}%
\bibitem [{\citenamefont {Briegel}\ and\ \citenamefont
  {Raussendorf}(2001)}]{Briegel2001Persistent}%
  \BibitemOpen
  \bibfield  {author} {\bibinfo {author} {\bibfnamefont {H.~J.}\ \bibnamefont
  {Briegel}}\ and\ \bibinfo {author} {\bibfnamefont {R.}~\bibnamefont
  {Raussendorf}},\ }\bibfield  {title} {\bibinfo {title} {Persistent
  entanglement in arrays of interacting particles},\ }\href
  {https://doi.org/10.1103/PhysRevLett.86.910} {\bibfield  {journal} {\bibinfo
  {journal} {Phys. Rev. Lett.}\ }\textbf {\bibinfo {volume} {86}},\ \bibinfo
  {pages} {910} (\bibinfo {year} {2001})}\BibitemShut {NoStop}%
\bibitem [{\citenamefont {Mi}\ \emph {et~al.}(2022{\natexlab{b}})\citenamefont
  {Mi}, \citenamefont {Ippoliti}, \citenamefont {Quintana}, \citenamefont
  {Greene}, \citenamefont {Chen}, \citenamefont {Gross}, \citenamefont {Arute},
  \citenamefont {Arya}, \citenamefont {Atalaya}, \citenamefont {Babbush} \emph
  {et~al.}}]{mi2022time}%
  \BibitemOpen
  \bibfield  {author} {\bibinfo {author} {\bibfnamefont {X.}~\bibnamefont
  {Mi}}, \bibinfo {author} {\bibfnamefont {M.}~\bibnamefont {Ippoliti}},
  \bibinfo {author} {\bibfnamefont {C.}~\bibnamefont {Quintana}}, \bibinfo
  {author} {\bibfnamefont {A.}~\bibnamefont {Greene}}, \bibinfo {author}
  {\bibfnamefont {Z.}~\bibnamefont {Chen}}, \bibinfo {author} {\bibfnamefont
  {J.}~\bibnamefont {Gross}}, \bibinfo {author} {\bibfnamefont
  {F.}~\bibnamefont {Arute}}, \bibinfo {author} {\bibfnamefont
  {K.}~\bibnamefont {Arya}}, \bibinfo {author} {\bibfnamefont {J.}~\bibnamefont
  {Atalaya}}, \bibinfo {author} {\bibfnamefont {R.}~\bibnamefont {Babbush}},
  \emph {et~al.},\ }\bibfield  {title} {\bibinfo {title} {Time-crystalline
  eigenstate order on a quantum processor},\ }\href
  {https://doi.org/https://doi.org/10.1038/s41586-021-04257-w} {\bibfield
  {journal} {\bibinfo  {journal} {Nature}\ }\textbf {\bibinfo {volume} {601}},\
  \bibinfo {pages} {531} (\bibinfo {year} {2022}{\natexlab{b}})}\BibitemShut
  {NoStop}%
\bibitem [{\citenamefont {Else}\ \emph
  {et~al.}(2017{\natexlab{b}})\citenamefont {Else}, \citenamefont {Bauer},\
  and\ \citenamefont {Nayak}}]{Else2017Prethermal2}%
  \BibitemOpen
  \bibfield  {author} {\bibinfo {author} {\bibfnamefont {D.~V.}\ \bibnamefont
  {Else}}, \bibinfo {author} {\bibfnamefont {B.}~\bibnamefont {Bauer}},\ and\
  \bibinfo {author} {\bibfnamefont {C.}~\bibnamefont {Nayak}},\ }\bibfield
  {title} {\bibinfo {title} {Prethermal phases of matter protected by
  time-translation symmetry},\ }\href
  {https://doi.org/10.1103/PhysRevX.7.011026} {\bibfield  {journal} {\bibinfo
  {journal} {Phys. Rev. X}\ }\textbf {\bibinfo {volume} {7}},\ \bibinfo {pages}
  {011026} (\bibinfo {year} {2017}{\natexlab{b}})}\BibitemShut {NoStop}%
\bibitem [{\citenamefont {Yin}\ and\ \citenamefont
  {Lucas}(2023)}]{Yin2023Prethermalization}%
  \BibitemOpen
  \bibfield  {author} {\bibinfo {author} {\bibfnamefont {C.}~\bibnamefont
  {Yin}}\ and\ \bibinfo {author} {\bibfnamefont {A.}~\bibnamefont {Lucas}},\
  }\bibfield  {title} {\bibinfo {title} {Prethermalization and the local
  robustness of gapped systems},\ }\href
  {https://doi.org/10.1103/PhysRevLett.131.050402} {\bibfield  {journal}
  {\bibinfo  {journal} {Phys. Rev. Lett.}\ }\textbf {\bibinfo {volume} {131}},\
  \bibinfo {pages} {050402} (\bibinfo {year} {2023})}\BibitemShut {NoStop}%
\bibitem [{\citenamefont {Roushan}\ \emph {et~al.}(2017)\citenamefont
  {Roushan}, \citenamefont {Neill}, \citenamefont {Tangpanitanon} \emph
  {et~al.}}]{Roushan2017Spectroscopic}%
  \BibitemOpen
  \bibfield  {author} {\bibinfo {author} {\bibfnamefont {P.}~\bibnamefont
  {Roushan}}, \bibinfo {author} {\bibfnamefont {C.}~\bibnamefont {Neill}},
  \bibinfo {author} {\bibfnamefont {J.}~\bibnamefont {Tangpanitanon}}, \emph
  {et~al.},\ }\bibfield  {title} {\bibinfo {title} {Spectroscopic signatures of
  localization with interacting photons in superconducting qubits},\ }\href
  {https://doi.org/10.1126/science.aao1401} {\bibfield  {journal} {\bibinfo
  {journal} {Science}\ }\textbf {\bibinfo {volume} {358}},\ \bibinfo {pages}
  {1175} (\bibinfo {year} {2017})}\BibitemShut {NoStop}%
\bibitem [{\citenamefont {Xu}\ \emph {et~al.}(2023)\citenamefont {Xu},
  \citenamefont {Sun}, \citenamefont {Wang} \emph {et~al.}}]{Xu2023CPL}%
  \BibitemOpen
  \bibfield  {author} {\bibinfo {author} {\bibfnamefont {S.}~\bibnamefont
  {Xu}}, \bibinfo {author} {\bibfnamefont {Z.-Z.}\ \bibnamefont {Sun}},
  \bibinfo {author} {\bibfnamefont {K.}~\bibnamefont {Wang}}, \emph {et~al.},\
  }\bibfield  {title} {\bibinfo {title} {{Digital Simulation of Projective
  Non-Abelian Anyons with 68 Superconducting Qubits}},\ }\href
  {https://doi.org/10.1088/0256-307X/40/6/060301} {\bibfield  {journal}
  {\bibinfo  {journal} {Chinese Phys. Lett.}\ }\textbf {\bibinfo {volume}
  {40}},\ \bibinfo {pages} {060301} (\bibinfo {year} {2023})}\BibitemShut
  {NoStop}%
\bibitem [{\citenamefont {Wildeboer}\ \emph {et~al.}(2022)\citenamefont
  {Wildeboer}, \citenamefont {Iadecola},\ and\ \citenamefont
  {Williamson}}]{Wildeboer2022Symmetry}%
  \BibitemOpen
  \bibfield  {author} {\bibinfo {author} {\bibfnamefont {J.}~\bibnamefont
  {Wildeboer}}, \bibinfo {author} {\bibfnamefont {T.}~\bibnamefont
  {Iadecola}},\ and\ \bibinfo {author} {\bibfnamefont {D.~J.}\ \bibnamefont
  {Williamson}},\ }\bibfield  {title} {\bibinfo {title} {Symmetry-protected
  infinite-temperature quantum memory from subsystem codes},\ }\href
  {https://doi.org/10.1103/PRXQuantum.3.020330} {\bibfield  {journal} {\bibinfo
   {journal} {PRX Quantum}\ }\textbf {\bibinfo {volume} {3}},\ \bibinfo {pages}
  {020330} (\bibinfo {year} {2022})}\BibitemShut {NoStop}%
\bibitem [{\citenamefont {Jiang}\ \emph {et~al.}(2024)\citenamefont {Jiang},
  \citenamefont {Yuan}, \citenamefont {Jiang}, \citenamefont {Deng},\ and\
  \citenamefont {Machado}}]{Jiang2024Prethermal}%
  \BibitemOpen
  \bibfield  {author} {\bibinfo {author} {\bibfnamefont {S.}~\bibnamefont
  {Jiang}}, \bibinfo {author} {\bibfnamefont {D.}~\bibnamefont {Yuan}},
  \bibinfo {author} {\bibfnamefont {W.}~\bibnamefont {Jiang}}, \bibinfo
  {author} {\bibfnamefont {D.-L.}\ \bibnamefont {Deng}},\ and\ \bibinfo
  {author} {\bibfnamefont {F.}~\bibnamefont {Machado}},\ }\href@noop {}
  {\bibinfo {title} {Prethermal time-crystalline corner modes}} (\bibinfo
  {year} {2024}),\ \Eprint {https://arxiv.org/abs/2406.01686} {arXiv:2406.01686
  [quant-ph]} \BibitemShut {NoStop}%
\bibitem [{\citenamefont {Else}\ \emph {et~al.}(2020)\citenamefont {Else},
  \citenamefont {Ho},\ and\ \citenamefont {Dumitrescu}}]{Else2020Long-Lived}%
  \BibitemOpen
  \bibfield  {author} {\bibinfo {author} {\bibfnamefont {D.~V.}\ \bibnamefont
  {Else}}, \bibinfo {author} {\bibfnamefont {W.~W.}\ \bibnamefont {Ho}},\ and\
  \bibinfo {author} {\bibfnamefont {P.~T.}\ \bibnamefont {Dumitrescu}},\
  }\bibfield  {title} {\bibinfo {title} {Long-lived interacting phases of
  matter protected by multiple time-translation symmetries in quasiperiodically
  driven systems},\ }\href {https://doi.org/10.1103/PhysRevX.10.021032}
  {\bibfield  {journal} {\bibinfo  {journal} {Phys. Rev. X}\ }\textbf {\bibinfo
  {volume} {10}},\ \bibinfo {pages} {021032} (\bibinfo {year}
  {2020})}\BibitemShut {NoStop}%
\bibitem [{\citenamefont {Friedman}\ \emph {et~al.}(2022)\citenamefont
  {Friedman}, \citenamefont {Ware}, \citenamefont {Vasseur},\ and\
  \citenamefont {Potter}}]{Friedman2022Topological}%
  \BibitemOpen
  \bibfield  {author} {\bibinfo {author} {\bibfnamefont {A.~J.}\ \bibnamefont
  {Friedman}}, \bibinfo {author} {\bibfnamefont {B.}~\bibnamefont {Ware}},
  \bibinfo {author} {\bibfnamefont {R.}~\bibnamefont {Vasseur}},\ and\ \bibinfo
  {author} {\bibfnamefont {A.~C.}\ \bibnamefont {Potter}},\ }\bibfield  {title}
  {\bibinfo {title} {Topological edge modes without symmetry in
  quasiperiodically driven spin chains},\ }\href
  {https://doi.org/10.1103/PhysRevB.105.115117} {\bibfield  {journal} {\bibinfo
   {journal} {Phys. Rev. B}\ }\textbf {\bibinfo {volume} {105}},\ \bibinfo
  {pages} {115117} (\bibinfo {year} {2022})}\BibitemShut {NoStop}%
\bibitem [{\citenamefont {Koch}\ \emph {et~al.}(2007)\citenamefont {Koch},
  \citenamefont {Yu}, \citenamefont {Gambetta}, \citenamefont {Houck},
  \citenamefont {Schuster}, \citenamefont {Majer}, \citenamefont {Blais},
  \citenamefont {Devoret}, \citenamefont {Girvin},\ and\ \citenamefont
  {Schoelkopf}}]{KochPRA2007}%
  \BibitemOpen
  \bibfield  {author} {\bibinfo {author} {\bibfnamefont {J.}~\bibnamefont
  {Koch}}, \bibinfo {author} {\bibfnamefont {T.~M.}\ \bibnamefont {Yu}},
  \bibinfo {author} {\bibfnamefont {J.}~\bibnamefont {Gambetta}}, \bibinfo
  {author} {\bibfnamefont {A.~A.}\ \bibnamefont {Houck}}, \bibinfo {author}
  {\bibfnamefont {D.~I.}\ \bibnamefont {Schuster}}, \bibinfo {author}
  {\bibfnamefont {J.}~\bibnamefont {Majer}}, \bibinfo {author} {\bibfnamefont
  {A.}~\bibnamefont {Blais}}, \bibinfo {author} {\bibfnamefont {M.~H.}\
  \bibnamefont {Devoret}}, \bibinfo {author} {\bibfnamefont {S.~M.}\
  \bibnamefont {Girvin}},\ and\ \bibinfo {author} {\bibfnamefont {R.~J.}\
  \bibnamefont {Schoelkopf}},\ }\bibfield  {title} {\bibinfo {title}
  {Charge-insensitive qubit design derived from the cooper pair box},\ }\href
  {https://doi.org/10.1103/PhysRevA.76.042319} {\bibfield  {journal} {\bibinfo
  {journal} {Phys. Rev. A}\ }\textbf {\bibinfo {volume} {76}},\ \bibinfo
  {pages} {042319} (\bibinfo {year} {2007})}\BibitemShut {NoStop}%
\bibitem [{\citenamefont {Yan}\ \emph {et~al.}(2018)\citenamefont {Yan},
  \citenamefont {Krantz}, \citenamefont {Sung}, \citenamefont {Kjaergaard},
  \citenamefont {Campbell}, \citenamefont {Orlando}, \citenamefont
  {Gustavsson},\ and\ \citenamefont {Oliver}}]{Fei2018PRB}%
  \BibitemOpen
  \bibfield  {author} {\bibinfo {author} {\bibfnamefont {F.}~\bibnamefont
  {Yan}}, \bibinfo {author} {\bibfnamefont {P.}~\bibnamefont {Krantz}},
  \bibinfo {author} {\bibfnamefont {Y.}~\bibnamefont {Sung}}, \bibinfo {author}
  {\bibfnamefont {M.}~\bibnamefont {Kjaergaard}}, \bibinfo {author}
  {\bibfnamefont {D.~L.}\ \bibnamefont {Campbell}}, \bibinfo {author}
  {\bibfnamefont {T.~P.}\ \bibnamefont {Orlando}}, \bibinfo {author}
  {\bibfnamefont {S.}~\bibnamefont {Gustavsson}},\ and\ \bibinfo {author}
  {\bibfnamefont {W.~D.}\ \bibnamefont {Oliver}},\ }\bibfield  {title}
  {\bibinfo {title} {Tunable coupling scheme for implementing high-fidelity
  two-qubit gates},\ }\href {https://doi.org/10.1103/PhysRevApplied.10.054062}
  {\bibfield  {journal} {\bibinfo  {journal} {Phys. Rev. Appl.}\ }\textbf
  {\bibinfo {volume} {10}},\ \bibinfo {pages} {054062} (\bibinfo {year}
  {2018})}\BibitemShut {NoStop}%
\bibitem [{\citenamefont {Foxen}\ \emph {et~al.}(2020)\citenamefont {Foxen},
  \citenamefont {Neill}, \citenamefont {Dunsworth} \emph
  {et~al.}}]{PhysRevLett.125.120504}%
  \BibitemOpen
  \bibfield  {author} {\bibinfo {author} {\bibfnamefont {B.}~\bibnamefont
  {Foxen}}, \bibinfo {author} {\bibfnamefont {C.}~\bibnamefont {Neill}},
  \bibinfo {author} {\bibfnamefont {A.}~\bibnamefont {Dunsworth}}, \emph
  {et~al.} (\bibinfo {collaboration} {Google AI Quantum}),\ }\bibfield  {title}
  {\bibinfo {title} {Demonstrating a continuous set of two-qubit gates for
  near-term quantum algorithms},\ }\href
  {https://doi.org/10.1103/PhysRevLett.125.120504} {\bibfield  {journal}
  {\bibinfo  {journal} {Phys. Rev. Lett.}\ }\textbf {\bibinfo {volume} {125}},\
  \bibinfo {pages} {120504} (\bibinfo {year} {2020})}\BibitemShut {NoStop}%
\bibitem [{\citenamefont {Magnus}(1954)}]{Magnus1954exponential}%
  \BibitemOpen
  \bibfield  {author} {\bibinfo {author} {\bibfnamefont {W.}~\bibnamefont
  {Magnus}},\ }\bibfield  {title} {\bibinfo {title} {On the exponential
  solution of differential equations for a linear operator},\ }\href
  {https://doi.org/https://doi.org/10.1002/cpa.3160070404} {\bibfield
  {journal} {\bibinfo  {journal} {Commun. Pure Appl. Math.}\ }\textbf {\bibinfo
  {volume} {7}},\ \bibinfo {pages} {649} (\bibinfo {year} {1954})}\BibitemShut
  {NoStop}%
\bibitem [{\citenamefont {Blanes}\ \emph {et~al.}(2009)\citenamefont {Blanes},
  \citenamefont {Casas}, \citenamefont {Oteo},\ and\ \citenamefont
  {Ros}}]{Blanes2009Magnus}%
  \BibitemOpen
  \bibfield  {author} {\bibinfo {author} {\bibfnamefont {S.}~\bibnamefont
  {Blanes}}, \bibinfo {author} {\bibfnamefont {F.}~\bibnamefont {Casas}},
  \bibinfo {author} {\bibfnamefont {J.}~\bibnamefont {Oteo}},\ and\ \bibinfo
  {author} {\bibfnamefont {J.}~\bibnamefont {Ros}},\ }\bibfield  {title}
  {\bibinfo {title} {The magnus expansion and some of its applications},\
  }\href {https://doi.org/10.1016/j.physrep.2008.11.001} {\bibfield  {journal}
  {\bibinfo  {journal} {Phys. Rep.}\ }\textbf {\bibinfo {volume} {470}},\
  \bibinfo {pages} {151} (\bibinfo {year} {2009})}\BibitemShut {NoStop}%
\end{thebibliography}%


\begin{thebibliography}{21}%
\makeatletter
\providecommand \@ifxundefined [1]{%
 \@ifx{#1\undefined}
}%
\providecommand \@ifnum [1]{%
 \ifnum #1\expandafter \@firstoftwo
 \else \expandafter \@secondoftwo
 \fi
}%
\providecommand \@ifx [1]{%
 \ifx #1\expandafter \@firstoftwo
 \else \expandafter \@secondoftwo
 \fi
}%
\providecommand \natexlab [1]{#1}%
\providecommand \enquote  [1]{``#1''}%
\providecommand \bibnamefont  [1]{#1}%
\providecommand \bibfnamefont [1]{#1}%
\providecommand \citenamefont [1]{#1}%
\providecommand \href@noop [0]{\@secondoftwo}%
\providecommand \href [0]{\begingroup \@sanitize@url \@href}%
\providecommand \@href[1]{\@@startlink{#1}\@@href}%
\providecommand \@@href[1]{\endgroup#1\@@endlink}%
\providecommand \@sanitize@url [0]{\catcode `\\12\catcode `\$12\catcode
  `\&12\catcode `\#12\catcode `\^12\catcode `\_12\catcode `\%12\relax}%
\providecommand \@@startlink[1]{}%
\providecommand \@@endlink[0]{}%
\providecommand \url  [0]{\begingroup\@sanitize@url \@url }%
\providecommand \@url [1]{\endgroup\@href {#1}{\urlprefix }}%
\providecommand \urlprefix  [0]{URL }%
\providecommand \Eprint [0]{\href }%
\providecommand \doibase [0]{https://doi.org/}%
\providecommand \selectlanguage [0]{\@gobble}%
\providecommand \bibinfo  [0]{\@secondoftwo}%
\providecommand \bibfield  [0]{\@secondoftwo}%
\providecommand \translation [1]{[#1]}%
\providecommand \BibitemOpen [0]{}%
\providecommand \bibitemStop [0]{}%
\providecommand \bibitemNoStop [0]{.\EOS\space}%
\providecommand \EOS [0]{\spacefactor3000\relax}%
\providecommand \BibitemShut  [1]{\csname bibitem#1\endcsname}%
\let\auto@bib@innerbib\@empty
\bibitem [{\citenamefont {Fendley}(2016)}]{Fendley2016Strong}%
  \BibitemOpen
  \bibfield  {author} {\bibinfo {author} {\bibfnamefont {P.}~\bibnamefont
  {Fendley}},\ }\bibfield  {title} {\bibinfo {title} {Strong zero modes and
  eigenstate phase transitions in the xyz/interacting majorana chain},\ }\href
  {https://doi.org/10.1088/1751-8113/49/30/30LT01} {\bibfield  {journal}
  {\bibinfo  {journal} {J. Phys. A}\ }\textbf {\bibinfo {volume} {49}},\
  \bibinfo {pages} {30LT01} (\bibinfo {year} {2016})}\BibitemShut {NoStop}%
\bibitem [{\citenamefont {Kemp}\ \emph {et~al.}(2017)\citenamefont {Kemp},
  \citenamefont {Yao}, \citenamefont {Laumann},\ and\ \citenamefont
  {Fendley}}]{Kemp2017Long}%
  \BibitemOpen
  \bibfield  {author} {\bibinfo {author} {\bibfnamefont {J.}~\bibnamefont
  {Kemp}}, \bibinfo {author} {\bibfnamefont {N.~Y.}\ \bibnamefont {Yao}},
  \bibinfo {author} {\bibfnamefont {C.~R.}\ \bibnamefont {Laumann}},\ and\
  \bibinfo {author} {\bibfnamefont {P.}~\bibnamefont {Fendley}},\ }\bibfield
  {title} {\bibinfo {title} {Long coherence times for edge spins},\ }\href
  {https://doi.org/10.1088/1742-5468/aa73f0} {\bibfield  {journal} {\bibinfo
  {journal} {J. Stat. Mech.}\ }\textbf {\bibinfo {volume} {2017}},\ \bibinfo
  {pages} {063105} (\bibinfo {year} {2017})}\BibitemShut {NoStop}%
\bibitem [{\citenamefont {Kemp}\ \emph {et~al.}(2020)\citenamefont {Kemp},
  \citenamefont {Yao},\ and\ \citenamefont {Laumann}}]{Kemp2020Symmetry}%
  \BibitemOpen
  \bibfield  {author} {\bibinfo {author} {\bibfnamefont {J.}~\bibnamefont
  {Kemp}}, \bibinfo {author} {\bibfnamefont {N.~Y.}\ \bibnamefont {Yao}},\ and\
  \bibinfo {author} {\bibfnamefont {C.~R.}\ \bibnamefont {Laumann}},\
  }\bibfield  {title} {\bibinfo {title} {Symmetry-enhanced boundary qubits at
  infinite temperature},\ }\href
  {https://doi.org/10.1103/physrevlett.125.200506} {\bibfield  {journal}
  {\bibinfo  {journal} {Phys. Rev. Lett.}\ }\textbf {\bibinfo {volume} {125}},\
  \bibinfo {pages} {200506} (\bibinfo {year} {2020})}\BibitemShut {NoStop}%
\bibitem [{\citenamefont {Else}\ \emph
  {et~al.}(2017{\natexlab{a}})\citenamefont {Else}, \citenamefont {Bauer},\
  and\ \citenamefont {Nayak}}]{Else2017Prethermal2}%
  \BibitemOpen
  \bibfield  {author} {\bibinfo {author} {\bibfnamefont {D.~V.}\ \bibnamefont
  {Else}}, \bibinfo {author} {\bibfnamefont {B.}~\bibnamefont {Bauer}},\ and\
  \bibinfo {author} {\bibfnamefont {C.}~\bibnamefont {Nayak}},\ }\bibfield
  {title} {\bibinfo {title} {Prethermal phases of matter protected by
  time-translation symmetry},\ }\href
  {https://doi.org/10.1103/PhysRevX.7.011026} {\bibfield  {journal} {\bibinfo
  {journal} {Phys. Rev. X}\ }\textbf {\bibinfo {volume} {7}},\ \bibinfo {pages}
  {011026} (\bibinfo {year} {2017}{\natexlab{a}})}\BibitemShut {NoStop}%
\bibitem [{\citenamefont {Abanin}\ \emph {et~al.}(2017)\citenamefont {Abanin},
  \citenamefont {Roeck}, \citenamefont {Ho},\ and\ \citenamefont
  {Huveneers}}]{abanin2017rigorous}%
  \BibitemOpen
  \bibfield  {author} {\bibinfo {author} {\bibfnamefont {D.}~\bibnamefont
  {Abanin}}, \bibinfo {author} {\bibfnamefont {W.~D.}\ \bibnamefont {Roeck}},
  \bibinfo {author} {\bibfnamefont {W.~W.}\ \bibnamefont {Ho}},\ and\ \bibinfo
  {author} {\bibfnamefont {F.}~\bibnamefont {Huveneers}},\ }\bibfield  {title}
  {\bibinfo {title} {A rigorous theory of many-body prethermalization for
  periodically driven and closed quantum systems},\ }\href
  {https://doi.org/10.1007/s00220-017-2930-x} {\bibfield  {journal} {\bibinfo
  {journal} {Commun. Math. Phys.}\ }\textbf {\bibinfo {volume} {354}},\
  \bibinfo {pages} {809} (\bibinfo {year} {2017})}\BibitemShut {NoStop}%
\bibitem [{\citenamefont {Else}\ \emph
  {et~al.}(2017{\natexlab{b}})\citenamefont {Else}, \citenamefont {Fendley},
  \citenamefont {Kemp},\ and\ \citenamefont {Nayak}}]{Else2017Prethermal}%
  \BibitemOpen
  \bibfield  {author} {\bibinfo {author} {\bibfnamefont {D.~V.}\ \bibnamefont
  {Else}}, \bibinfo {author} {\bibfnamefont {P.}~\bibnamefont {Fendley}},
  \bibinfo {author} {\bibfnamefont {J.}~\bibnamefont {Kemp}},\ and\ \bibinfo
  {author} {\bibfnamefont {C.}~\bibnamefont {Nayak}},\ }\bibfield  {title}
  {\bibinfo {title} {Prethermal strong zero modes and topological qubits},\
  }\href {https://doi.org/10.1103/PhysRevX.7.041062} {\bibfield  {journal}
  {\bibinfo  {journal} {Phys. Rev. X}\ }\textbf {\bibinfo {volume} {7}},\
  \bibinfo {pages} {041062} (\bibinfo {year} {2017}{\natexlab{b}})}\BibitemShut
  {NoStop}%
\bibitem [{\citenamefont {Machado}\ \emph {et~al.}(2020)\citenamefont
  {Machado}, \citenamefont {Else}, \citenamefont {Kahanamoku-Meyer},
  \citenamefont {Nayak},\ and\ \citenamefont {Yao}}]{machado2020long}%
  \BibitemOpen
  \bibfield  {author} {\bibinfo {author} {\bibfnamefont {F.}~\bibnamefont
  {Machado}}, \bibinfo {author} {\bibfnamefont {D.~V.}\ \bibnamefont {Else}},
  \bibinfo {author} {\bibfnamefont {G.~D.}\ \bibnamefont {Kahanamoku-Meyer}},
  \bibinfo {author} {\bibfnamefont {C.}~\bibnamefont {Nayak}},\ and\ \bibinfo
  {author} {\bibfnamefont {N.~Y.}\ \bibnamefont {Yao}},\ }\bibfield  {title}
  {\bibinfo {title} {Long-range prethermal phases of nonequilibrium matter},\
  }\href {https://doi.org/10.1103/PhysRevX.10.011043} {\bibfield  {journal}
  {\bibinfo  {journal} {Phys. Rev. X}\ }\textbf {\bibinfo {volume} {10}},\
  \bibinfo {pages} {011043} (\bibinfo {year} {2020})}\BibitemShut {NoStop}%
\bibitem [{\citenamefont {Else}\ \emph {et~al.}(2020)\citenamefont {Else},
  \citenamefont {Ho},\ and\ \citenamefont {Dumitrescu}}]{Else2020Long-Lived}%
  \BibitemOpen
  \bibfield  {author} {\bibinfo {author} {\bibfnamefont {D.~V.}\ \bibnamefont
  {Else}}, \bibinfo {author} {\bibfnamefont {W.~W.}\ \bibnamefont {Ho}},\ and\
  \bibinfo {author} {\bibfnamefont {P.~T.}\ \bibnamefont {Dumitrescu}},\
  }\bibfield  {title} {\bibinfo {title} {Long-lived interacting phases of
  matter protected by multiple time-translation symmetries in quasiperiodically
  driven systems},\ }\href {https://doi.org/10.1103/PhysRevX.10.021032}
  {\bibfield  {journal} {\bibinfo  {journal} {Phys. Rev. X}\ }\textbf {\bibinfo
  {volume} {10}},\ \bibinfo {pages} {021032} (\bibinfo {year}
  {2020})}\BibitemShut {NoStop}%
\bibitem [{\citenamefont {Lieb}\ \emph {et~al.}(1961)\citenamefont {Lieb},
  \citenamefont {Schultz},\ and\ \citenamefont {Mattis}}]{Lieb1961Two}%
  \BibitemOpen
  \bibfield  {author} {\bibinfo {author} {\bibfnamefont {E.}~\bibnamefont
  {Lieb}}, \bibinfo {author} {\bibfnamefont {T.}~\bibnamefont {Schultz}},\ and\
  \bibinfo {author} {\bibfnamefont {D.}~\bibnamefont {Mattis}},\ }\bibfield
  {title} {\bibinfo {title} {Two soluble models of an antiferromagnetic
  chain},\ }\href
  {https://doi.org/https://doi.org/10.1016/0003-4916(61)90115-4} {\bibfield
  {journal} {\bibinfo  {journal} {Ann. Phys.}\ }\textbf {\bibinfo {volume}
  {16}},\ \bibinfo {pages} {407} (\bibinfo {year} {1961})}\BibitemShut
  {NoStop}%
\bibitem [{\citenamefont {Magnus}(1954)}]{Magnus1954exponential}%
  \BibitemOpen
  \bibfield  {author} {\bibinfo {author} {\bibfnamefont {W.}~\bibnamefont
  {Magnus}},\ }\bibfield  {title} {\bibinfo {title} {On the exponential
  solution of differential equations for a linear operator},\ }\href
  {https://doi.org/https://doi.org/10.1002/cpa.3160070404} {\bibfield
  {journal} {\bibinfo  {journal} {Commun. Pure Appl. Math.}\ }\textbf {\bibinfo
  {volume} {7}},\ \bibinfo {pages} {649} (\bibinfo {year} {1954})}\BibitemShut
  {NoStop}%
\bibitem [{\citenamefont {Kuwahara}\ \emph {et~al.}(2016)\citenamefont
  {Kuwahara}, \citenamefont {Mori},\ and\ \citenamefont
  {Saito}}]{Kuwahara2016Floquet}%
  \BibitemOpen
  \bibfield  {author} {\bibinfo {author} {\bibfnamefont {T.}~\bibnamefont
  {Kuwahara}}, \bibinfo {author} {\bibfnamefont {T.}~\bibnamefont {Mori}},\
  and\ \bibinfo {author} {\bibfnamefont {K.}~\bibnamefont {Saito}},\ }\bibfield
   {title} {\bibinfo {title} {Floquet{\textendash}magnus theory and generic
  transient dynamics in periodically driven many-body quantum systems},\ }\href
  {https://doi.org/10.1016/j.aop.2016.01.012} {\bibfield  {journal} {\bibinfo
  {journal} {Ann. Phys.}\ }\textbf {\bibinfo {volume} {367}},\ \bibinfo {pages}
  {96} (\bibinfo {year} {2016})}\BibitemShut {NoStop}%
\bibitem [{\citenamefont {Heyl}\ \emph {et~al.}(2019)\citenamefont {Heyl},
  \citenamefont {Hauke},\ and\ \citenamefont {Zoller}}]{Heyl2019Quantum}%
  \BibitemOpen
  \bibfield  {author} {\bibinfo {author} {\bibfnamefont {M.}~\bibnamefont
  {Heyl}}, \bibinfo {author} {\bibfnamefont {P.}~\bibnamefont {Hauke}},\ and\
  \bibinfo {author} {\bibfnamefont {P.}~\bibnamefont {Zoller}},\ }\bibfield
  {title} {\bibinfo {title} {Quantum localization bounds trotter errors in
  digital quantum simulation},\ }\href {https://doi.org/10.1126/sciadv.aau8342}
  {\bibfield  {journal} {\bibinfo  {journal} {Sci. Adv.}\ }\textbf {\bibinfo
  {volume} {5}},\ \bibinfo {pages} {eaau8342} (\bibinfo {year}
  {2019})}\BibitemShut {NoStop}%
\bibitem [{\citenamefont {Thakurathi}\ \emph {et~al.}(2013)\citenamefont
  {Thakurathi}, \citenamefont {Patel}, \citenamefont {Sen},\ and\ \citenamefont
  {Dutta}}]{Thakurathi2013Floquet}%
  \BibitemOpen
  \bibfield  {author} {\bibinfo {author} {\bibfnamefont {M.}~\bibnamefont
  {Thakurathi}}, \bibinfo {author} {\bibfnamefont {A.~A.}\ \bibnamefont
  {Patel}}, \bibinfo {author} {\bibfnamefont {D.}~\bibnamefont {Sen}},\ and\
  \bibinfo {author} {\bibfnamefont {A.}~\bibnamefont {Dutta}},\ }\bibfield
  {title} {\bibinfo {title} {Floquet generation of majorana end modes and
  topological invariants},\ }\href {https://doi.org/10.1103/PhysRevB.88.155133}
  {\bibfield  {journal} {\bibinfo  {journal} {Phys. Rev. B}\ }\textbf {\bibinfo
  {volume} {88}},\ \bibinfo {pages} {155133} (\bibinfo {year}
  {2013})}\BibitemShut {NoStop}%
\bibitem [{\citenamefont {Akila}\ \emph {et~al.}(2016)\citenamefont {Akila},
  \citenamefont {Waltner}, \citenamefont {Gutkin},\ and\ \citenamefont
  {Guhr}}]{Akila2016Particle}%
  \BibitemOpen
  \bibfield  {author} {\bibinfo {author} {\bibfnamefont {M.}~\bibnamefont
  {Akila}}, \bibinfo {author} {\bibfnamefont {D.}~\bibnamefont {Waltner}},
  \bibinfo {author} {\bibfnamefont {B.}~\bibnamefont {Gutkin}},\ and\ \bibinfo
  {author} {\bibfnamefont {T.}~\bibnamefont {Guhr}},\ }\bibfield  {title}
  {\bibinfo {title} {Particle-time duality in the kicked ising spin chain},\
  }\href {https://doi.org/10.1088/1751-8113/49/37/375101} {\bibfield  {journal}
  {\bibinfo  {journal} {J. Phys. A Math. Theor.}\ }\textbf {\bibinfo {volume}
  {49}},\ \bibinfo {pages} {375101} (\bibinfo {year} {2016})}\BibitemShut
  {NoStop}%
\bibitem [{\citenamefont {Bertini}\ \emph {et~al.}(2018)\citenamefont
  {Bertini}, \citenamefont {Kos},\ and\ \citenamefont
  {Prosen}}]{Bertini2018Exact}%
  \BibitemOpen
  \bibfield  {author} {\bibinfo {author} {\bibfnamefont {B.}~\bibnamefont
  {Bertini}}, \bibinfo {author} {\bibfnamefont {P.}~\bibnamefont {Kos}},\ and\
  \bibinfo {author} {\bibfnamefont {T.~c.~v.}\ \bibnamefont {Prosen}},\
  }\bibfield  {title} {\bibinfo {title} {Exact spectral form factor in a
  minimal model of many-body quantum chaos},\ }\href
  {https://doi.org/10.1103/PhysRevLett.121.264101} {\bibfield  {journal}
  {\bibinfo  {journal} {Phys. Rev. Lett.}\ }\textbf {\bibinfo {volume} {121}},\
  \bibinfo {pages} {264101} (\bibinfo {year} {2018})}\BibitemShut {NoStop}%
\bibitem [{\citenamefont {Lerose}\ \emph {et~al.}(2021)\citenamefont {Lerose},
  \citenamefont {Sonner},\ and\ \citenamefont {Abanin}}]{Lerose2021Scaling}%
  \BibitemOpen
  \bibfield  {author} {\bibinfo {author} {\bibfnamefont {A.}~\bibnamefont
  {Lerose}}, \bibinfo {author} {\bibfnamefont {M.}~\bibnamefont {Sonner}},\
  and\ \bibinfo {author} {\bibfnamefont {D.~A.}\ \bibnamefont {Abanin}},\
  }\bibfield  {title} {\bibinfo {title} {Scaling of temporal entanglement in
  proximity to integrability},\ }\href
  {https://doi.org/10.1103/PhysRevB.104.035137} {\bibfield  {journal} {\bibinfo
   {journal} {Phys. Rev. B}\ }\textbf {\bibinfo {volume} {104}},\ \bibinfo
  {pages} {035137} (\bibinfo {year} {2021})}\BibitemShut {NoStop}%
\bibitem [{\citenamefont {Ren}\ \emph {et~al.}(2022)\citenamefont {Ren},
  \citenamefont {Li}, \citenamefont {Xu}, \citenamefont {Wang} \emph
  {et~al.}}]{renExperimentalQuantumAdversarial2022}%
  \BibitemOpen
  \bibfield  {author} {\bibinfo {author} {\bibfnamefont {W.}~\bibnamefont
  {Ren}}, \bibinfo {author} {\bibfnamefont {W.}~\bibnamefont {Li}}, \bibinfo
  {author} {\bibfnamefont {S.}~\bibnamefont {Xu}}, \bibinfo {author}
  {\bibfnamefont {K.}~\bibnamefont {Wang}}, \emph {et~al.},\ }\bibfield
  {title} {\bibinfo {title} {Experimental quantum adversarial learning with
  programmable superconducting qubits},\ }\href
  {https://doi.org/10.1038/s43588-022-00351-9} {\bibfield  {journal} {\bibinfo
  {journal} {Nat. Comp. Sci.}\ }\textbf {\bibinfo {volume} {2}},\ \bibinfo
  {pages} {711} (\bibinfo {year} {2022})}\BibitemShut {NoStop}%
\bibitem [{\citenamefont {Neill}\ \emph {et~al.}(2021)\citenamefont {Neill},
  \citenamefont {McCourt}, \citenamefont {Mi} \emph
  {et~al.}}]{neillAccuratelyComputingElectronic2021}%
  \BibitemOpen
  \bibfield  {author} {\bibinfo {author} {\bibfnamefont {C.}~\bibnamefont
  {Neill}}, \bibinfo {author} {\bibfnamefont {T.}~\bibnamefont {McCourt}},
  \bibinfo {author} {\bibfnamefont {X.}~\bibnamefont {Mi}}, \emph {et~al.},\
  }\bibfield  {title} {\bibinfo {title} {Accurately computing the electronic
  properties of a quantum ring},\ }\href
  {https://doi.org/10.1038/s41586-021-03576-2} {\bibfield  {journal} {\bibinfo
  {journal} {Nature}\ }\textbf {\bibinfo {volume} {594}},\ \bibinfo {pages}
  {508} (\bibinfo {year} {2021})}\BibitemShut {NoStop}%
\bibitem [{\citenamefont {Mi}\ \emph {et~al.}(2022)\citenamefont {Mi},
  \citenamefont {Ippoliti}, \citenamefont {Quintana} \emph
  {et~al.}}]{miTimecrystallineEigenstateOrder2022}%
  \BibitemOpen
  \bibfield  {author} {\bibinfo {author} {\bibfnamefont {X.}~\bibnamefont
  {Mi}}, \bibinfo {author} {\bibfnamefont {M.}~\bibnamefont {Ippoliti}},
  \bibinfo {author} {\bibfnamefont {C.}~\bibnamefont {Quintana}}, \emph
  {et~al.},\ }\bibfield  {title} {\bibinfo {title} {Time-crystalline eigenstate
  order on a quantum processor},\ }\href
  {https://doi.org/10.1038/s41586-021-04257-w} {\bibfield  {journal} {\bibinfo
  {journal} {Nature}\ }\textbf {\bibinfo {volume} {601}},\ \bibinfo {pages}
  {531} (\bibinfo {year} {2022})}\BibitemShut {NoStop}%
\bibitem [{\citenamefont {Foxen}\ \emph {et~al.}(2020)\citenamefont {Foxen},
  \citenamefont {Neill}, \citenamefont {Dunsworth} \emph
  {et~al.}}]{foxenDemonstratingContinuousSet2020}%
  \BibitemOpen
  \bibfield  {author} {\bibinfo {author} {\bibfnamefont {B.}~\bibnamefont
  {Foxen}}, \bibinfo {author} {\bibfnamefont {C.}~\bibnamefont {Neill}},
  \bibinfo {author} {\bibfnamefont {A.}~\bibnamefont {Dunsworth}}, \emph
  {et~al.},\ }\bibfield  {title} {\bibinfo {title} {Demonstrating a
  {Continuous} {Set} of {Two}-qubit {Gates} for {Near}-term {Quantum}
  {Algorithms}},\ }\href {https://doi.org/10.1103/PhysRevLett.125.120504}
  {\bibfield  {journal} {\bibinfo  {journal} {Physical Review Letters}\
  }\textbf {\bibinfo {volume} {125}},\ \bibinfo {pages} {120504} (\bibinfo
  {year} {2020})}\BibitemShut {NoStop}%
\bibitem [{\citenamefont {Smolin}\ \emph {et~al.}(2012)\citenamefont {Smolin},
  \citenamefont {Gambetta},\ and\ \citenamefont
  {Smith}}]{smolinEfficientMethodComputing2012}%
  \BibitemOpen
  \bibfield  {author} {\bibinfo {author} {\bibfnamefont {J.~A.}\ \bibnamefont
  {Smolin}}, \bibinfo {author} {\bibfnamefont {J.~M.}\ \bibnamefont
  {Gambetta}},\ and\ \bibinfo {author} {\bibfnamefont {G.}~\bibnamefont
  {Smith}},\ }\bibfield  {title} {\bibinfo {title} {Efficient {Method} for
  {Computing} the {Maximum}-{Likelihood} {Quantum} {State} from {Measurements}
  with {Additive} {Gaussian} {Noise}},\ }\href
  {https://doi.org/10.1103/PhysRevLett.108.070502} {\bibfield  {journal}
  {\bibinfo  {journal} {Phys. Rev. Lett.}\ }\textbf {\bibinfo {volume} {108}},\
  \bibinfo {pages} {070502} (\bibinfo {year} {2012})}\BibitemShut {NoStop}%
\end{thebibliography}%

\clearpage
\noindent\textbf{\large{}Methods}
\vspace{2mm}
\\ {\bf Experimental setup}
\\Our experiments are performed on a two-dimensional flip-chip superconducting quantum processor, which possesses $125$ frequency-tunable transmon qubits~\cite{KochPRA2007} and $218$ tunable couplers~\cite{Fei2018PRB} between the adjacent qubits (Fig.~\ref{fig:1}a of the main text). In our experiments, we actively use $100$ qubits and $100$ couplers of them to simulate the many-body dynamics of one-dimensional disorder-free ``cluster'' Hamiltonian $H$ in Eq. (\ref{eq:1}) of the main text. Each time step of its evolution unitary $U(\delta t)$ is decomposed into combinations of single-qubit rotations and two-qubit gates.
For each qubit, a single-qubit rotation is implemented by applying a microwave pulse or a fast flux pulse, which are combined by a combiner at room temperature and then transmitted to the qubit at low temperature ($20$ mK) to rotate the qubit state along longitudinal or latitudinal lines of the Bloch sphere. Two-qubit interaction between the nearest-neighbor two qubits can be dynamically controlled by applying a fast flux pulse to the corresponding coupler, which also enables the implementation of high-fidelity two-qubit controlled-phase (CPhase) gates~\cite{PhysRevLett.125.120504}. Each qubit is capacitively coupled to a readout resonator for dispersive readout, which is designed at the frequency of around $6.4$ GHz. The processor is integrated into a printed circuit board package using the wire bonding technique. This package is further protected by magnetic shields before being mounted on the mixing chamber plate of a dilution refrigerator. See Supplementary Fig. S4 for the wiring information of the dilution refrigerator and room-temperature control electronics.
\vspace{2mm}
\\ {\bf Initial state preparation}
\\ In our experiments, the system is initialized to the ground-state manifold $\{\ket{\Psi_{\rm g}}\}$, an excited-state manifold $\{\ket{\Psi_{\rm e}}\}$, or the product states $|\bullet 00...0 \bullet\rangle$, each with a predetermined bulk state and varying edge modes.
To measure the temporal dependence of the logical operators $\tilde{Z}$ and $\tilde{X}$, we prepare the edge modes into their eigenstates, which are denoted as $\tilde{\ket{0}}, \tilde{\ket{1}}$ for $\tilde{Z}$, and $\tilde{\ket +}$, $\tilde{\ket -}$ for $\tilde{X}$.
For the ground-state case, such states are defined as 
\begin{eqnarray}
    \tilde{\ket{0}}_{\rm L}\tilde{\ket{0}}_{\rm R} &=& \prod_{i=1}^{99} {\rm CZ}_{i,i+1}\left[\ket{0}_1\left(\bigotimes_{i=2}^{99}\ket{+}_i\right)\ket{0}_{100}\right] ,  \\
    \tilde{\ket{+}}_{\rm L}\tilde{\ket{+}}_{\rm R} &=& \prod_{i=1}^{99} {\rm CZ}_{i,i+1}\left(\bigotimes_{i=1}^{100}\ket{+}_i\right) ,  
\end{eqnarray}
and the circuits for preparing these states are shown in Extended Data Fig.~\ref{fig:ex2}{a-b}. For the excited-state case, excitations are induced into the bulk by applying $X_i(\pi)$ [$Z_i(\pi)$ in Extended Data Fig.~\ref{fig:ex5}] gates on the qubit $i$. 
For the product-state case, we prepare the $\ket{000\dots 00}$ state for measuring $\{\Tilde{Z}_{\rm L}, \Tilde{Z}_{\rm R}\}$ and the $\ket{+00\dots 0+}$ state for measuring  $\{\Tilde{X}_{\rm L} , \Tilde{X}_{\rm R}\}$.

The preparation for the logical Bell state $\Tilde{\ket{0}}_{\rm L}\Tilde{\ket{0}}_{\rm R}+\rm{i}\Tilde{\ket{1}}_{\rm L}\Tilde{\ket{1}}_{\rm R}$ is more involved.
This is done by first applying a logical $\tilde{X}(-\pi/2)$ rotation on $\tilde{\ket{0}}_{\rm L}\tilde{\ket{0}}_{\rm R}$, 
and then a combination of two-qubit gates and single-qubit gates on two edge modes to effectively implement the logical controlled-NOT gate.
The total circuit for preparing the logical Bell state is shown in Extended Data Fig.~\ref{fig:ex4}{a}. 
\vspace{2mm}
\\ {\bf Characterization of Trotter errors}
\\ Quantum simulation of time evolution for the Hamiltonian $H$ using the digital circuit $U$ is prone to an accumulation of Trotter errors.
However, the finite-temperature edge modes in our model are quite robust and can be observed in the presence of these Trotter errors, making the exact simulation of $H$ unnecessary. 
Note that the evolution unitary $U$ corresponds to a Floquet Hamiltonian $H_{\rm F}$ at stroboscopic time, defined by $\exp(-iH_{\rm F}T)\equiv U$.
While $H_{\rm F}$, in general, is difficult to analyze, it can be constructed order by order via the Floquet-Maguns expansion~\cite{Magnus1954exponential, Blanes2009Magnus}, whereby the lower-order terms are sufficient to describe the short-term evolution on current noisy intermediate-scale quantum devices. 
The zeroth-order term gives $H_0+H_1$, and the first-order terms present additional many-body terms (Supplementary Section 1E). 
Hence, the Trotter errors can be considered as extra interaction terms that make the edge-bulk interaction in our model more general. 
\vspace{2mm}
\\ {\bf Transformation to Majorana fermions}
\\ The spin Hamiltonian $H=H_0+H_1$ can be transformed into two Kitaev chains of Majorana fermions.
This is done by first applying the Jordan-Wigner transformation, which maps Pauli spin operators into fermionic creation/annihilation operators, and then transforming the latter into Majorana operators $\alpha_i$, $\beta_i$ (Supplementary Section 1D). 
The total transformation reads
\begin{equation}
    \sigma_i^x = -i\alpha_i\beta_i, \quad \sigma^z_i = - \left[\prod_{j=1}^{i-1}(-i\alpha_j\beta_j)\right]\alpha_i. \label{eq:total_map}
\end{equation}
Besides $\sigma_i^x$, the three-body stabilizers and two-body interactions in $H$ are mapped into the following forms,
\begin{equation}
    \sigma_{i-1}^z\sigma_{i}^x\sigma_{i+1}^z = -i\beta_{i-1}\alpha_{i+1}, \;\sigma_i^x\sigma_{i+1}^x = -\alpha_i\beta_i\alpha_{i+1}\beta_{i+1}.
\end{equation}
Notably, the three-body stabilizers at even (odd) sites are mapped into coupling terms involving Majorana operators only at odd (even) sites, giving rise to two Kitaev chains.
In addition, the logical operators for edge modes become 
\begin{equation}
    \tilde{Z}_{\rm L} = -\alpha_1, \; \tilde{X}_{\rm L}=-\alpha_2, \; \tilde{Z}_{\rm R} = -iG\beta_N, \; \tilde{X}_{\rm R} = -i G\beta_{N-1},
\end{equation}
where $G=\prod_{j=1}^{N}(-i\alpha_j\beta_j)=\prod_{j=1}^N\sigma_i^x$, is the generator for the total $\mathbb{Z}_2$ symmetry. 
As $H$ preserves the $\mathbb{Z}_2\times \mathbb{Z}_2$ symmetry generated by $\prod_{i=1}^{\frac{N}{2}}\sigma_{2i}^{x}$ and $\prod_{i=1}^{\frac{N}{2}}\sigma_{2i-1}^{x}$, $G$ is also preserved during the evolution. 
Therefore, the logical operators $\tilde{Z}_{\rm L}$, $\tilde{X}_{\rm L}$, $\tilde{Z}_{\rm R}$, and $\tilde{X}_{\rm R}$ are determined by Majorana edge modes $\alpha_1$, $\alpha_2$, $\beta_{N-1}$, and $\beta_{N}$, respectively.

\vspace{.5cm}
\noindent\textbf{\large Acknowledgements} 
\\
We thank A. Gorshkov, F. L. Liu, and Z. X. Gong for helpful discussion. The device was fabricated at the Micro-Nano Fabrication Center of Zhejiang University. We acknowledge the support from the National Natural Science Foundation of China (Grant Nos. 92365301, 12274368, 12274367, 12174342, 12322414, 12404570, 12404574, U20A2076, T2225008, 12075128, 123B2072), 
the Innovation Program for Quantum Science and Technology (Grant Nos. 2021ZD0300200 and 2021ZD0302203), 
the Zhejiang Provincial Natural Science Foundation of China (Grant Nos. LR24A040002 and LDQ23A040001), 
and the National Key Research and Development Program of China (Grant No. 2023YFB4502600). T.I. acknowledges support from the National Science Foundation under Grant No. DMR-2143635. F.M. acknowledges support from the NSF through a grant for ITAMP (Award No. 2116679) at Harvard University. J.K.  acknowledges support from the Army Research Office (Grant No.~W911NF-24-1-0079).
N.Y.Y acknowledges support from the U.S. Department of Energy via the QuantISED 2.0 program
and from a Simons Investigator award. S.J., W.L., Z.L., Z.-Z.S., and D.-L.D. acknowledge in addition support from the Tsinghua University Dushi Program and the Shanghai Qi Zhi Institute Innovation Program SQZ202318.

\vspace{.5cm}
\noindent\textbf{\large Author contributions}
\\
F.J., X.Z. and Z.B. carried out the experiments and analyzed the experimental data under the supervision of Q.G. and H.W.;   S.J., J.K., N.Y., T.I., F.M., W.L., Z.L., Z.-Z. S., D. Y., and D.-L.D. conducted the theoretical analysis; H.L. and J.C. fabricated the device supervised by H.W.; D.-L.D., Q.G., S.J., F.J., X.Z., H.W., J.K., N.Y., T.I., and F.M. co-wrote the manuscript; 
H.W., Q.G., Z.W., C.S, J.Z., F.J., X.Z., Z.B., F.S., K.W., Z.Z., S.X., Z.S., J.C., Z.T., Y.W., C.Z., Y.G., N.W., Y.Z., A.Z., T.L., J.Z., Z.C., Y.-h.H., Y.-y.H., Han.W., J.Y., Y.W., J.S., G.L., J.D., H.D. and P.Z. contributed to experimental setup. All authors contributed to the discussions of the results.

\beginExtendedData

\begin{figure*}[ht]
    \centering
    \includegraphics[width=1\linewidth]{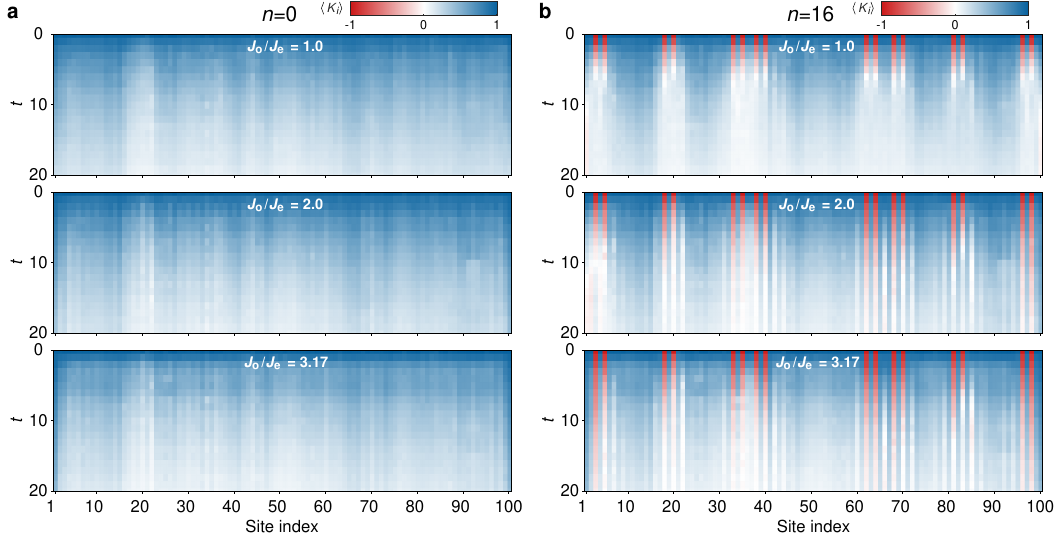}
    \caption{\textbf{Raw data of excitation dynamics.}
    {\bf a}, Measured site-resolved dynamics $\Braket{\Psi_{\rm{g}}|{K}_i(t)|\Psi_{\rm{g}}}$ with the initial state being within the ground-state manifold with no excitation ($n=0$).
    The top, middle, and bottom panels show the data obtained from the system in the homogeneous, dimerized but resonant, and dimerized and off-resonant regimes, respectively.
    {\bf b}, Measured site-resolved dynamics $\Braket{\Psi_{\rm{e}}|{K}_i(t)|\Psi_{\rm{e}}}$ with the initial state being within the excited-state manifold with $n=16$ excitations.
    The excitation dynamics in Fig.~\ref{fig:3} are normalized by $\bar{\braket{{{K}_i}}}={\Braket{\Psi_{\rm e}|{K}_i(t)|\Psi_{\rm e}}}/{\Braket{\Psi_{\rm g}|{K}_i(t)|\Psi_{\rm g}}}$ to reveal the effect caused by the bulk excitations. 
    }
    \label{fig:ex3}
\end{figure*}

\end{document}


\nolinenumbers	
\title{Supplementary Information for\\
``Observation of topological prethermal strong zero modes''
}
\maketitle

\tableofcontents
\beginsupplement

\section{Theoretical analysis}
In this section, we first analyze the $1$D symmetry-protected topological (SPT) spin Hamiltonian and corresponding edge modes in the ground-state manifold. 
We explain why it is ruined by thermal excitations at finite temperatures, and how it presents as prethermal strong zero modes with dimerized parameters in the system Hamiltonian.
Such strong zero modes can be understood as a consequence of U(1)$\times$U(1) symmetry, which is approximately conserved in the system's prethermal regime.
We introduce how to map our spin Hamiltonian into two Kitaev chains by applying the Jordan-Wigner transformation.
In addition, we analyze the evolution circuit applied in our work, obtained from Trotterizing the Hamiltonian to the first-order.
Finally, we explain how energy spectroscopy is carried out and the form of energy gaps in an integrable chain. 

\subsection{1D SPT spin chain and edge modes at zero temperature}
Our 1D Hamiltonian in the main text comprises two ingredients as $H=H_0(J_{\rm e}, J_{\rm o})+H_1(h_x, V_{xx})$. 
The first part $H_0$ includes strong interaction among neighboring sites, which introduces SPT phases in our system. 
The second part $H_1$ considers perturbation terms, which include a transverse field in $\hat x$ direction and two-body $XX$ interactions for neighboring qubits.

We first investigate the properties of the SPT Hamiltonian $H_0$:
\begin{equation}
    H_0(J_{\rm e}, J_{\rm o}) =- J_{\rm e} \sum_{i=1}^{N/2-1} \sigma_{2i-1}^z \sigma^x_{2i}\sigma^z_{2i+1} -J_{\rm o}\sum_{i=1}^{N/2-1} \sigma^z_{2i}\sigma^x_{2i+1}\sigma^z_{2i+2},
\end{equation}
which contains $N$ qubits in total. 
In our study, we consider $N$ to be an even number for simplicity of notation, while all results can readily be extended to odd-number cases.
The Hamiltonian preserves a $\mathbb{Z}_2\times \mathbb{Z}_2$ symmetry, which is generated by the parity operators on even sites $G_{\rm{e}} = \prod_{i=1}^{N/2}\sigma_{2i}^{x}$ and odd sites $G_{\rm{o}} = \prod_{i=1}^{N/2}\sigma_{2i-1}^{x}$. 
The three-body interacting terms $K_i=\sigma_{i-1}^z\sigma_{i}^x\sigma_{i+1}^z$ in $H_0$ commutes with each others and are called stabilizers. 
Note that there are only $N-2$ stabilizers in $H_0$, while the system degree of freedom is $N$. 
This leads to the system having a four-fold degenerate ground-state manifold. 
At zero temperature, the system stays in the ground states, which can be distinguished by two edge modes induced by the $\mathbb{Z}_2\times \mathbb{Z}_2$ symmetry.
To see this, first note that the parity operators can be decomposed into the product of stabilizers and operators at edges:
\begin{equation}
    G_{\rm{e}} = \sigma_1^z\left(\prod_{i=1}^{N/2-1} K_{2i}\right)\sigma_{N-1}^z\sigma_{N}^x,\quad G_{\rm{o}} = \sigma_1^x\sigma_2^z\left(\prod_{i=1}^{N/2-1} K_{2i+1}\right)\sigma_{N}^z.
    \label{eq:Z2xZ2}
\end{equation}
For ground states of $H_0$, all stabilizers $K_i$ take constant values. In our case with positive $J_{\rm e}, J_{\rm o}$, all $K_i=1$ in the ground states, leading to the parity operators further projected into the edges:
\begin{eqnarray}
    G_{\rm{e}} = \sigma_1^z\left(\sigma_{N-1}^z\sigma_{N}^x\right)=\tilde{Z}_{\rm L}\tilde{X}_{\rm R},\quad G_{\rm{o}} = \left(\sigma_1^x\sigma_2^z\right)\sigma_{N}^z=\tilde{X}_{\rm L}\tilde{Z}_{\rm R},\quad \begin{cases}
        \tilde{Z}_{\rm L}\equiv \sigma_1^z \\
        \tilde{X}_{\rm L}\equiv \sigma_1^x\sigma_1^z
    \end{cases}, \quad 
    \begin{cases}
        \tilde{Z}_{\rm R}\equiv \sigma_{N}^z \\
        \tilde{X}_{\rm R}\equiv \sigma_{N-1}^z\sigma_{N}^x
    \end{cases}.\label{eq:projected_Z2xZ2}
\end{eqnarray}
As $H_0$ is local, the preserved symmetry $[H_0, G_{\rm e}]=[H_0, \tilde{Z}_{\rm L}\tilde{X}_{\rm R}]=0$ gives rise to both $[H_0, \tilde{Z}_{\rm L}]=0$ and $[H_0, \tilde{X}_{\rm R}]=0$. 
Similar results are obtained from $[H_0, G_{\rm o}]=0$, together gives four conserved quantities $\tilde{Z}_{\rm L}$, $\tilde{X}_{\rm L}$, $\tilde{Z}_{\rm R}$, and $\tilde{X}_{\rm R}$ at the edges.
In addition, since $\tilde Z$ and $\tilde X$ are anti-commuted at both left and right edges, we conclude that the ground-state degeneracy is four-fold, with two effectively spin-1/2 edge modes described by $\tilde Z$ and $\tilde X$ residing at two ends of the chain.
These edge operators connect between the different sectors of the ground-state manifold.

When generic perturbations are added into $H_0$, the $\mathbb{Z}_2\times \mathbb{Z}_2$ symmetry is broken as the parity operators $G_{\rm e}$, $G_{\rm o}$ are no longer preserved, destroying the edge modes. 
However, if the perturbations also preserve the $\mathbb{Z}_2\times \mathbb{Z}_2$ symmetry with considerably small strength compared with the strength of stabilizers in $H_0$, which is the case in our work, the system remains deep in the SPT phase with the original localized edge modes now extend to the bulk of the system. 
These extensions make the left and right edge modes hybridize with each other, resulting in the original degenerated ground states now opening energy gaps $\delta\propto \exp(-N)$ exponentially small in the system size.
This gives rise to the exponentially long lifespan for edge modes at zero temperature under symmetry-preserved perturbations. 

\subsection{Edge modes as prethermal strong zero modes at finite temperatures}
\begin{figure}
    \centering
    \includegraphics[width=0.8\linewidth]{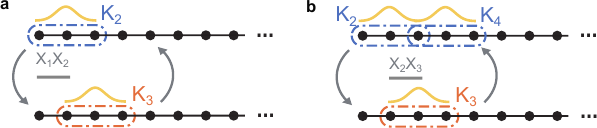}
    \caption{\textbf{First-order resonances.} 
    {\bf a}, In the system with homogeneous stabilizer strength $J_{\rm o}=J_{\rm e}$, the two-body interaction $V_{xx}\sigma_1^x\sigma_2^x$ in $H_1$ can resonantly exchange one excitation between $J_{\rm e}K_2$ and $J_{\rm o}K_3$, which also flips $\tilde{Z}_{\rm L}$ and $\tilde{X}_{\rm L}$.
    {\bf b}, In the system with stabilizer strength $J_{\rm o}=2J_{\rm e}$, the $V_{xx}\sigma_2^x\sigma_3^x$ term resonantly transfers two excitations $J_{\rm e}K_2$, $J_{\rm e}K_4$ into $J_{\rm o}K_3$, and vice versa.
    This process makes $\tilde{X}_{\rm L}$ rapidly decohered.
    A similar process can happen at the right edge with $J_{\rm o}=0.5J_{\rm e}$ and is not shown.
    }
    \label{fig:resonance}
\end{figure}
The above discussion is restricted to the system being in the ground state. 
At finite temperatures, thermal excitations emerge within the systems, interacting with edge modes and causing them to rapidly decohere. 
The vulnerability of edge modes against thermal excitations is uncovered by the fact that $\left(\prod_{i=1}^{N/2-1} K_{2i}\right)$ and $\left(\prod_{i=1}^{N/2-1} K_{2i+1}\right)$ in Eq.~\eqref{eq:Z2xZ2} are no longer conserved quantities for excited states, and hence $G_{\rm e}$, $G_{\rm o}$ cannot be projected into the edges.
For example, for the perturbations in our work:
\begin{equation}
    H_1(h_x, V_{xx})=h_x \sum_{i=1}^{N}\sigma^x_i+  V_{xx} \sum_{i=1}^{N-1} \sigma^x_i\sigma^x_{i+1},
\end{equation}
directly applying $\sigma^x_1\sigma^x_2$ or considering the second-order process of $\sigma^x_1+\sigma^x_2$ will flip $\tilde{Z}_{\rm L}$,  $\tilde{X}_{\rm L}$, $K_2$, and $K_3$, while keeping $G_{\rm e}$ and $G_{\rm o}$ unchanged.
If exactly one of $K_2$ and $K_3$ is equal to $-1$, and both exhibit identical strength (i.e. $J_{\rm e}=J_{\rm o}$), this process further preserves the system energy and becomes a resonant perturbation, leading to rapid decoherence of both $\tilde{Z}_{\rm L}$ and  $\tilde{X}_{\rm L}$.
Physically, this represents transferring an excitation between the even and odd sites through the edge (Fig.~\ref{fig:resonance}a). 
Notably, the first-order resonance can happen at edges even for unequal stabilizer strength. 
For example, in the main text, we observe the left edge operator $\tilde{X}_{\rm L}$ is rapidly decohered when $J_{\rm o}/J_{\rm e}=2.0$.
This is caused by the $\sigma^x_2\sigma^x_3$ term which flips $\tilde{X}_{\rm L}$, $K_2$, $K_3$, and $K_4$, and hence resonantly transferring two excitations with energy $J_{\rm e}$ to one excitation with energy $J_{\rm o}$ (Fig.~\ref{fig:resonance}b).
Similar process can happen for $\tilde{X}_{\rm R}$ at right edge when $J_{\rm o}/J_{\rm e}=0.5$.
For systems taking other dimerized stabilizer strengths, the resonance only happens for higher-order processes that are hard to observe within the current experimental time scale.
This leads to the prolonged edge mode lifetime observed in our experiment. 

Theoretically, a local operator being approximately conserved for arbitrary system configurations is characterized by a prethermal strong zero mode (PSZM)~\cite{Fendley2016Strong, Kemp2017Long, Kemp2020Symmetry}.
Such an operator almost commutes with the system Hamiltonian and maps system eigenstates from one symmetry sector to another. 
By definition, a PSZM must satisfy the following three conditions: 
(1) Squares to the identity.
(2) Almost commutes with the Hamiltonian with an error term exponentially small in the system size.  
(3) Anti-commutes with the system's symmetry.
In practice, such an operator is constructed from perturbation theory order by order and is cut off at some finite order to obtain a bounded commutator with the Hamiltonian.
In our setting, the PSZM can be constructed for all $\tilde{Z}_{\rm L}$,  $\tilde{X}_{\rm L}$, $\tilde{Z}_{\rm R}$, and $\tilde{X}_{\rm R}$, which leads to robust edge modes locally encoding spin-1/2 degrees of freedom at both ends of the chain under finite temperatures.
Up to the first order in $h_x$ and $V_{xx}$, such PSZMs for the left edge reads~\cite{Kemp2020Symmetry},
\begin{eqnarray}
    \Psi^z_{\rm L} &=& \tilde{Z}_{\rm L} - \frac{h_x}{J_{\rm e}}\sigma_1^x\sigma_2^x\sigma_3^z+\frac{V_{xx}}{J_{\rm o}^2-J_{\rm e}^2}(J_{\rm e}\sigma_1^x\sigma_3^z+J_{\rm o}\sigma_1^y\sigma_2^y\sigma_3^x\sigma_4^z),\\
    \Psi^x_{\rm L} &=& \tilde{X}_{\rm L} -\frac{h_x}{J_{\rm o}}\sigma_1^x\sigma_2^x\sigma_3^x\sigma_4^z-\frac{V_{xx}}{J_{\rm o}^2-J_{\rm e}^2}(J_{\rm o}\sigma_2^x\sigma_3^x\sigma_4^z+J_{\rm e}\sigma_1^z\sigma_2^z\sigma_3^z) \nonumber\\
    &&+\frac{V_{xx}J_{\rm e}}{J_{\rm o}^2-4J_{\rm e}^2}\left[\sigma_1^y\sigma_2^z\sigma_3^y+\left(\frac{2J_{\rm e}}{J_{\rm o}}-\frac{J_{\rm o}}{J_{\rm e}}\right)\sigma_1^x\sigma_2^x\sigma_4^z-\sigma_1^x\sigma_2^y\sigma_3^y\sigma_4^x\sigma_5^z-\frac{2J_{\rm e}}{J_{\rm o}}\sigma_1^y\sigma_4^y\sigma_5^z\right],
\end{eqnarray}
and similar PSZMs can be constructed for the right edge. 
We identify that both $\Psi_{\rm L}^z$ and $\Psi_{\rm L}^x$ commute with $H_0+H_1$ and squares to the identity up to error terms with order $O\left(\max\{h_x^2, V_{xx}^2\}\right)$.
For a homogeneous system with $J_{\rm o}= J_{\rm e}$, the first-order terms in both $\Psi_{\rm L}^z, \Psi_{\rm R}^x$ diverge, resulting from the resonant process of $V_{xx}\sigma_1^x\sigma_2^x$. 
In addition, we find $\Psi_{\rm L}^x$ also diverges at $J_{\rm o}= 2J_{\rm e}$, characterizing the effect of $V_{xx}\sigma_2^x\sigma_3^x$.
Besides these divergent points, $\Psi_{\rm L}^z, \Psi_{\rm R}^x$ keep finite and satisfy $\{\Psi^z_{\rm L}, G_{\rm o}\}=[\Psi^z_{\rm L}, G_{\rm e}]=0$ and $\{\Psi^x_{\rm L}, G_{\rm e}\}=[\Psi^x_{\rm L}, G_{\rm o}]=0$.
This induces the almost conserved degeneracy throughout the entire spectrum, thus giving rise to robust edge modes for arbitrary system configurations. 
Given that both $h_x$ and $V_{xx}$ remain small, $\Psi^z_{\rm L}$ and $\Psi^x_{\rm L}$ have large overlaps with $\tilde{Z}_{\rm L}$ and $\tilde{X}_{\rm L}$.
This makes the latter operators, which we measure in experiments, good approximations to describe the edge modes.

\subsection{Prethermalization and emergent U(1)$\times$U(1) symmetry}
It was recently shown in Ref.~\cite{Else2017Prethermal2} that PSZMs can be understood as a phenomenon of prethermalization with emergent symmetries. 
The authors proved that there was an additional U(1) symmetry in the prethermal regime of a Kitaev chain, given that the perturbations were much smaller than the interaction strength.
This gave rise to robust Majorana edge modes at finite temperatures. 
For the SPT chain in our experiments, a similar U(1) symmetry is observed in the homogeneous regime (Fig.~3\textbf{d} in the main text), however, it is insufficient for protecting the spin-1/2 edge modes. 
Instead, we observe that the robust edge modes only occur under the protection of the U(1)$\times$U(1) symmetry in the dimerized and off-resonant case. 

We start with first investigating the emergent U(1) symmetry in the homogeneous case, which comes from considering the structure of the stabilizer terms in the Hamiltonian $H$.
It has shown that after applying local Schrieffer-Wolff transformations order by order and stopping at a certain order to restrict the growth of perturbation terms, the symmetry-breaking perturbations can be eliminated in a rotated frame. 
In particular, if the system Hamiltonian takes the form
\begin{equation}
    H = -JN+ V,\label{eq:continuous_H}
\end{equation}
with $N$ being a sum of mutual-commuting local terms and having integer eigenvalues, and $V$ being a sum of local perturbations with energy scale $J_0$, the evolution generated by $H$ can be approximated by the following equation~\cite{abanin2017rigorous, Else2017Prethermal, machado2020long}:
\begin{equation}
    \exp(-iHt) = \mathcal{V}\exp\left[-i(-JN+V_p+E)t\right]\mathcal{V}^\dagger, \quad [N, V_p]=0, \quad ||E|| = O\left[\exp\left(-\frac{J}{J_0}\right)\right]. \label{eq:u1}
\end{equation}
Note that the system now preserves $N$ up to an exponentially small error term $E$, indicating that an additional U(1) symmetry generated by $N$ emerges.
The system will eventually thermalize to an infinite-temperature state due to the errors $E$.
However, before that, the effective Hamiltonian $-JN+V_p$ and the U(1) symmetry will survive for an exponentially long lifespan, referred to as the prethermal regime.
In homogeneous system with $J_{\rm e}=J_{\rm o}$, the Hamiltonian $H = - J_{\rm e} \sum_{i=2}^{N-1} K_{i} +H_1(h_x, V_{xx})$ exactly fits into Eq.~\eqref{eq:continuous_H}. 
Given that $J_{\rm e}\gg \max\{h_x, V_{xx}\}$, the sum of stabilizers in bulk $\sum_{i=2}^{N-1}K_i$ is approximately conserved.
This leads to the conservation law on the total excitation number $(N-2-\sum_{i=2}^{N-1}K_i)/2$, which generates the U(1) symmetry in the prethermal regime lasting for $t = O\left[\exp(J_{\rm e}/\max\{h_x, V_{xx}\})\right]$.

The edge mode of spin-1/2 fails to maintain robustness at finite temperatures, even in the presence of this additional U(1) symmetry. 
This is exemplified by the first-order resonant process depicted in Fig.~\ref{fig:resonance}\textbf{a}, where transferring an excitation between sites with different parity ruins the edge modes without modifying the total excitation number. 
The robust edge mode requires a larger symmetry group U(1)$\times$U(1), denoting the conservation laws on excitation numbers within even and odd sites. Such a U(1)$\times$U(1) symmetry can emerge within the system with dimerized stabilizer strength $J_{\rm e}\neq J_{\rm o}$. 
Intuitively, this separates the energy scales of stabilizers on the even and odd sites, leading to large energy obstacles on exchanging excitations. 
However, if the $J_{\rm e}/J_{\rm o}$ is a rational number, there could be resonant processes in the perturbation theory, which happens at a finite order independent with $J_{\rm o}$, $J_{\rm e}$ and the system size. 
The first-order resonance in the $J_{\rm o}=2J_{\rm e}$ case is an example~(Fig.~\ref{fig:resonance}\textbf{b}). 
To avoid these resonances, the stabilizer strength should be irrational multiples of each other. 
Then, the U(1)$\times$U(1) symmetry is maintained within the exponentially long prethermal regime. 
Formally, for the Hamiltonian taking the form
\begin{equation}
    H = -\sum_{i=1}^m J_i N_i +V,
\end{equation}
with $\{N_i\}$ be mutually commuting operators taking integer eigenvalues, and $\{J_i\}$ being irrational multiples of each other, and $V$ being a sum of local perturbations with energy scale $J_0$, the evolution generated by $H$ can be approximated by the following equation ~\cite{Else2020Long-Lived}:
\begin{equation}
    \exp(-iHt) = \mathcal{V}\exp\left[-i\left(-\sum_{i=1}^mJ_iN_i+V_p+E\right)t\right]\mathcal{V}^\dagger, \quad \forall i, \;[N_i, V_p]=0, \quad ||E|| = O\left[\exp\left(-\left(\frac{\min\{J_i\}}{J_0}\right)^{1/(m+\epsilon)}\right)\right], \label{eq:u1xu1}
\end{equation}
with $\epsilon$ being a small constant. Within the rotated frame $\mathcal{V}$, the system now presents $m$ approximate conservation laws on all $N_i$, leading to the emergent U$(1)^{\times m}$ symmetry. Note that for $m=1$ the equation aligns with the one in Eq.~\eqref{eq:u1}, and for our case with $m=2$, the prethermal lifetime is proportional to $\exp\left[\left({\min\{J_{\rm e}, J_{\rm o}\}}/{\max\{h_x, V_{xx}\}}\right)^{1/2}\right]$.

The key point to understand how this emergent U(1)$\times$U(1) symmetry gives rise to the robust edge modes is that the conserved sum of stabilizers on even and odd sites will also lead to the conserved parity of stabilizers on even and odd sites, i.e. $\prod_{i=1}^{N/2-1} K_{2i}$ and $\prod_{i=1}^{N/2-1} K_{2i+1}$, which are originally not conserved when there are thermal excitations in the homogeneous regime. 
With these conserved parity operators, together with the system preserved $\mathbb{Z}_2\times \mathbb{Z}_2$ symmetry, we can again project $G_{\rm e}, G_{\rm o}$ in Eq.~\eqref{eq:Z2xZ2} into the edges and obtain Eq.~\eqref{eq:projected_Z2xZ2} without requiring the system being in ground states, leading to robust edge operators $\tilde{Z}, \tilde{X}$ for arbitrary system configurations. 

\subsection{Jordan-Wigner transformation and Majorana fermion picture}
\begin{figure}[t]
    \centering
    \includegraphics[width=0.9\linewidth]{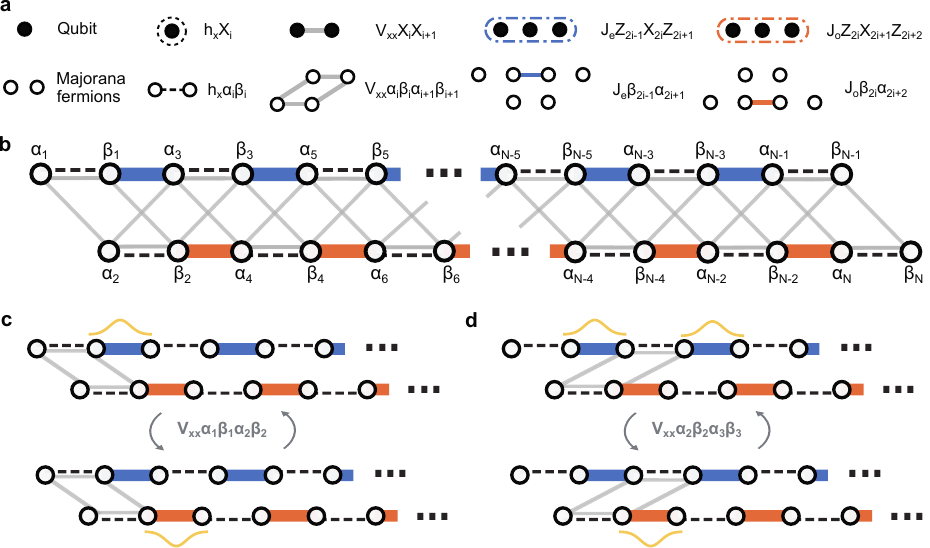}
    \caption{{\bf The SPT spin chain in the Majorana fermion picture.} 
    {\bf a,} The correspondence between each term in spin Hamiltonian $H$ and Majorana operators $\alpha, \beta$. 
    Each qubit is non-locally transformed into two Majorana fermions. Single-body $\sigma_i^x$ and two-body $\sigma_i^x\sigma_{i+1}^x$ terms are transformed into onsite and inter-chain couplings involving two and four Majoranas, respectively. The three-body stabilizers on even (odd) sites are mapped into couplings between two Majoranas on different sites. 
    {\bf b,} The SPT spin chain is transformed into two Kitaev chains with inter-site coupling strengths $J_{\rm e}$ (upper chain) and $J_{\rm o}$ (lower chain).
    {\bf c-d,} The first-order resonances between two Kitaev chains for (c) $J_{\rm o}=J_{\rm e}$ and (d) $J_{\rm o}=2J_{\rm e}$.
    }
    \label{fig:majorana}
\end{figure}

In the main text, we argue that our SPT qubit chain can be transformed into two Kitaev chains in the Majorana picture. 
To see this, we first consider applying the Jordan-Wigner transformation~\cite{Lieb1961Two} to map the spin Hamiltonian $H=H_0+H_1$ into fermionic creation/annihilation operators $c_i^\dagger, c_i$:
\begin{equation}
    \sigma_i^x = 1-2c_i^\dagger c_i, \quad \sigma^z_i = - \left[\prod_{j=1}^{i-1}(1-2c_j^\dagger c_j)\right](c_i^\dagger + c_i),\label{eq:WJT}
\end{equation}
and the inverse transformation is $c_i = -\frac{1}{2}\left(\prod_{j=1}^{i-1}\sigma_j^x\right) \sigma_i^z(1-\sigma_i^x)$, $c_i^\dagger = -\frac{1}{2}\left(\prod_{j=1}^{i-1}\sigma_j^x\right) (1-\sigma_i^x)\sigma_i^z$. This readily gives the canonical fermionic algebra $\{c_k, c_l^\dagger\}=\delta_{kl},  \{c_k, c_l\}=0$.
After applying Eq.~\eqref{eq:WJT}, $H$ is transformed into the following form:
\begin{eqnarray}
    H_{\rm f} &=& -J_{\rm e} \sum_{i=1}^{N/2-1} (c_{2i-1}^\dagger -c_{2i-1})(c_{2i+1}^\dagger+c_{2i+1}) -J_{\rm o} \sum_{i=1}^{N/2-1} (c_{2i}^\dagger -c_{2i})(c_{2i}^\dagger+c_{2i}) +h_x\sum_{i=1}^N (1-2c_i^\dagger c_i) \nonumber\\
    &&+V_{xx}\sum_{i=1}^{N-1}(1-2c_i^\dagger c_i)(1-2c_{i+1}^\dagger c_{i+1}) \\
    &=& \underbrace{\sum_{i=1}^{N/2-1} \left[-J_{\rm e}(c_{2i-1}^\dagger -c_{2i-1})(c_{2i+1}^\dagger+c_{2i+1})+h_x(1-2c_{2i-1}^\dagger c_{2i-1})\right]+h_x(1-2c_{N-1}^\dagger c_{N-1})}_{\text{Upper Kitaev chain}} \nonumber\\
    && + \underbrace{\sum_{i=1}^{N/2-1} \left[-J_{\rm o}(c_{2i}^\dagger -c_{2i})(c_{2i+2}^\dagger+c_{2i+2})+h_x(1-2c_{2i}^\dagger c_{2i})\right]+h_x(1-2c_{N}^\dagger c_{N})}_{\text{Lower Kitaev chain}} \nonumber\\
    && + \underbrace{V_{xx}\sum_{i=1}^{N-1}(1-2c_i^\dagger c_i)(1-2c_{i+1}^\dagger c_{i+1})}_{\text{Inter-chain coupling}}. \label{eq:fermionic H}
\end{eqnarray}
Here, we already obtain the result that the spin Hamiltonian is transformed into two Kitaev chains with distinct coupling strengths $J_{\rm e}$ and $J_{\rm o}$, where onsite and inter-chain couplings are present with strengths $h_x$, $V_{xx}$, respectively. 
To see this result in the Majorana picture, we further transform each $c_i^\dagger, c_i$ into Majorana fermionic operators $\alpha_i, \beta_i$ by:
\begin{equation}
    c_i^\dagger = \frac{\alpha_i-i\beta_i}{2}, \quad c_i = \frac{\alpha_i+i\beta_i}{2}.
\end{equation}
Then, the fermionic Hamiltonian in Eq.~\eqref{eq:fermionic H} is mapped into:
\begin{equation}
    H_{\rm {mf}} = \sum_{i=1}^{N/2-1} \left(iJ_{\rm e}\beta_{2i-1}\alpha_{2i+1}-ih_x\alpha_{2i-1}\beta_{2i-1}\right)-ih_x\alpha_{N-1}^\dagger \beta_{N-1}  + \sum_{i=1}^{N/2-1} \left(iJ_{\rm o}\beta_{2i}\alpha_{2i+2}-ih_x\alpha_{2i}\beta_{2i}\right)-ih_x\alpha_{N}^\dagger \beta_{N} + V_{xx}\sum_{i=1}^{N-1}\alpha_{i}\beta_{i}\alpha_{i+1}\beta_{i+1}. \label{eq:MF_H}
\end{equation}
In summary, the total transformation is given by
\begin{equation}
     \sigma_i^x = -i\alpha_i\beta_i, \quad \sigma^z_i = - \left[\prod_{j=1}^{i-1}(-i\alpha_j\beta_j)\right]\alpha_i. \label{eq:total_transform}
\end{equation}
We illustrate how each term in the spin Hamiltonian $H$ mapped into Majorana fermionic operators in Fig.~\ref{fig:majorana}\textbf{a}, and the two coupled Kitaev chains after the transformation is shown in Fig.~\ref{fig:majorana}\textbf{b}. 

A few remarks are in order. 
First, the edge operators originally defined in the spin Hamiltonian are now mapped into the following form:
\begin{equation}
    \tilde{Z}_{\rm L} = -\alpha_1, \quad \tilde{X}_{\rm L}=-\alpha_2, \quad \tilde{Z}_{\rm R} = -i\left[\prod_{j=1}^{N}(-i\alpha_j\beta_j)\right] \beta_N, \quad \tilde{X}_{\rm R} = -i \left[\prod_{j=1}^{N}(-i\alpha_j\beta_j)\right]\beta_{N-1}.
\end{equation}
While the left edge operators are directly represented by Majorana edge modes $\alpha_1, \alpha_2$ in each Kitaev chain, there is an additional term $\prod_{j=1}^{N}(-i\alpha_j\beta_j)$ at right edges.
Notably, this is the generator for the total $\mathbb{Z}_2$ symmetry: 
\begin{equation}
    G=\prod_{j=1}^{N}(-i\alpha_j\beta_j) = G_{\rm e}G_{\rm o}, \quad G_{\rm e}=\prod_{i=1}^{N/2}(-i\alpha_{2i}\beta_{2i}),\quad G_{\rm o} = \prod_{i=1}^{N/2}(-i\alpha_{2i-1}\beta_{2i-1}).
\end{equation}
As the system preserves $G_{\rm e}$ and $G_{\rm o}$, it also preserves $G$.
Therefore, the right edge operator $\tilde{Z}_{\rm R}$, $\tilde{X}_{\rm R}$ are sorely determined by the state of $\beta_N$, $\beta_{N-1}$ during the evolution.

Second, the single-body and two-body terms in $H_1$ are mapped into onsite and inter-chain coupling terms. 
Given that $h_x$ and $V_{xx}$ are small, the system keeps in the topological phase with $\alpha_1, \alpha_2, \beta_{2N-1}, \beta_{2N}$ nearly unpaired. 
In addition, the first-order resonance we discussed above is better understood in the Majorana fermion picture.
The $\sigma_1^x\sigma_2^x$ term now becomes $\alpha_{1}\beta_{1}\alpha_{2}\beta_{2}$, pairing $\alpha_1, \alpha_2$ and exchanging the occupation between $J_{\rm e}\beta_1\alpha_3$ and $J_{\rm o}\beta_2\alpha_4$ (Fig.~\ref{fig:majorana}\textbf{c}).
The $\sigma_2^x\sigma_3^x$ term reads $\alpha_{2}\beta_{2}\alpha_{3}\beta_{3}$, which involves $\alpha_2$ and transfers both occupied $J_{\rm e}\beta_1\alpha_3$ and $J_{\rm e}\beta_3\alpha_5$ into $J_{\rm o}\beta_2\alpha_4$ (Fig.~\ref{fig:majorana}\textbf{d}). 
These readily give the first-order resonant conditions $J_{\rm o}=J_{\rm e}$ and $J_{\rm o}=2J_{\rm e}$.

Third, in the Majorana picture, the U(1)$\times$U(1) symmetry now represents the conservation laws for occupation numbers of inter-site couplings within each Kitaev chain, i.e. $\sum_{i=1}^{N-1}\beta_{2i-1}\alpha_{2i+1}$ and $\sum_{i=1}^{N-1}\beta_{2i}\alpha_{2i+2}$.  As a result, the $\mathbb{Z}_2$ charges in the bulk of each chain are also conserved:
\begin{eqnarray}
    \mathcal{F}_{\rm o} = \prod_{i=1}^{N/2-1}\beta_{2i-1}\alpha_{2i+1} = \beta_1\left(\prod_{i=2}^{N/2-1} \alpha_{2i-1}\beta_{2i-1} \right)\alpha_{N-1},\quad 
    \mathcal{F}_{\rm e} = \prod_{i=1}^{N/2-1}\beta_{2i}\alpha_{2i+2} = \beta_2\left(\prod_{i=2}^{N/2-1} \alpha_{2i}\beta_{2i} \right)\alpha_{N}. 
\end{eqnarray}
As the U(1)$\times$U(1) symmetry emerges in the dimerized and off-resonant region, the system commutes with all of the symmetries $G_{\rm e}$, $G_{\rm o}$, $\mathcal{F}_{\rm e}$, and $\mathcal{F}_{\rm o}$. 
This gives the conserved $\mathcal{F}_{\rm o}G_{\rm o} = (-i)^{N/2}\alpha_1\beta_{2N-1}$ and $\mathcal{F}_{\rm e}G_{\rm e} = (-i)^{N/2} \alpha_2\beta_{2N}$. 
Since $\alpha_1$ ($\alpha_2$) and $\beta_{2N-1}$ ($\beta_{2N}$) are at two ends of the chain separated by $N$ sites, and perturbations are local, we conclude that each of $\alpha_1$, $\alpha_2$, $\beta_{2N-1}$ and $\beta_{2N}$ is conserved.

\begin{figure}[t]
    \centering
    \includegraphics[width=1\linewidth]{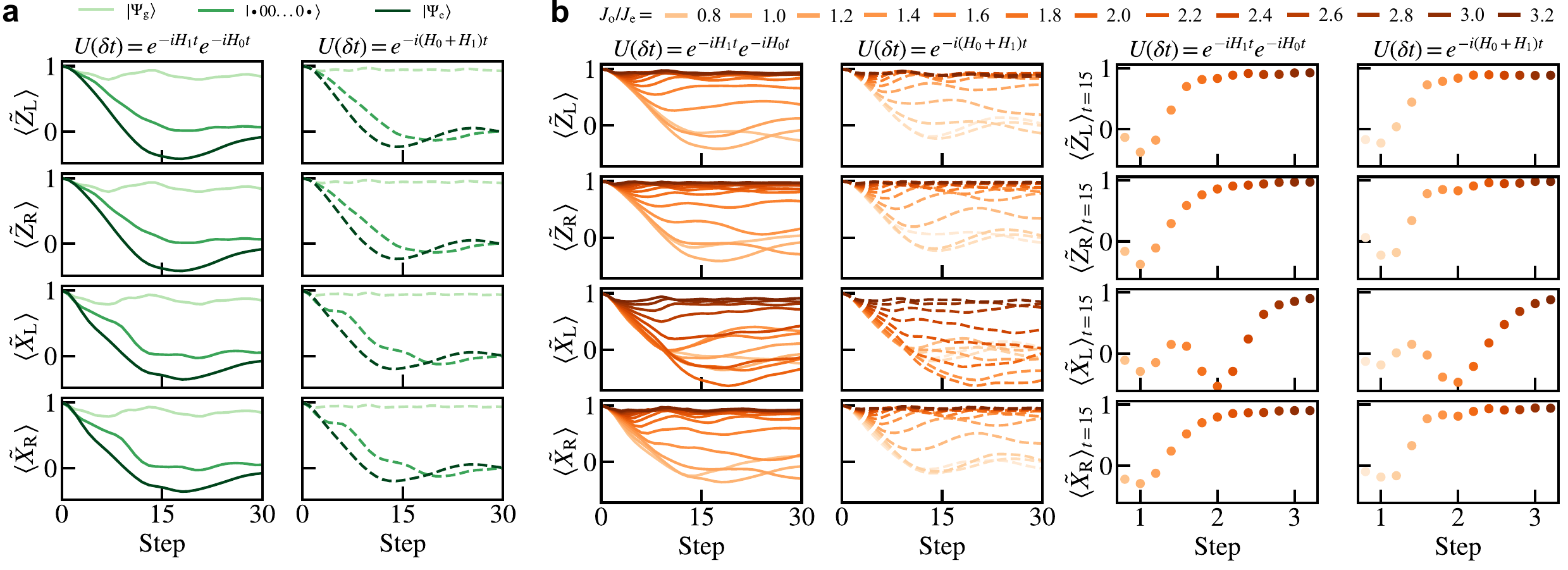}
    \caption{{\bf Comparison between the exact evolution and the evolution obtained from the first-order Trotterization.} 
    {\bf a,} Measured time dynamics for edge operators under the Trotter circuit (left) and the exact evolution (right) in the homogeneous regime ($J_{\rm o} = J_{\rm e} = \pi/5$).
    {\bf b,} Measured time dynamics for edge operators with fixed $J_{\rm e} = \pi/5$ and varying $J_{\rm o}$. Data in the first (second) and third (fourth) columns are obtained from the Trotter circuit (exact evolution).
    The numerical calculations are carried out on a $N=14$ system with $h_x=0.11$ and $V_{xx}=0.2$.
    }
    \label{fig:trotter}
\end{figure}

\subsection{Characterization of the evolution circuit}
So far, our discussion has been focused on the system with Hamiltonian $H=H_0+H_1$. 
Such a Hamiltonian is suitable for theoretical analysis, yet it poses considerable challenges for experimental implementation.
This is due to the three-body interactions inherent in $H_0$, and the fact that $H_0$ and $H_1$ do not commute.
We address this problem through digitally implementing the evolution of $H_0$, $H_1$ with duration $\delta t$ by quantum circuits $U_0(\delta t)$, $U_1(\delta t)$. 
The evolution of $H_0+H_1$ is then approximated by first-order Trotter decomposition $U(\delta t)=U_1(\delta t)U_0(\delta t)$.
The robust edge modes in the dimerized regime can be observed with the presence of these Trotter errors in our experiments.
As shown in Fig.~\ref{fig:trotter}, we carry out the noiseless simulation on the temporal dependence of the edge modes under the Trottered circuit $e^{-iH_1 \delta t}e^{-iH_0 \delta t}$ and the exact evolution $e^{-i(H_0+H_1)\delta t}$. 
We observe that the behavior of edge modes are qualitatively similar in two cases, as they decay rapidly when the system is starting with finite-temperature states, and become resilient against excitations with dimerized $J_{\rm o}/J_{\rm e}$. 
Strikingly, the Trotter errors do not change the first-order resonance point at $J_{\rm o}/J_{\rm e}=2$.

In addition, Trotter errors can be considered as additional perturbations added to the system, making the interactions between the edges and the bulk more general. 
To show this, We apply the Floquet-Magnus expansion~\cite{Magnus1954exponential, Kuwahara2016Floquet, Heyl2019Quantum} on the Trotter decomposition circuit $U$.
Note that $U$ is generated by the following time-dependent Hamiltonian:
\begin{equation}
    H(t) =\begin{cases}
        2H_{0}, \quad & 0<t\leq \delta t/2, \\
        2H_{1}, & \delta t/2<t\leq \delta t.
    \end{cases},\quad U(\delta t) = \mathcal{T}e^{-i\int_0^{\delta t}H(t)\text{d} t} = e^{-i\delta t H_1}e^{-i\delta t H_0}.
\end{equation}
This is a Floquet process with periodicity $\delta t$. At the stroboscopic time $t=n\delta t$, the system evolution can be described by an effective Floquet Hamiltonian: $U(n\delta t)=\exp[{-i(n\delta t)H_{\rm F}}]$.
When $H_0$ and $H_1$ do not commute, $H_{\rm F}$ becomes non-local, which is constructed order by order through the Floquet-Magnus expansion:
\begin{equation}
    H_{\text{F}} = \sum_{n=0}^\infty (\delta t)^n \Omega_n.
\end{equation}
This series expansion typically does not converge, potentially signaling the ergodicity of the Floquet process. 
However, given the scaling of the coefficient of the $n$-th order term proportional to $(\delta t)^n$, we expect that higher-order terms will not take effect until a considerable delayed time, provided that $\delta t$ is relatively small. 
Consequently, truncating the series expansion after the first few orders usually suffices to depict the system behavior within an experimental timescale.
In our scenario, the first two orders read,
\begin{eqnarray}
    \Omega_0 &=& \frac{1}{\delta t}\int_0^{\delta t}H(t_1)\text{d} t_1= H_0+H_1, \\
    \Omega_1 &=& \frac{1}{2i(\delta t)^2}\int_{0}^{\delta t}\text{d} t_1 \int_{0}^{t_1}\text{d} t_2[H(t_1), H(t_2)] \nonumber\\
    &=& - J_{\rm e}h_x \sum_{i=1}^{N/2-1} \left(\sigma_{2i-1}^y \sigma^x_{2i}\sigma^z_{2i+1}+\sigma_{2i-1}^z\sigma^x_{2i}\sigma^y_{2i+1}\right) 
    -J_{\rm o}h_x \sum_{i=1}^{N/2-1}\left(\sigma^y_{2i}\sigma^x_{2i+1}\sigma^z_{2i+2}+\sigma^y_{2i}\sigma^x_{2i+1}\sigma^y_{2i+2}\right) \nonumber\\
    &&-J_{\rm e}V_{xx}\left[ \sum_{i=2}^{N/2-1}\sigma_{2i-2}^x\sigma_{2i-1}^y\sigma_{2i}^x\sigma_{2i+1}^z+\sum_{i=1}^{N/2-1}\left(\sigma_{2i-1}^y\sigma_{2i+1}^z+\sigma_{2i-1}^z\sigma_{2i+1}^y+\sigma_{2i-1}^z\sigma_{2i}^x\sigma_{2i+1}^y\sigma_{2i+2}^x\right)\right] \nonumber\\
    &&-J_{\rm o}V_{xx}\left[ \sum_{i=1}^{N/2-1}\left(\sigma_{2i-1}^x\sigma_{2i}^y\sigma_{2i+1}^x\sigma_{2i+2}^z+\sigma_{2i}^y\sigma_{2i+2}^z+\sigma_{2i}^z\sigma_{2i+2}^y\right)+\sum_{i=1}^{N/2-2}\sigma_{2i}^z\sigma_{2i+1}^x\sigma_{2i+2}^y\sigma_{2i+3}^x\right].
\end{eqnarray}
This gives $H_{\rm F}^{(2)} = H + (\delta t)\Omega_1$. 
Strikingly, besides $H$, additional interactions occur in $H_{\rm F}^{(2)}$.
As both $H_0$ and $H_1$ preserve the $\mathbb{Z}_2\times \mathbb{Z}_2$ symmetry, $\Omega_1$ also preserves it. 
Therefore, provided that $\delta t$ is small, we expect $H_{\rm F}^{(2)}$ with these additional perturbations will remain in the SPT phase. 
\subsection{Energy spectroscopy in an integrable chain}
Here, we briefly explain how we measure the single-particle spectrum. First, we note that the Jordan-Wigner transformation in Eq.~\eqref{eq:WJT} can also be applied to the quantum circuits in our experiments. In the limit with $V_{xx}=0$, the unitary after transformation reads,
\begin{eqnarray}
    U_{\rm f} &=& \left[\prod_{i=1}^{N/2-1}e^{i\delta tJ_{\rm e}(c_{2i-1}^\dagger -c_{2i-1})(c_{2i+1}^\dagger+c_{2i+1})}\prod_{i=1}^{N/2}e^{-i\delta t h_x(1-2c_{2i-1}^\dagger c_{2i-1})}\right] \left[\prod_{i=1}^{N/2-1}e^{i\delta tJ_{\rm o}(c_{2i}^\dagger -c_{2i})(c_{2i+2}^\dagger+c_{2i+2})}\prod_{i=1}^{N/2}e^{-i\delta t h_x(1-2c_{2i}^\dagger c_{2i})}\right] = U_{\rm {K,e}}U_{\rm {K,o}},
\end{eqnarray}
with mutually commuting $U_{\rm {K,e}}$ and $U_{\rm {K,o}}$. 
$U_{\rm {K,e}}$, $U_{\rm {K,o}}$, which are called kicked Kitaev models, are widely studied in theory~\cite{Thakurathi2013Floquet, Akila2016Particle, Bertini2018Exact, Lerose2021Scaling}.
This model is exactly solvable with Bogoliubov eigenmodes $\nu^\dagger, \nu$. For $U_{\rm {K,e}}$ and $U_{\rm {K,o}}$, we have 
\begin{eqnarray}
        U_{\rm {K,e}}^\dagger\nu_{2i-1}U_{\rm{K,e}} &=&e^{-i\epsilon_{2i-1}}\nu_{2i-1}, \quad \nu_{2i-1}=\sum_{j=1}^{N/2}u_{2i-1, 2j-1} c_{2j-1}^\dagger+v_{2i-1, 2j-1} c_{2j-1}. \\
        U_{\rm {K,o}}^\dagger\nu_{2i}U_{\rm{K,o}} &=&e^{-i\epsilon_{2i}}\nu_{2i}, \quad \quad \quad \nu_{2i}=\sum_{j=1}^{N/2}u_{2i, 2j} c_{2j}^\dagger+v_{2i, 2j} c_{2j}. \\
\end{eqnarray}
Notably, as two kicked Kitaev chains are decoupled, the Bogoliubov eigenmodes are constructed from the fermionic operators within each chain, leading to
\begin{equation}
    [U_{\rm {K,e}}, \nu_{2i}]=0, \quad [U_{\rm {K,o}}, \nu_{2i-1}]=0.
\end{equation}

In our experiments, we measure the logical operators $\tilde{Z}_{\rm L}, \tilde{Z}_{\rm R}$, which can be decomposed as following:
\begin{equation}
    \tilde{Z}_{\rm L} = -(c_1+c_1^\dagger) = \sum_{i=1}^{N/2} l_{ 2i-1} \nu_{2i-1} + l_{2i-1}^* \nu_{2i-1}^\dagger,\quad \tilde{Z}_{\rm R} = -G(c_N-c_N^\dagger) = G\sum_{i=1}^{N/2} l_{2i} \nu_{2i} + l_{2i}^* \nu_{2i}^\dagger.
\end{equation}
Again, we observe that the decomposition of $\tilde{Z}_{\rm L}$ ($\tilde{Z}_{\rm R}$) only involves the Bogoliubov eigenmodes within each Kitaev chain. Then, the dynamics of these operators from an arbitrary initial state $\ket{\psi_0}$ read
\begin{eqnarray}
    \braket{\psi(t)|\tilde{Z}_{\rm L}|\psi(t)} &=& \sum_{i=1}^{N/2}\Braket{\psi_0|(U_{\rm{K,o}}^\dagger U_{\rm{K,e}}^\dagger)^t (l_{ 2i-1} \nu_{2i-1} + l_{2i-1}^* \nu_{2i-1}^\dagger)(U_{\rm{K,e}}U_{\rm{K,o}})^t|\psi_0} = \sum_{i=1}^{N/2}l_{2i-1}\braket{\psi_0|\nu_{2i-1}|\psi_0}e^{-it\epsilon_{2i-1}} +\text{h.c.},\\
    \braket{\psi(t)|\tilde{Z}_{\rm R}|\psi(t)} &=& \sum_{i=1}^{N/2}\Braket{\psi_0|(U_{\rm{K,o}}^\dagger U_{\rm{K,e}}^\dagger)^t G(l_{ 2i} \nu_{2i} + l_{2i}^* \nu_{2i}^\dagger)(U_{\rm{K,e}}U_{\rm{K,o}})^t|\psi_0} = \sum_{i=1}^{N/2}l_{2i}\braket{\psi_0|G\nu_{2i}|\psi_0}e^{-it\epsilon_{2i}} +\text{h.c.},
\end{eqnarray}
where h.c. denotes Hermitian conjugate terms. Provided that the initial state $\ket{\psi_0}$ has finite overlaps with each eigenmode $\nu$, the Fourier transform of $\braket{\psi(t)|\tilde{Z}_{\rm L}|\psi(t)}$ ($\braket{\psi(t)|\tilde{Z}_{\rm R}|\psi(t)}$) reveals the spectrum $\{\epsilon_{2i-1}\}$ ($\{\epsilon_{2i}\}$) of the kicked Kitaev chain with inter-site coupling strength $J_{\rm e}$ ($J_{\rm o}$). 
Consequently, their combination leads to the complete spectrum $\{\epsilon_1, \epsilon_2, \dots, \epsilon_N\}$ of the system.

\section{Experimental Information}

\begin{figure}
    \centering
    \includegraphics[width=1\linewidth]{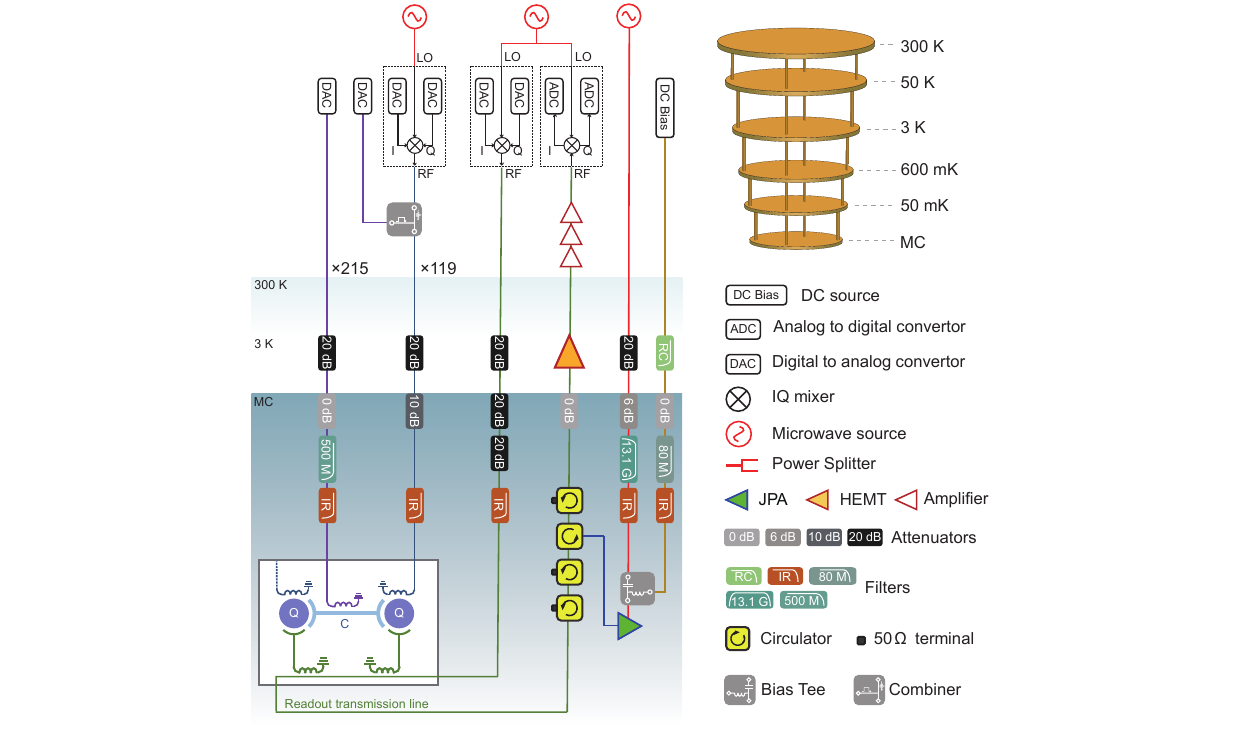}
    \caption{\textbf{Experimental setup.} Quantum processor, denoted by the white box at the bottom left corner, is mounted on the mixing chamber plate (MC) of the dilution refrigerator, whose base temperature is around $20$ mK. Fast Z-pulse lines (purple), microwave-drive lines (blue), and readout lines (green) connect room-temperature electronics to the processor for control and measurement. Details of the microwave components are provided in the legend on the right.}
    \label{fig:wiring}
\end{figure}

\begin{figure}
    \centering
    \includegraphics[width=1\linewidth]{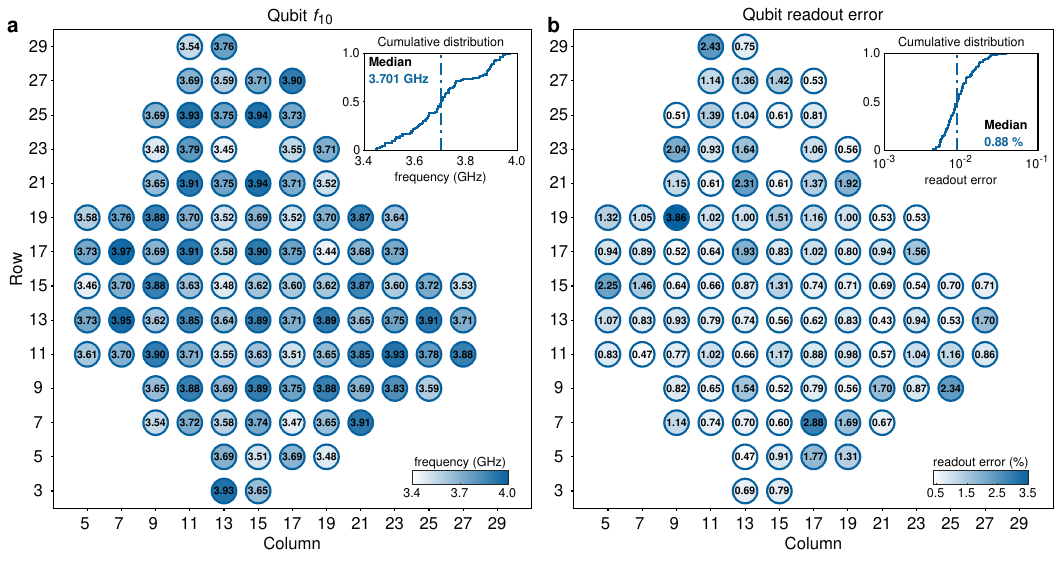}
    \caption{\textbf{Qubit idle frequency and readout error.} {\bf a}, Idle frequencies for all $100$ qubits used in our experiments. Inset shows the cumulative distribution, with the dashed line indicating the median value. {\bf b}, Qubit readout errors measured at idle frequencies in {\bf a}. The data for each qubit is the average of the errors when qubit in state $\ket{0}$ and $\ket{1}$.}
    \label{fig:f10_readout}
\end{figure}

\begin{figure}
    \centering
    \includegraphics[width=1\linewidth]{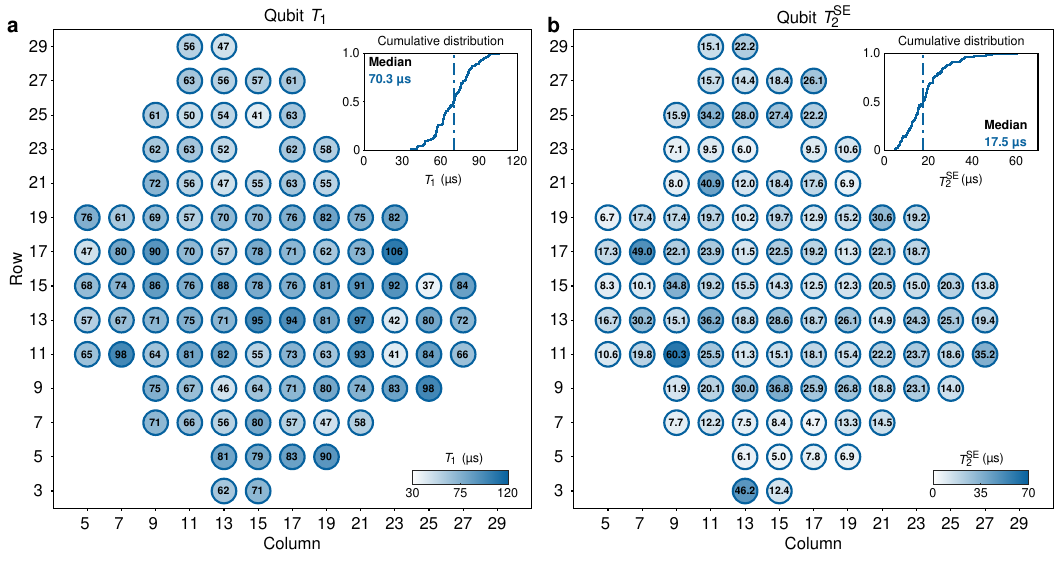}
    \caption{\textbf{Qubit coherence time.} {\bf a}, Energy relaxation time $T_1$ of the 100 qubits measured at idle frequencies, with median value of $70.3$ $\mu$s.  {\bf b}, Spin-echo dephasing time measured at idle frequencies, with median value of $17.5$ $\mu$s.}
    \label{fig:coherence_time}
\end{figure}

\subsection{Device performance}
As shown in Fig.~1a of the main text, we construct a one-dimensional chain with $100$ qubits on our 125-qubit quantum processor to implement the theoretical model. The wiring information for our device and room-temperature control electronics are sketched in Fig.~\ref{fig:wiring}. Fig.~\ref{fig:f10_readout}a displays the idle frequencies of the $100$ qubits, where we apply single-qubit gates in our experiments. The measured energy relaxation time $T_1$ and spin-echo dephasing time $T_{2}^{\rm SE}$ at idle frequencies in Fig.~\ref{fig:f10_readout}a are listed in Fig.~\ref{fig:coherence_time}, whose median values are $70.3$ $\mu$s and $17.5$ $\mu$s, respectively.
Fig.~\ref{fig:f10_readout}b shows the readout error for each qubit, which is defined by the average error of measuring $|0\rangle$ state and $|1\rangle$ state, which are measured by preparing the 100 qubits in random product states $\{|0\rangle, |1\rangle\}^{\otimes 100}$ and averaging for each qubit. We note that an extra microwave pulse that yields $|1\rangle\leftrightarrow|2\rangle$ transition is applied to each qubit before the readout microwave pulse to improve readout fidelity. The median value of readout errors is $0.88\%$.

\subsection{Gate calibration}
In our experiments, single-qubit gates are realized using $20$-ns microwave pulses with Gaussian envelope modulated with the derivative reduction by adiabatic gate (DRAG) pulse. We compile consecutive single-qubit gates into a single-qubit rotation $U3(\theta,\varphi,\lambda)$ with the following matrix form
\begin{equation}
    U3(\theta,\varphi,\lambda)=e^{-i\frac{\varphi}{2}\sigma_z}e^{-i\frac{\theta}{2}\sigma_y}e^{-i\frac{\lambda}{2}\sigma_z}=\left(\begin{array}{cc}
        \cos{\frac{\theta}{2}} & -e^{i\lambda}\sin{\frac{\theta}{2}}  \\
        e^{i\varphi}\sin{\frac{\theta}{2}} & e^{i(\varphi+\lambda)}\cos{\frac{\theta}{2}}
    \end{array}\right).
\end{equation}
In practice, $U3(\theta,\varphi,\lambda)$ gate is realized by a virtual phase gate and a subsequent XY rotation. CPhase($\phi$) gates ($\phi$ is the conditional phase; $\phi\in\{\pi,-0.4\}$ in our experiments) are realized by tuning the $|11\rangle$ and $|20\rangle$ states of the two interacting qubits near resonance and switching on the coupling between them for a certain duration. Experimentally, we achieve a specific $\phi$ by tuning the Z-pulse amplitudes (amplitudes of the flux pulses input from fast Z-pulse lines in Fig.~\ref{fig:wiring}) of the qubits and coupler. The pulse durations for CZ and CPhase($-0.4$) gates are $40$ ns and $34$ ns, respectively. As shown in Fig.~2a of the main text, implementing a single Trotter step $U(\delta t)$ requires four layers of CZ gates, two layer of CPhase($-0.4$) gates, and three layers of single-qubit gates, corresponding to a sequence of $288$-ns duration, which places high demands on gate fidelity.

\begin{figure}
    \centering
    \includegraphics[width=1\linewidth]{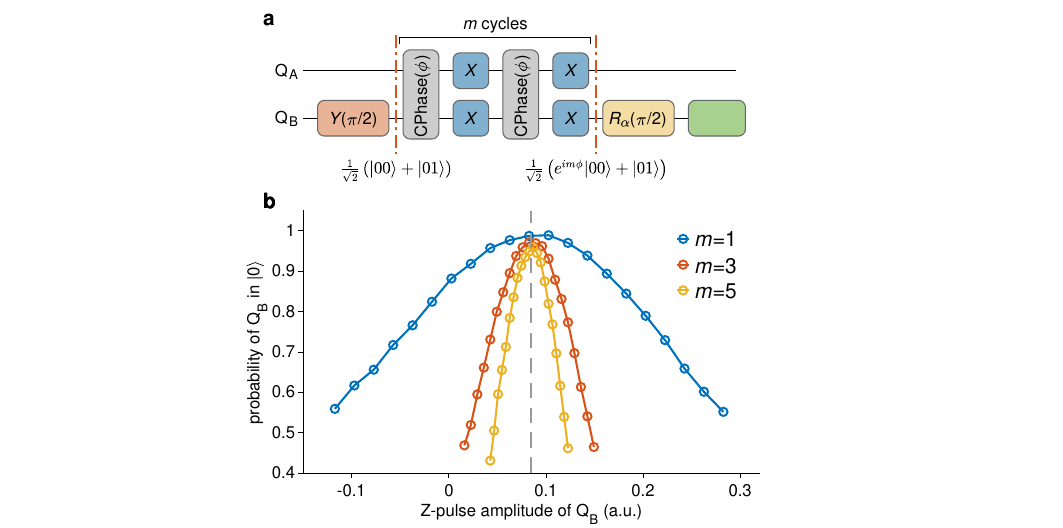}
    \caption{\textbf{CPhase gate calibration.} 
    {\bf a}, Quantum circuit for calibrating CPhase($\phi$) gate. The circuit begins with a $Y(\pi/2)$ gate acting on $Q_B$, which prepares a superposition state $\left(|00\rangle+|01\rangle\right)/\sqrt{2}$. Then, $m$ cycles of the interleaved circuit (including two layers of CPhase($\phi$) gates and two layers of single-qubit $X$ gates) are applied, where each cycle accumulates a phase of $\phi$ on $|00\rangle$ state. Finally, an $R_\alpha(\pi/2)$ gate (a single-qubit rotation around the $\alpha$-axis of Bloch sphere, where $\alpha$ refers to an equatorial rotation axis that has an angle $\alpha$ with respect to the $x$-axis) is applied to $Q_B$ before readout. The measured probability of $Q_B$ in $|0\rangle$ state is given by $\left[1+\cos{(\pi/2+\alpha+m\phi)}\right]/2$. In the ideal case without errors, by choosing $\alpha=-m\phi-\pi/2$, we should expect $Q_B$ to be in $|0\rangle$ state with probability of 1. 
    {\bf b}, Measured $|0\rangle$ state probability of $Q_B$ as a function of the Z-pulse amplitude of $Q_B$. Gray dashed line indicates the calibrated Z-pulse amplitude we choose for $Q_B$.}
    \label{fig:cphase_cal}
\end{figure}

For CPhase($\phi$) gates with $\phi$ close to $\pi$ (typically $|\phi-\pi|<2$ in our case), we usually calibrate Z-pulse amplitudes of qubits either by maximizing qubit entanglement~\cite{renExperimentalQuantumAdversarial2022} or Floquet calibration~\cite{neillAccuratelyComputingElectronic2021, miTimecrystallineEigenstateOrder2022}, and the Z-pulse amplitude of coupler is determined by minimizing leakage via measuring the probability of $|2\rangle$ state. While for the cases of $|\phi-\pi|\geq2$, the low-leakage area becomes broad, making it hard to identify the right parameters~\cite{foxenDemonstratingContinuousSet2020}. Here, we propose a sequence to calibrate CPhase gates with arbitrary conditional phase $\phi$ (Fig.~\ref{fig:cphase_cal}), which we use to calibrate the Z-pulse amplitudes of qubits for CZ gates. Specifically, for CPhase($-0.4$) gates, we start from the control parameters of CZ gates and use Floquet calibration to initialize the Z-pulse amplitudes of qubits so that the conditional phase is around $4.5$. Then, we use the sequence described in Fig.~\ref{fig:cphase_cal} to calibrate the Z-pulse amplitude of the coupler.

\subsection{Experiment circuits}
\begin{figure}
    \centering
    \includegraphics[width=1\linewidth]{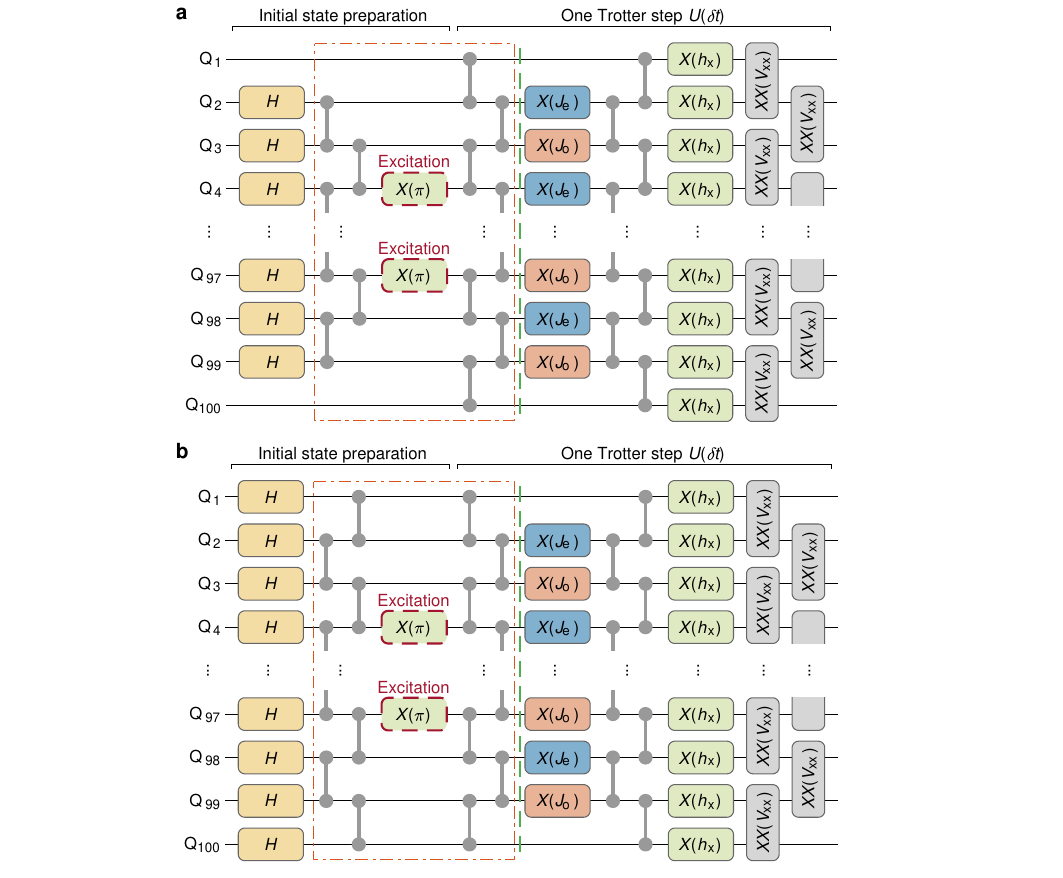}
    \caption{\textbf{Experimental circuits.} {\bf a}, Quantum circuit for measuring $\Tilde{Z}$ operator in Fig.~2 of the main text when $t=1$. The CZ gates in the orange dashed frame are canceled out after compilation in our experiments. The circuit on the right of the green dashed line corresponds to the unitary $U^t$ as the building block of the echo circuit $U_{\text{echo}}(t)=(U^\dagger)^t U^t$. {\bf b}, Quantum circuit for measuring $\Tilde{X}$ operator in Fig.~2 and Fig.~3 of the main text when $t=1$.}
    \label{fig:exp_circuit}
\end{figure}

In this section, we illustrate details about the main quantum circuits used in our experiments. Fig.~\ref{fig:exp_circuit}a and b show the quantum circuit for measuring $\Tilde{Z}$ and $\Tilde{X}$ operators in Fig.~2 and Fig.~3 of the main text at $t=1$ (circuits with $t > 1$ are constructed by repeating the Trotter step circuit). In the experimental realization, the circuit will be further compiled to reduce the circuit depth. For example, the excitation gate $X(\pi)$ acting on $Q_i$ in Fig.~\ref{fig:exp_circuit}a 
is compiled as three gates $\{Z(\pi),X(\pi),Z(\pi)\}$ acting on $\{Q_{i-1},Q_{i},Q_{i+1}\}$ before the first layer of CZ gates. Then, these gates are merged into the Hadamard gate layer. Further, the CZ layers surrounded by the orange dashed frame in Fig.~\ref{fig:exp_circuit}a are eliminated because the two CZ gates on edges are redundant (edge qubits are in $\ket{0}$ state) and the rest CZ gates in the bulk can cancel out with each other. This similar elimination is also applied to the circuit in Fig.~\ref{fig:exp_circuit}b as well. Note that the two-qubit gates in Trotter steps of $t>1$ can not be eliminated.

After compilation, we utilize Pauli twirling technique to suppress the damaging coherent noise, which is realized by inserting  random single-qubit gates before and after each two adjacent CZ layers. For the two edge qubits $Q_1$ and $Q_{100}$, we apply an extra $\pi$-rotation in the single-qubit rotation layer sandwiched by the four CZ layers in each Trotter step if there is no $\pi$-rotation in this layer, in order to protect the qubits from dephasing.

The echo evolution $U_{\text{echo}}(t)=(U^\dagger)^t U^t$ in the main text consists of $t$ steps of forward time evolution $U^t$ and the followed $t$ steps of backward time evolution $(U^\dagger)^t$. Thus, the decay of echo evolution characterizes the accumulated circuit errors after initial state preparation. In our experiments, the circuit of $U^t$ is defined as the circuit after the green dashed line in Fig.~\ref{fig:exp_circuit}, where the Pauli twirling gates are also integrated. 

\subsection{Mitigation of leakage error}
The experimental circuit for measuring the energy spectrum in Fig.~4 of the main text contains up to $150$ Trotter steps with $900$ layers of two-qubit gates. Such a long sequence makes the leakage to $\ket{2}$ state non-negligible, which systematically causes a slightly higher probability of qubit in $|1\rangle$ state. To suppress this effect, we also measure the $|2\rangle$ state probability in the many-body spectroscopy experiment and correct the experimental data with the following procedure.
\begin{itemize}
    \item First, we run the experimental circuits and measure the probability of the qubit $\vec{P}_{\rm exp}=\left(p_{0, {\rm exp}},p_{1, {\rm exp}},p_{2, {\rm exp}}\right)^{\rm T}$, where $p_{\alpha, {\rm exp}}$ is the measured probability of qubit in $\ket{\alpha}$ state $(\alpha\in\{0, 1, 2\})$.
    \item Then, we use three-level readout correction matrix $\mathcal{C}$ to mitigate readout errors, which satisfies $\mathcal{C}\vec{P}_{\rm ideal}=\vec{P}_{\rm exp}$ with the definition as below
    \begin{equation}
        \mathcal{C}=\left(\begin{array}{ccc}
            1-\epsilon_{0\rightarrow1}-\epsilon_{0\rightarrow2} &  \epsilon_{1\rightarrow0} & \epsilon_{2\rightarrow0}  \\
            \epsilon_{0\rightarrow1}  &1-\epsilon_{1\rightarrow0}-\epsilon_{1\rightarrow2} & \epsilon_{2\rightarrow1} \\
            \epsilon_{0\rightarrow2} & \epsilon_{1\rightarrow2} & 1-\epsilon_{2\rightarrow0}-\epsilon_{2\rightarrow1}
        \end{array}\right),
    \end{equation}
    where $\epsilon_{i\rightarrow j}$ refers to the measured probability of a qubit in $|j\rangle$ when it is prepared in $|i\rangle$ state. The matrix elements of $\mathcal{C}$ are benchmarked with a separate experiment. Thus, the estimated $\vec{P}_{\rm ideal}$ after readout correction is given by $\vec{P}_{\rm corr}=\mathcal{C}^{-1}_{\rm exp}\vec{P}_{\rm exp}$, where $\mathcal{C}^{-1}_{\rm exp}$ is experimentally measured readout correction matrix.
    \item To eliminate the state leakage error, we discard the probability in $|2\rangle$ state and normalize the measured probability in the computational space with the following equation
    \begin{equation}
        p_{0,{\rm norm}}=\frac{p_{0,{\rm corr}}}{p_{0, {\rm corr}}+p_{1, {\rm corr}}}, p_{1,{\rm norm}}=\frac{p_{1, {\rm corr}}}{p_{0, {\rm corr}}+p_{1, {\rm corr}}}.
    \end{equation}
    \item Finally, we calculate the expectation value $\Tilde{\braket{Z}}=p_{0,{\rm norm}}-p_{1, {\rm norm}}$. In Fig.~4a of the main text, we show three representative time-domain signals of the measured $\braket{\Tilde{Z}_{\rm L}}$ and $\braket{\Tilde{Z}_{\rm R}}$.
\end{itemize}

\subsection{Quantum state tomography}
In this section, we provide details about how we obtain the density matrix in the main text. To investigate the dynamics of logical state fidelity (Fig.~5a of the main text), it is natural to perform two-qubit logical state tomography on the two edge states and reconstruct the full logical density matrix $\rho_{\rm logic}$, then calculate the logical state fidelity $F\left(\rho_{\rm logic},\rho_{\rm ideal}\right)$ ($\rho_{\rm ideal}$ is the ideal density matrix of logical state) by the following formula
\begin{equation}
    \rho_{\rm logic}=\frac{1}{4}\left(\sum_{\Tilde{P}_{\rm L},\Tilde{P}_{\rm R}\in \{I,\Tilde{X},\Tilde{Y},\Tilde{Z}\}}\braket{\Tilde{P}_{\rm L}\Tilde{P}_{\rm R}}\Tilde{P}_{\rm L}\Tilde{P}_{\rm R}\right),
    \label{eq:dm}
\end{equation}
\begin{equation}
    F\left(\rho_{\rm logic},\rho_{\rm ideal}\right)={\rm tr}\left(\sqrt{\sqrt{\rho_{\rm logic}}\rho_{\rm ideal}\sqrt{\rho_{\rm logic}}}\right)^2,
\end{equation}
where $\Tilde{X},\Tilde{Y},\Tilde{Z}$ are logical Pauli operators. To obtain the fidelity of logical Bell state ($\Tilde{\ket{0}}_{\rm L}\Tilde{\ket{0}}_{\rm R}+\rm{i}\Tilde{\ket{1}}_{\rm L}\Tilde{\ket{1}}_{\rm R}$) in Fig.~5a of the main text, we simplify the measurement of $F\left(\rho_{\rm logic},\rho_{\rm ideal}\right)$ by only probing three logical Pauli strings
\begin{equation}
\begin{aligned}
    F_{\rm Bell}& =\frac{1}{4}\left(1+\braket{\Tilde{X}_{\rm L}\Tilde{Y}_{\rm R}}+\braket{\Tilde{Y}_{\rm L}\Tilde{X}_{\rm R}}+\braket{\Tilde{Z}_{\rm L}\Tilde{Z}_{\rm R}}\right).
\end{aligned}
\end{equation}
However, in Fig.~5b of the main text and Extended Data Fig.~4b, we measure all the logical Pauli operators in Eq. \ref{eq:dm} to reconstruct all the elements of the full density matrices. In Extended Data Fig.~4c, we perform full quantum state tomography of the four physical qubits $Q_1,Q_2,Q_{99},Q_{100}$ at edges and reconstruct its density matrix with a similar method of Eq. \ref{eq:dm}. The reconstructed full density matrices are further validated in the constraints of Hermitian, unit trace, and positive semi-definite with the method described in Ref.~\cite{smolinEfficientMethodComputing2012}.

\bibliography{suppRef.bib}